\documentclass[11pt]{arxiv-article}

\numberwithin{equation}{section}

\bibliography{References}

\usepackage{hyperref}
\newcommand{\OfficialTitle}{
  Selected Topics in the Large Quantum Number Expansion
}

\title{\setstretch{1.4}
  {\color{Thoughtless}\Huge\textbf{\dosserif\OfficialTitle}}
}

\hypersetup{pdfauthor={Luis Àlvarez Gaumé and Domenico Orlando and Susanne Reffert},pdftitle={\OfficialTitle},%
  colorlinks=true,linkcolor=ThoughtYouWere,citecolor=ThoughtYouWere,urlcolor=ThoughtYouWere}

\usepackage{Definitions}
\usepackage{slashed}
\newcommand*{\Jexp}{\mathcal{J}}

\allowdisplaybreaks[1]

\RequirePackage{ytableau}
\RequirePackage{booktabs}
\RequirePackage{array}
\newcolumntype{L}[1]{>{\raggedright\let\newline\\\arraybackslash\hspace{0pt}}m{#1}}

\makeatletter
\renewcommand*\env@matrix[1][*\c@MaxMatrixCols c]{%
  \hskip -\arraycolsep
  \let\@ifnextchar\new@ifnextchar
  \array{#1}}
\makeatother
\RequirePackage{tikz-cd}
\RequirePackage{pgfplots}

\author{%
  \begin{minipage}{.97\linewidth}
    \vspace{1cm}
    \begin{center} \dosserif%
      {\small
         \textbf{Luis~Alvarez-Gaume}\textsuperscript{\ding{71}\ding{95}},
         \textbf{Domenico Orlando}\textsuperscript{\ding{72}\ding{73}} and
         \textbf{Susanne Reffert}\textsuperscript{\ding{73}} 
         }
    \end{center}
    \vspace{1cm}
     \authorBlock{\ding{71}}{\dosserif{} Simons Center for Geometry and Physics,\\ State University of New York
       Stony Brook,\\
       NY--11794--3636, USA}
     \authorBlock{\ding{95}}{Theory Department -- CERN,\\ 
 \textsc{ch}-1211 Geneva 23, Switzerland}
     \authorBlock{\ding{72}}{\dosserif{} INFN sezione di Torino | Arnold--Regge Center\\
      via Pietro Giuria 1, 10125 Turin, Italy}
    \authorBlock{\ding{73}}{\dosserif{} Albert Einstein Center for Fundamental Physics\\
      Institute for Theoretical Physics, University of Bern,\\
      Sidlerstrasse 5, CH-3012 Bern, Switzerland}
  \end{minipage}
}

\date{}

\begin{document}

\begin{titlepage}

  \newgeometry{top=23.1mm,bottom=46.1mm,left=34.6mm,right=34.6mm}

  \maketitle

  \thispagestyle{empty}

  \vfill\dosserif{}

  \abstract{\normalfont{}\noindent{}In this review we study quantum field theories and conformal field theories with global symmetries in the limit of large charge for some of the generators of the symmetry group. At low energy the sectors of the theory with large charge are described by a hybrid form of Goldstone’s theorem, involving its relativistic and non-relativistic forms. The associated effective field theory in the infrared allows the computation of anomalous dimensions, and operator product expansion coefficients in a well defined expansion in inverse powers of the global charge. This applies even when the initial theory does not have a reliable semiclassical approximation. The large quantum number expansion complements, and may provide an alternative approach to the bootstrap and numerical treatments. We will present some general features of the symmetry breaking patterns and the low-energy effective actions, and a fairly large number of examples exhibiting the salient features of this method.
  }

\vfill

\end{titlepage}

\restoregeometry{}

\tableofcontents 
 
\setstretch{1.2}

\section{Introduction}%
\label{sec:introduction}

The last decades have witnessed great progress in our understanding of the structure of the space of \acp{qft} -- generic theories, generic amplitudes, generic spectra.
Unravelling its geometry and topology will very likely 
have deep implications in mathematics, quantum gravity, string theory, cosmology and
condensed matter physics.
One of the basic tools to investigate its properties
is the study of the flows of the Wilsonian \ac{rg}
(for a review and references see~\cite{Komargodski:2011xv}).
If we consider the space of unitary, local, relativistic \acp{qft} in various dimensions, the 
limiting behavior in the \ac{ir} and the \ac{uv} is believed to be given by fixed
points.  Without those assumptions, the long-term behavior of the flows could
be more elaborate and display features such as limit cycles or strange attractors. 

The fixed points are \acp{cft} -- they represent the beacons of light orienting us in the uncharted and hostile ocean of \acp{qft}.
Unfortunately most fixed points correspond to theories at strong coupling away from any simplifying limit that would allow
some form of semiclassical analysis.
They may be studied using numerical techniques like the Monte Carlo (\ac{rg}~\cite{Campostrini:2000iw,Campostrini:2002ky} and references therein).
The conformal bootstrap~\cite{Polyakov:1970xd,Ferrara:1973vz,Belavin:1984vu,Dolan:2000ut,Dolan:2003hv} had a major renaissance after~\cite{Rattazzi:2008pe} that led to many deep results in \acp{cft} beyond two spacetime dimensions~\cite{Simmons-Duffin:2016gjk}.

\medskip

The approach we present in this review is the large quantum number expansion which, although limited to theories with global symmetries, allows the analytic treatment of otherwise inaccessible systems. In many cases, it complements bootstrap techniques, and can offer results not easily accessible using bootstrap or numerical methods in others.\footnote{A very interesting study of the bootstrap approach for theories with global charges can be found in~\cite{Jafferis:2017zna}.}
There is a long history to this approach.\footnote{Simeon Hellerman often starts his talks by referring all the way back to Democritus and his atomic hypothesis.}
One could start with Bohr's correspondence principle and the \ac{wkb} approximation in quantum mechanics, or the Regge limit in the hadron spectrum, or the large $N$-limit in \ac{qft}~\cite{Moshe:2003xn}.
More recently, examples abound: the large $R$-charge limit in $\mathcal{N}=4$ super-Yang Mills theory~\cite{Berenstein:2002jq}, the large-spin limit in \ac{qcd}~\cite{Basso:2006nk} and conformal gauge theories~\cite{Alday:2007mf}, large-spin expansions in general \acp{cft}~\cite{Komargodski:2012ek,Fitzpatrick:2012yx}, large-spin expansions in hadrons~\cite{Hellerman:2013kba,Caron-Huot:2016icg}, or extremal correlators in $\mathcal{N}=2$ theories and matrix theory~\cite{Grassi:2019txd}, just to mention a few.

\medskip

In~\cite{Hellerman:2015nra}, a systematic large-charge expansion was presented for generic systems with Abelian global symmetries, and the authors obtained general results for the scaling dimensions of spinless charged operators in the large-charge limit at the three-dimensional \ac{wf} fixed point.
A new form of Goldstone's theorem appears as a consequence of restricting to the large-charge sector of the Hilbert space of states.
The minimal-energy state of large charge and zero spin (see below) corresponds to a state breaking boost invariance, and for which the conserved effective time translation operator includes also specific time-dependent charge rotations in field space.
This combined breaking of boosts and charge invariance by the charged ground state leads to a novel application of Goldstone's theorem.

\medskip

The large-$Q$ limit of the $O(2)$ \acl{wf} theory in the \ac{ir} is dominated by a Goldstone-like state whose dispersion relation is $ E=\frac{1}{\sqrt{d}} k$ (where $d$ is the space dimension -- this case is explored in Sections~\ref{sec:original-paper} and~\ref{sec:O2n}), and with a well-defined \ac{eft}, where most couplings are suppressed by inverse powers of $Q^{\alpha}$, where $\alpha>0$ depends on the spacetime dimension.
In~\cite{Monin:2016jmo} the results of~\cite{Hellerman:2015nra} were interpreted in terms of a finite-density superfluid phase for the theory quantized on the cylinder.
The results of~\cite{Hellerman:2015nra} were extended in~\cite{Alvarez-Gaume:2016vff} to the non-Abelian case, in particular, the $O(2n)$ theory.
It was found that the large-charge sector of the theory presents a hybrid version between the relativistic and non-relativistic forms of Goldstone's theorem.
In particular the low-energy \ac{eft} contains a ``relativistic'' Goldstone boson with speed of light $\frac{1}{\sqrt{d}} $ as above, together with $n-1$ non-relativistic Goldstone bosons with a quadratic dispersion relation $E \sim k^2$.
This is in part a realization of the results of~\cite{Nielsen:1975hm} in the context of the nonrelativistic form of Goldstone's theorem.
Similar variations on this theorem in non-relativistic contexts were found in the description of kaon condensation in \ac{qcd}~\cite{Schafer:2001bq,Miransky:2001tw}, and in the phenomenon of \ac{ssp}~\cite{Nicolis:2011pv,Nicolis:2012vf}.

The breaking of boost invariance in the charge sectors of a \ac{qft} leads to a very rich phenomenology, including the derivation of \acp{eft} in the non-relativistic domain pioneered by Leutwyler in~\cite{Leutwyler:1993gf}, whose equations were finally solved in full generality by Murayama and Watanabe~\cite{Watanabe:2014fva}.
The interplay between the lack of relativistic invariance, ground-state charge densities, the special counting of Goldstone states, the properties of the \ac{eft} and its generic geometric make-up is an intricate and fascinating subject that we review in Section~\refstring{sec:Goldstone}, where many results which have appeared over decades scattered through the literature are collected.
The subsequent sections will show explicitly how some low-energy properties of the theories we work with incorporate this generic analysis.

\medskip

A useful way of looking at the charged sector of \acp{cft} is to use the state-operator correspondence and radial quantisation~\cite{Hellerman:2015nra,Monin:2016jmo}. In that context, we can insert a primary spinless operator with the lowest dimension and charge $Q$ at the origin, (and another with the opposite charge at infinity). In the spacelike sphere with radius $R$ this will generate a state describing a homogeneous charge density with lowest energy in the \ac{cft}.
In the conformal group acting on the spacetime $\setR^{d+1}$, the group of translations fixes infinity but moves the origin, while the group of special conformal transformations fixes the origin but translates the point at infinity.
In the cylinder picture, the state corresponding to the operator inserted at the center of the sphere $S^d$ will be manifestly invariant under the standard $SO(d)$ group of rotations; moreover, since the operator at the origin is primary, it is annihilated by the special conformal transformations, and the state will be invariant under the full isometry group of the sphere, $SO(d+1)$.
In the symmetry-breaking pattern we observe in the case $O(2n)$ global symmetry, the kinematical part will exhibit the breaking of the conformal group $SO(d+2,1) \rightarrow SO(d+1)$.
For the internal symmetry an interesting pattern appears.
The time translations of the problem are redefined as a linear combination of the dilatation operator and a combination of the broken generators.
The natural clock in this sector involves uniform rotations in field space along symmetry generators (this is reminiscent of \ac{ssp}~\cite{Nicolis:2011pv}). The details will be spelled out in Sections~\refstring{sec:original-paper}~and~\refstring{sec:O2n}.

Since we want to invoke the state-operator correspondence of \ac{cft}, we will consider the microscopic \ac{qft} on a space $R \times \Sigma_d$, with conformal coupling, where $\Sigma_d$ is a convenient $d$-dimensional manifold.
Minimal-energy states with a given charge are in the case of $\Sigma_d=S^d$ related to scaling dimensions via radial quantization.
For constant charge density, the computations are essentially the same for any compact $\Sigma_d$ once we include the conformal coupling to the curvature of the surface. Hence, to get the leading behavior and some generic properties of the semiclassical large-charge ground states, it is often convenient to work with $\Sigma_d$ being a flat torus, or a box with adequate boundary conditions.

\medskip

We can briefly illustrate the previous discussion by considering an $O(2n)$-invariant theory with a set of scalar fields in the vector representation.
The details appear in Sections~\ref{sec:original-paper} and \ref{sec:O2n}.
For $\Sigma_d$ we can take a box of characteristic size $L$ and volume $L^d$.
In a state with total charge $Q$, we have a new scale given by the charge density, $\Lambda_Q= Q^{1/d}/L$.
In most cases the properties of the theory at scales in between the \ac{uv} scales and $\Lambda_Q$ are not accessible to semiclassical expansions or an \ac{eft} treatment.
However, as the Wilsonian scale $\Lambda$ moves below $\Lambda_Q$, and the theory moves towards the \ac{ir} fixed point, things simplify substantially.
In the $O(2n)$ case we can independently fix the charge of the generators of the Cartan subalgebra.
For this it is convenient to represent $\setR^{2n}$ as a collection of $n$ $2$-planes.
The independent $O(2)$ rotations in each plane generate the Cartan subalgebra.
We charge each plane with a large charge $q_i$, $Q=\sum_i \abs{q_i}$, and describe 
the fields in the vector representation of $O(2n)$ with $n$ complex fields $\phi_i = a_i e^{i \chi_i}/\sqrt{2}$.
As we move $\Lambda \ll \Lambda_Q$ and look for the minimal energy configurations under these conditions, the centrifugal barrier in field space freezes the radial modes $a_i$ to some \ac{vev} \(A_i\) determined by the charge assignments, and synchronizes the circular motion in each plane, $\chi_i = \mu t$, which is the same for all $i$.
Hence, the motion in the projected planes corresponds to circles of different radii, but equal periods.
The $O(2n)$-invariant combination $\rho=\sum_i A_i^2$ and $\mu$ take simple forms as we approach the scale-invariant region (the scale-invariant potential has the form $(\phi_a \phi_a)^{\frac{d+1}{d-1}}$).
Their leading large-charge behavior is $\rho \sim Q^{(d-1)/d}$, $ \mu\sim Q^{1/d}$.
The phases represent the Goldstone bosons of the low-energy \ac{eft}, where the count of relativistic vs non-relativistic \acp{dof} is non-trivial.
We will see that there is one relativistic-like Goldstone boson as in the $O(2)$ theory, and $n-1$ non-relativistic ones.
The \ac{eft} describing these low-energy excitations has universal properties, as in the case of pions in \ac{qcd}, and there is a number of non-trivial predictions that can be made in such theories.
For more elaborate models with various vector or matrix fields, the results are very similar.
As we move well below $\Lambda_Q$ the many ``moduli'' fields freeze in their \acp{vev} determined by the centrifugal barrier, and the low-energy excitations are described by the hybrid version of Goldstone's theorem.
Notice we are not considering the most general formulation of symmetry breaking in non-relativistic theories.
The \acp{qft} we study are all local and relativistic in the \ac{uv}: it is the restriction to the study of large-charge sectors that breaks boost invariance.

It is also possible to consider charged states where space homogeneity does not provide those with the lowest energy.
This happens for charged states with spin~\cite{Cuomo:2017vzg}, where the lowest-energy state is represented by a charged superfluid with vortices, or when in the $O(2n)$ theory with $n > 1$ one considers states charged along independent directions in the symmetry group.
This case is presented in Section~\ref{sec:NLSM-O2n} for an $O(4)$ theory, where the ground state has cylindrical symmetry on the sphere with two vortex-like configurations at the poles.
We have not studied more general configurations and groups, but this is a subject worth pursuing.
Furthermore, there are some interesting, preliminary results for matrix models that we present in Section~\refstring{sec:matrixModels}.
Also this subject remains so far largely unexplored.

The large charge or large spin limit of \acp{qft} has been studied in theories with supercharges.
The larger their number, the sharper the results.
We cannot do justice to the extensive literature on supersymmetric models.
A sampling of results appears in Section~\ref{sec:susyModels}.

\medskip

This review is not exhaustive.
We concentrate on some aspects of the theory that we know better, and feel more comfortable with.
In the last section we will list a series of other approaches that will not be covered here.
We apologize to all those authors whose work is not reviewed or cited.
It is not a lack of appreciation, but simply an expression of our limited competence in this vast subject.

\bigskip

The detailed plan of this review is as follows.
In Section~\refstring{sec:Goldstone}, we revisit the basics of \ac{ssb}.
In Section~\ref{sec:operator-approach}, we follow the standard operator approach to develop \ac{ssb} in non-relativistic systems.
In order to be able to move on to the second approach to \ac{ssb} via effective actions, we next discuss the non-linear realization of broken symmetries in Section~\ref{sec:non-linear-realization}.
This is followed by a brief review of the \ac{ccwz} construction (Section~\ref{sec:CCWZ}).
Finally, we discuss the Leutwyler equations (Section~\ref{sec:Leutwyler}), including the Murayama--Watanabe solution to them, the counting of the Goldstone bosons, and its geometrical interpretation.
In Section~\refstring{sec:original-paper}, we develop the large-charge expansion using the simplest example of the O(2) model.
We first present a \ac{rg} flow argument in~Section~\ref{sec:top-down} and then move on to a bottom-up argument using the dilaton dressing in Section~\ref{sec:bottom-up-approach}.
In Section~\ref{sec:linear-sigma-model}, we identify the classical ground state, the fluctuations around it and the symmetry-breaking pattern in the \ac{lsm} approach.
In Section~\ref{sec:non-linear-sigma-model}, we develop the \ac{nlsm} based on our insights from the \ac{lsm}.
The quantum corrections to the semi-classical results are discussed in Section~\ref{sec:conformal-goldstone}.
In Section~\ref{sec:observables}, we finally compute the conformal data for the lowest states of charge $Q$.
In Section~\refstring{sec:O2n}, we generalize our approach to the O(2n) vector model, which displays a much richer structure due to its non-Abelian symmetry group.
In a first step, we discuss the meaning of charge fixing for a non-Abelian symmetry (Section~\ref{sec:charge-fixing-O(2n)}).
We again discuss the \ac{lsm} description in Section~\ref{sec:O2n-lsm} and then the fluctuations, which here comprise both type-I and type-II Goldstones (Section~\ref{sec:O2n-Goldstones}).
In Section~\ref{sec:algebraic-homo-On}, we briefly reinterpret our results in algebraic terms.
In Section~\ref{sec:NLSM-O2n}, we comment on the \ac{nlsm} approach to the O(2n) vector model, albeit not in full generality.
A new feature at large charge due to the larger number of degrees of freedom is the existence of ground states with a spatial inhomogeneity, which we discuss in Section~\ref{sec:inhomGS}.
Lastly, we discuss the O(2n) vector model in the limit of large $n$ (Section~\ref{sec:O2n-largeN}).
After having developed the large-charge expansion of the O(2n) model in great detail, we move on to some other notable applications, but without the same level of detail and generality.
In Section~\refstring{sec:matrixModels}, we discuss examples of matrix models at large charge, which display a richer phenomenology than the vector models.
Section~\ref{sec:SU2SU2} deals with the SU(2) $\times$ SU(2) matrix model, which is equivalent to the O(4) vector model, in Section~\ref{sec:SU(N)matrix} we discuss SU(N) matrix models, %
and in Section~\ref{sec:asymptotically-safe} we discuss a phenomenological model with a matrix-valued scalar field.
In Section~\refstring{sec:susyModels}, we discuss three instances of \ac{scft} at large charge, namely the $W=\Phi^3$ model in 2+1 dimensions (Section~\ref{sec:Phi3}), the XYZ model in 2+1 dimensions (Section~\ref{sec:XYZ}), and $\mathcal{N}=2$ models in 3+1 dimensions with a one-dimensional moduli space (Section~\ref{sec:4d-1dmoduli}).
Here, we find in particular that the behavior of models with and without a moduli space is qualitatively very different.
In Section~\refstring{sec:alternative}, we comment briefly on alternative approaches to the large-charge expansion.%
Finally, we give some mathematical background needed for Section~\refstring{sec:Goldstone} in Appendix~\ref{sec:MathematicalBackground}; discuss how finite-volume effects are controlled at large charge in Appendix~\ref{sec:finite-volume}; and comment on the difference between fixing the charge and fixing the chemical potential in Appendix~\ref{sec:finite-mu-vs-Q}.

\section{Revisiting Goldstone's theorem}
\label{sec:Goldstone}

The notion of symmetries in physical systems and their realization at the classical and quantum level plays a central role in modern physics~\cite{Weinberg:1996kr}.
In this review we study \acp{qft} invariant under a global symmetry in the limit of large charge.
As we will show, there is an intimate interplay with the phenomena of symmetry breaking in non-relativistic systems (see for instance~\cite{Watanabe:2019xul,Burgess:1998ku} and references therein).
This is a vast subject which has not yet revealed all of its secrets.
There are new forms of symmetry breaking that are discovered constantly~\cite{Sachdev}.
In our case we consider theories that in the \ac{uv} are local relativistic field theories, and restrict our attention to the sector where some of the global charges have fixed, large values~\cite{Hellerman:2015nra}.
In the low-energy limit those sectors are described by non-relativistic \acp{eft} with a surprisingly rich phenomenology.
We will show under rather general assumptions that the low-energy behavior in this limit leads to interesting implementations of Goldstone's theorem~\cite{Nambu:1960tm,Nambu:1961tp,Nambu:1961fr,Goldstone:1961eq,Goldstone:1962es}, namely a hybrid of the relativistic and non-relativistic versions~\cite{Gilbert:1964iy,Lange:1965zz,Lange:1966zz,Guralnik:1967zz,Nielsen:1975hm,Brauner:2010wm,Leutwyler:1993gf,Watanabe:2014fva}.

The study of field theories in fixed-charge sectors is similar to the case of theories with finite density and chemical potential (see for example~\cite{Kapusta:1981aa,Schafer:2001bq,Miransky:2001tw,Blaschke:2004cs}). As we will show using a simple example in Appendix~\ref{sec:finite-mu-vs-Q}, there are however differences between these cases.
There are also many similarities with the study of \ac{ssp}~\cite{Nicolis:2011pv,Nicolis:2012vf}, where the breaking of Lorentz invariance is linked to the breaking of a global charge.
All these cases give rise to interesting variations on Goldstone's theorem.

\medskip
There is a basic difference between the implementation of symmetries in the relativistic and non-relativistic domains.
In the former, Coleman's theorem~\cite{doi:10.1063/1.1931207,DellAntonio} holds: \emph{the symmetries of the vacuum are the symmetries of the world}.
Coleman provided a rather convincing heuristic argument that was made rigorous in~\cite{DellAntonio}.
Given a vector current \(j^\mu(x)\) we can construct the charge:
\begin{equation}
  Q = \int_V \dd[3]{x} j^0(t,x).
\end{equation}
If one assumes that \(Q\) annihilates the ground state \(Q \ket{0} = 0\), then the postulates of local relativistic field theory imply that
\begin{equation}
  \del_\mu j^\mu(x) = 0,
\end{equation}
\emph{i.e.} the current is conserved and therefore the charge \(Q\) commutes with the Hamiltonian.
In proving this result it is necessary to use the Johnson--Federbush theorem~\cite{Federbush:1960zz} implying that if a local operator annihilates the vacuum \(A(x)\ket{0} = 0\) for all \(x\), then \(A(x) = 0\).
This property can be heuristically understood because for local operators in the interaction representation, \(A(x)\) always contains contributions with only creation operators.

Once Lorentz invariance no longer holds, the theorem does not apply, current
conservation does not follow from  $Q \ket{0}=0$.  This may also be related to the relative simplicity of Goldstone's theorem in relativistic
theories on the one hand, and the remarkable richness of \ac{ssb} in the non-relativistic domain on the other.  Relativistic invariance also poses stringent
constraints on the dispersion relation of the low-energy excitations.

In the traditional description of the implementation of continuous symmetries in \ac{qft}~\cite{Weinberg:1996kr}, one considers the Wigner--Weyl mode, where \(\comm{Q}{H} = 0\), \(Q \ket{0} = 0\), \emph{i.e.} the charge annihilates the ground state and commutes with the Hamiltonian.\footnote{To be more precise we could require that $|0\rangle$ be an eigenstate of $Q$.  In view of the
fact that we will be considering sectors of large charge, this detail is particularly pertinent.  We thank the referee for pointing this
out to us.}
The Nambu--Goldstone mode appears when \(\comm{H}{Q} =0\) but \(Q \ket{0} \neq 0\).
The study of theories with global symmetries in the large-charge sector limit falls in between: the full theory is symmetric under a global symmetry group, but in the sector with large charge, the lowest-energy state satisfies \(Q\ket{0} \neq 0\).
It is in this sector that the low-energy excitations can be described by a mixture of the relativistic and non-relativistic versions of symmetry breaking.
We find it instructive to divide the approach to \ac{ssb}. We first follow the traditional operator approach~\cite{Weinberg:1996kr}.
It is conceptually clear and easy to apply both in the relativistic and non-relativistic cases.
After that we turn to the theory of non-linear realization of symmetries and the construction of effective actions.
The two approaches complement each other, thus providing a more complete picture of symmetry breaking.

Before proceeding with the technical details, we would like to highlight two important issues.
\begin{description}
\item[Goldstone boson count.] A simplifying aspect of the relativistic description in a relativistic \ac{qft} is the fact that to every broken generator, a Goldstone boson is associated.
  This follows from the general properties of the generating function for 1\textsc{pi} diagrams or effective action \(\Gamma[\Phi]\).
  By translation and Lorentz invariance we can write \(\Gamma[\Phi]\) as an expansion in derivatives~\cite{Weinberg:1996kr}:
  \begin{equation}
    \Gamma[\Phi^a] = \int \dd[3]{x} \pqty{-V_{\text{eff}}(\Phi) + G_{ab}(\Phi) \del_\mu \Phi^a \del^\mu \Phi^b + \dots}.
  \end{equation}
  In the vacuum the minimum of the effective potential describes the \ac{vev} of the fields \(\Phi^a\).
  The \ac{vev} \(\ev{\Phi^a}\) are spacetime independent.
  If the theory is invariant under some symmetry transformation,
  \begin{align}
    \label{eq:effective-action-invariance}
    \delta \Phi^a &= \epsilon^A (T_A)\indices{^a_b} \Phi^b, & \pdv{V_{\text{eff}}}{\Phi^a} \delta \Phi^a &= 0,
  \end{align}
  where the \(T_A\)s are the generators of the symmetry group.
  The broken generators are such that \(T_A \ev{\Phi} \neq 0\).
  The mass matrix of the theory (up to a possible wavefunction renormalization) is then given by
  \begin{equation}
    M^2_{ab} = \eval{\pdv{V_{\text{eff}}}{\Phi^a}{\Phi^b}}_{\Phi^a = \ev{\Phi^a}} ,
  \end{equation}
  and it follows immediately from differentiating the second expression in Eq.~(\ref{eq:effective-action-invariance}) and evaluating it at \(\ev{\Phi^a}\) that for each broken generator there is an associated massless scalar field.
  The count of fields is more subtle in the non-relativistic case as we will see later when reviewing the work of Nielsen and Chadha~\cite{Nielsen:1975hm}.
  The actual counting rule appeared (with only a partial proof) in~\cite{Watanabe:2011ec}. It was eventually proven by Watanabe and Murayama~\cite{Watanabe:2014fva}, and simultaneously by Hidaka~\cite{Hidaka:2012ym}.
\item[Finite charge and Lagrange multipliers.] Many theories can be described by a set of variables that satisfy constraints.
  For instance to describe the motion of a particle on a sphere of radius \(a\) in \(n\) dimensions we can use the \(n\) variables \(x_1, \dots, x_n\) in \(\setR^n\) satisfying the condition \(x_1^2 + \dots + x_n^2 = a^2\), where \(a\) could generically depend on time.
  These constraints are called \emph{holonomic} and they can be implemented by including Lagrange multipliers in the theory.
  This method works both at the classical and quantum levels.
  If however the constraint is \emph{non-holonomic}, \emph{i.e.} if it also contains velocities \(f_{\alpha}(q, \dot q, t) = 0\), the naive implementation with Lagrange multipliers leads often to the wrong physics even at the classical level.
  In some cases, like in gauge theories, these constraints are implicitly contained in the very formulation of the theory, and in general Dirac's theory of constrained systems deals with them satisfactorily.
  If, on the other hand, these non-holonomic constraints are imposed in an \emph{ad hoc} manner with multipliers, the results are often wrong.
  First, the addition of such terms changes the canonical structure in an undesirable way.
  Second, when migrating from Newton's equation to a variational description of the dynamics, \emph{i.e.} to a Hamilton principle, we need to make sure that this is done by implementing d'Alembert's principle of virtual displacements (that can be shown to be equivalent to Newton's equations).
  In general, the Hamilton principle following from the use of Lagrange multipliers is not equivalent to d'Alembert's principle (for many details and examples on this very interesting and not yet fully resolved problem, see~\cite{prokhorov_shabanov_2011,Gantmacher}).
  
  We went through this long digression because the restriction to a subsector of the theory with a given value of the charge, if thought of as a constraint, would be of non-holonomic type, and implementing the restriction with a Lagrange multiplier would give a wrong result.
  A simple example of this phenomenon is the study of a particle on a plane under the action of a central force.
  If we want to find the trajectories with a given value of angular momentum \(M\), and find the minimum-energy configuration for fixed \(M\), the use of Lagrange multipliers, or even the use of Dirac's theory of constraints does not generate the correct answer, unless accompanied by ad hoc additional conditions that amount to knowing the correct answer from the beginning.
  The correct procedure will be explained in detail in Section~\ref{sec:O2n} for the example of the \(O(2n)\) theory, and it has similarities to the definition of the grand partition function.
\end{description}

We find it convenient to work in finite but large volume, eventually taking the infinite-volume limit.
In Appendix~\ref{sec:finite-volume} it is shown how transitions between different vacua are exponentially suppressed in the large-volume limit for spatial dimension \(d> 1\).
This restriction makes many of the technical arguments much simpler and in the end we can take the appropriate limit for the relevant quantities.

\subsection{First approach to SSB -- standard operator theory}
\label{sec:operator-approach}

As explained above, we work mostly at finite volume and when needed take the limit \(V \to \infty\).
Let us analyze a theory with \ac{ssb} for some conserved currents \(j^\mu_a\) (for the time being we will suppress the \(a\) index), and assume only homogeneity and isotropy.
Let us define the charge operator at finite volume,
\begin{equation}
  Q_V(t) = \int \dd[d]{x} j^0(t,x) .
\end{equation}
\ac{ssb} implies the existence of some local field \(\phi\), such that
\begin{equation}
  \delta \phi = \lim_{V \to \infty} \ev{\comm{Q_V(t)}{\phi}}{0} \neq 0 .
\end{equation}
The state \(Q\ket{0}\) is not normalizable in the infinite-volume limit, but commutators of \(Q\) with other operators and current correlation functions are well defined.
In the large-volume limit, \(\delta \phi\) is time-independent as a consequence of current conservation:
\begin{equation}
  \dv{t} \delta \phi = \lim_{V \to \infty} \ev{\comm{\int_V \dd[d]{x} \del_0 j^0(t,x)}{\phi}}{0} = - \lim_{V \to \infty} \ev{\int \dd{S} \cdot \comm{\mathbf{j}(t,x)}{\phi}}{0} = 0 .
\end{equation}
The vanishing depends on two assumptions: that there are no charges at infinity, and that commutators for space-like operators decay sufficiently rapidly as the distance between them increases (without these conditions there is no ``Goldstone''-like theorem~\cite{Lange:1965zz,Lange:1966zz}).
Now we follow the standard techniques used in the study of the Källén–Lehmann representations~\cite{Weinberg:1996kr}.
Since we have spacetime translational symmetry, we can write
\begin{equation}
  j^0(x)= e^{i p x} j^0(0) e^{-i p x}
\end{equation}
and compute the expectation value using the completeness relation of the theory and translation symmetry,
\begin{multline}
  \delta \phi = \lim_{V \to \infty} \ev{\comm{Q_V(t)}{\phi}}{0} = \sum_{n} (2 \pi)^d \delta^{d}(p_n) \left[ e^{-i E_n t}\mel{0}{j^0(0)}{n} \mel{n}{\phi}{0}\right. -\\
  \left.e^{i E_n t}\mel{0}{\phi}{n} \mel{n}{j^0(0)}{0} \right].
\end{multline}
Furthermore,
\begin{equation}
  \label{eq:ddt-delta-phi-expanded}
  \dv{t} \delta \phi = - i \sum_n (2\pi)^d\delta^{d}(p_n) E_n \bqty{ e^{-i E_n t}\mel{0}{j^0(0)}{n} \mel{n}{\phi}{0} + e^{i E_n t}\mel{0}{\phi}{n} \mel{n}{j^0(0)}{0} } . 
\end{equation}
We learn from these two equations that the only states contributing are those satisfying
\begin{align}
  \label{eq:Goldstone-states}
  E_n(p) &\to 0 \text{ as \(p \to 0\)}, & \mel{0}{\phi}{n} \mel{n}{j^0(0)}{0} & \neq 0 .
\end{align}
These states are the Goldstone states.

Next, it is important to explore the possible forms of the dispersion relations \(E_n = E_n(p)\).
For this it is useful to consider the Fourier transform of the commutator of the current \(j_\mu(x)\) and the fields getting \acp{vev} \(\Phi^a\).
Since we want to consider relativistic theories and some subset of non-relativistic theories, we will distinguish a time-like direction \(n^\mu = (1, \mathbf{0})\), and use the classical analysis of W.~Gilbert that goes back to~\cite{Gilbert:1964iy}.
Consider the Fourier transform
\begin{equation}
  \label{eq:j-phi-commutator-Fourier}
  \ev{\comm{j_\mu(x)}{\Phi^a(0)}}{0} = \int \frac{ \dd[d+1]{k}}{(2\pi)^{d+1}} e^{-i k x} j_\mu(k) .
\end{equation}
Current conservation and no charge at infinity imply that
\begin{equation}
  k^\mu j_\mu  = 0 .
\end{equation}
The general solution then takes the form
\begin{multline}
	\label{eq:j-Gilbert-expansion}
  j_\mu(k) = k_\mu \rho_1(k^2, k_\nu n^\nu) + ( k^2 n_\mu - k_\mu(n_\nu k^\nu)) \rho_2(k^2, k_\nu n^\nu) \\
  + n_\mu \delta(k_\nu n^\nu) \rho_3(k^2) + n_\nu \delta^{d+1}(k) \rho_4 .
\end{multline}
For the first term, current conservation implies \(k^\mu k_\mu \rho_1 = 0\), thus
\begin{equation}
  \label{eq:Gilbert-rho1}
  \rho_1 = i c_1 \delta(k^2) \sign(k^0) + i c_2 \delta(k^2) .
\end{equation}
As it is well known from basic \ac{qft}, \(c_2\) does not contribute to the commutator in Eq.~(\ref{eq:j-phi-commutator-Fourier}).
This implies the relativistic dispersion relation.
In fact, if we consider relativistic invariance, we can ignore  \(\rho_2\), \(\rho_3\), \(\rho_4\) and consider the intermediate states contributing in Eq.~(\ref{eq:Goldstone-states}).
Since the \(\Phi^a\) fields are the order parameters, they belong to a representation of the symmetry: if \(Q\) is the charge associated to the \(j_\mu\) current,
\begin{equation}
  \label{eq:Q-action-of-phil}
  \comm{Q}{\Phi^a} = i T\indices{^a_b} \Phi^b .
\end{equation}
In the relativistic case, Eq.~(\ref{eq:ddt-delta-phi-expanded}) to Eq.~(\ref{eq:Gilbert-rho1}) imply that the \ac{vev} of the action of the symmetry in Eq.~(\ref{eq:Q-action-of-phil}) will be dominated by  massless relativistic spinless particles.
In that case, using relativistic normalization, the relevant matrix elements are
\begin{align}
  \mel{0}{j^\lambda(x)}{GB} &= \frac{i F p^{\lambda} e^{-i p x}}{\sqrt{(2\pi)^3 2 p^0}}, & \mel{GB}{\Phi^a(x)}{0} &= \frac{Z^a e^{i p x}}{\sqrt{(2\pi)^3 2 p^0}},                                                                               
\end{align}
where \(p^\mu\) is the momentum of the Goldstone boson, $F$ is a constant of dimension 1 and $Z^a$ are dimensionless constants.
Then it can be shown that~\cite{Weinberg:1996kr}
\begin{equation}
  F Z^a = \sum_b T\indices{^a_b} \ev{\Phi^b(0)}{0} .
\end{equation}

Now that the relativistic case is settled, we can go back to the general expansion in Eq.~(\ref{eq:j-Gilbert-expansion}) and analyze the various terms.
The contribution from \(\rho_4\) would violate microscopic causality, since the \ac{vev} \(\ev{\comm{j^0}{\Phi^a}}{0}\) would get a contribution for all positions \(x\), and moreover \(\rho_4\) must scale as \( 1/V\) to get a finite value for the \ac{vev} \(\ev{\comm{Q}{\Phi^a}}{0}\). It is safe to set \(\rho_4 = 0\).
If \(\rho_3 \neq 0\), then there are states with \(k^0 = 0\), but \(\mathbf{k} \ne 0\), and this implies that translational symmetry in space is broken.
Since we are assuming translational invariance, we take \(\rho_3 = 0\).
The last term remaining is
\begin{equation}
  j_\mu(k) = (k^2 n_\mu - k_\mu n_\nu k^\nu) \rho_2(k^2, n_\nu k^\nu).
\end{equation}
Then
\begin{equation}
  \ev{ \comm{j^0(x)}{\Phi^a(0)}}{0} = \int \frac{\dd[d+1]{k}}{(2\pi)^{d+1}} e^{-i k x} \mathbf{k}^2 \rho_2(\mathbf{k}^2, k^0)^a.
\end{equation}
Integrating over finite volume,
\begin{equation}
  \ev{\comm{Q}{\phi(0)}}{0} = \int \frac{\dd[d+1]{k}}{(2\pi)^{d+1}} e^{-i k^0 x^0} \delta_V^{d}(\mathbf{k}) \mathbf{k}^2 \rho_2(\mathbf{k}^2, k^0) \neq 0
\end{equation}
because the \(\phi\) are the fields getting expectation values and \(\delta_V^{d}(\mathbf{k})\) is the finite-volume version of \(\delta^{d}(\mathbf{k})\).
Thus, for \(k\to 0\),
\begin{align}
  \rho_2(k) &= \frac{\bar \rho_2(k)}{\mathbf{k}^2}, & \bar \rho_2(0) \neq 0.
\end{align}
Taking the time derivative of the last \ac{vev}, that we know should vanish for \(V \to \infty\), we recover that the energy vanishes for zero momentum as expected:
\begin{equation}
  k^0 = k^0(\mathbf{k}^2) \to 0 \text{ for \(\mathbf{k} \to 0\).}
\end{equation}

In the early studies of \ac{ssb} in non-relativistic systems~\cite{Lange:1965zz,Lange:1966zz,Guralnik:1967zz}, it was known that the count of Nambu--Goldstone states did not follow the relativistic pattern.
This is the case for the Heisenberg ferromagnet, where \(SO(3) \to SO(2)\), but there is a single magnon with dispersion relation \(E \sim k^2\).
The authors of~\cite{Guralnik:1967zz} noted that the simplest example is the free non-relativistic particle
\begin{equation}\label{eq:SchroedingerParticle}
  \Lag = i \bar \psi \pdv{t} \psi - \frac{1}{2m} \nabla \bar \psi \nabla \psi
\end{equation}
invariant under \(ISO(2)\) symmetry. 
The symmetries are
\begin{align}
  \psi &\to \psi + \theta_1, & \psi &\to \psi + i \theta_2, & \psi &\to e^{i \theta_3} \psi,
\end{align}
whose charges satisfy the algebra
\begin{align}
  \comm{Q_3}{Q_1} &= - i Q_2, & \comm{Q_3}{Q_2} &= i Q_1, & \comm{Q_1}{Q_2} &= 2i V .
\end{align}
There are two broken generators but only one non-relativistic Goldstone boson (for details see~\cite{Miransky:2001tw,Brauner:2010wm}). We will recover this precise structure for the fluctuations around the homogeneous ground state in the O(2n) model in Section~\ref{sec:O2n-Goldstones}.

\paragraph{Nielsen--Chadha theorem}\label{par:NielsenChadha}
The first general result on the Goldstone boson count was obtained in~\cite{Nielsen:1975hm}.
The statement of the theorem is as follows.
\emph{Assume that
\begin{enumerate}
\item \(m\) of the Hermitian generators \(Q_a\), \(a = 1, \dots , m\) of the symmetry group are spontaneously broken.
  More precisely, there are \(m\) fields \(\phi_i\) and a vacuum state such that
  \begin{equation}
    \det (\ev{\comm{\phi_i}{Q_a}}{0}) \neq 0 .
  \end{equation}
\item The theory obeys some microcausality condition, so that for any two local operators \(A(t,x)\), \(B(t,x)\) their commutator vanishes exponentially with the distance,
  \begin{equation}
    \abs{ \ev{\comm{A(t,x)}{B(0)}}} \sim e^{-\tau \abs{x}} \text{ for \(\abs{x} \to \infty\).}
  \end{equation}
\item Translational invariance is not entirely broken.
\end{enumerate}
Under these circumstances there are two types of Goldstone bosons in the system: type I for which \(E \sim k^{2n +1}\) and type II such that \(E \sim k^{2n}\).
Moreover, the number of type-I and type-II Goldstone bosons satisfies \(n_I + 2n_{II} \ge m\).}

The proof is a further elaboration of the arguments used above.
Let us outline some of the salient features.
Define the matrix
\begin{equation}
  M_{ia} = \ev{\comm{\phi_i}{Q_a}}{0} = \eval{\sum_{n=1}^l e^{-i E_k t} \mel{0}{\phi_i}{n_k} \mel{n_k}{j^0_a}{0} - e^{i E_k t} \mel{0}{j^0_a}{n_k} \mel{n_k}{\phi_i}{0}}_{\mathbf{k} = 0}.
\end{equation}
Using previous arguments, the sum receives contributions only from the \(l\) states whose energy vanishes as \(\mathbf{k} \to 0\).

\begin{figure}
  \centering
  \begin{footnotesize}
  \begin{tikzpicture}
    \draw[-latex] (-3,0) -- (3,0) node[above]{\(\mathbf{k}\)};
    \draw[-latex] (0,-3) -- (0,1) node[left]{\(k^0\)};
    \draw[black, line width = 0.50mm] (-3,-3) -- (0,0) -- (3,-3);
    \node at (0, -4){(a)};
  \end{tikzpicture}
  \hfill
  \begin{tikzpicture}
    \draw[-latex] (-3,0) -- (3,0) node[above]{\(\mathbf{k}\)};
    \draw[-latex] (0,-3) -- (0,1) node[left]{\(k^0\)};
    \draw[black, line width = 0.50mm]   plot[smooth,domain=-3:3] (\x, {-\x*\x/4});
    \node at (0, -4){(b)};
  \end{tikzpicture}
\end{footnotesize}
  \caption{Type-I (a) and type-II (b) Goldstone dispersion relations in Fourier space}
  \label{fig:type-I-type-II-tangent}
\end{figure}
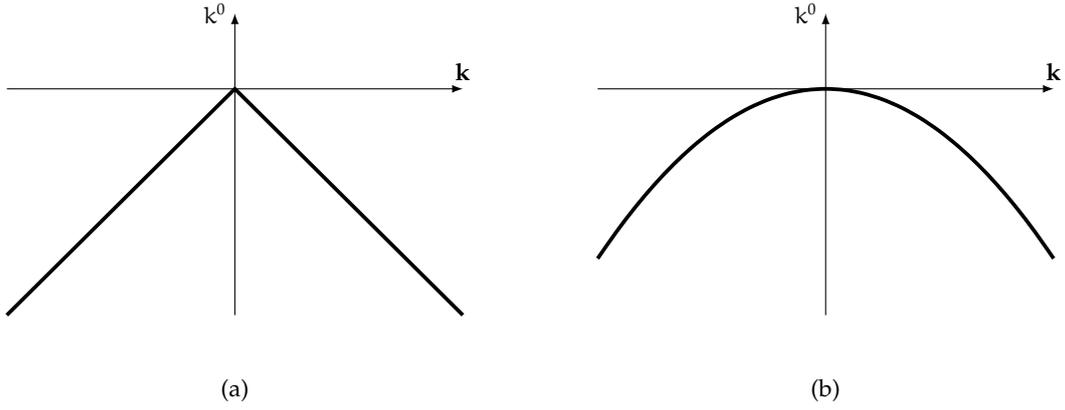

By assumption, the rank of \(M\) is \(m\), the number of broken generators.
Consider the following matrix, thought of as a collection of vectors:
\begin{equation}
  \nu_{ia} = (\nu_a)_i = \sum_{n=1}^l \mel{0}{\phi_i}{n} \mel{n}{j^0_a}{0} .
\end{equation}
Note that
\begin{equation}
  M_{ia} = 2 \Im(v_{ia}) .
\end{equation}
This, however does not mean that \(\nu_{ia}\) has rank \(m\).
In general the rank will be lower: if we think of \(\mel{0}{\phi_i}{n}\) and \(\mel{n}{j^0_A}{0}\) as vectors in \(\setC^l\), the matrix \(\nu_{ia}\) is a generalized Gram matrix and \(\rank(\nu) \le l\).
Since \(\nu_{ia}\) is a \(l \times m\) matrix, it is more convenient to construct \(l\) vectors in \(\setC^m\):
\begin{align}
  A_n &=
  \begin{pmatrix}
    \mel{0}{\phi_1}{n} \\
    \vdots \\
    \mel{0}{\phi_m}{n} \\
  \end{pmatrix}, & n &= 1, \dots, l
\end{align}
and
\begin{align}
  \nu_a &= \sum_{n=1}^l A_n \gamma_{a n}, & \gamma_{an} &= \mel{n}{j^0_a}{0}.
\end{align}
Writing \(\Im(\nu_a)\) in terms of the real vectors \(\Re(A_n)\) and \(\Im(A_n)\) and using the fact that \(\rank(Im(\nu)) = m\), we learn that \(l \ge m/2\).
Let \(p = \rank(\nu)\). Then there are \(m-p\) linear relations
\begin{align}
  \sum_{a=1}^m c^\alpha_a \nu_a &= 0, & \alpha &=1, \dots, m - p
\end{align}
with some of the \(c^\alpha_a\) non-vanishing. Furthermore,
\begin{equation}
  \sum_{a=1}^m (c^\alpha_a)^* \nu_a \neq 0,
\end{equation}
otherwise \(\rank(\Im(\nu)) < m\), contrary to assumption.
Then for every \(\alpha = 1, \dots, m - p\), Nielsen and Chadha show that there is one Goldstone boson whose energy goes like \(E(k) \sim k^{2n}\).
The proof is based on the standard analysis of the commutator \(\ev{\comm{\phi_i}{\sum_{a=1}^m c^\alpha_a j^0_a(x)}}{0}\) by inserting a complete set of states.
Then compute the spacetime Fourier transform: its \(\mathbf{k} \to 0\) limit, which is essentially computing \(\ev{\comm{\phi_i}{\sum_{a=1}^m c^\alpha_a Q_a(x)}}{0}\) contains a single sum, with the result
\begin{equation}
  F^\alpha_i(\mathbf{k} \to 0) = \eval{- 2 \pi \sum_{n=1}^l \delta(k^0 + E_{0k}) \sum_{a=1}^m c^\alpha_a \mel{0}{j^0_a}{n_{-k}} \mel{n_{-k}}{\phi_i}{0}}_{\mathbf{k} =0} .
\end{equation}
Since \(E_k > 0\), this expression is different from zero only for \(k^0 < 0\).
In \(k\)-space we have the plane \(k^0 =0\) and \(F^\alpha_i(\mathbf{k})\) is different from zero in a hypersurface tangent to the \(k^0 = 0\) plane from below.
The assumption (ii) about the exponential decay of the expectation value of commutators at large space-like separations implies that the Fourier transform is analytic in \(\mathbf{k}\).
Close to the origin, this hypersurface then is of the form \(E_k \sim k^2\) (or higher even power).
If instead of exponential decay, we have power-law decay, the point of tangency may have a cusp \(E_k \sim \abs{k}\) near \(\mathbf{k} = 0\) (see Figure~\ref{fig:type-I-type-II-tangent}).
Finally, we have \((m-p)\) type--II Goldstone bosons and \(l - (m-p)\) of type-I.
Hence
\begin{equation}
  n_I + 2 n_{II} = l + m - p \ge m .
\end{equation}
This completes the proof of the theorem.
There is not enough structure in the assumptions to determine when the bound is saturated.

To better understand when there is a mismatch between the number of broken generators and Goldstone bosons, we use a nice result in~\cite{Schafer:2001bq}.
The authors were studying the description of kaon condensation in \ac{qcd} (see also~\cite{Miransky:2001tw}) and noted an abnormal count of Goldstone states.
They proved the following lemma.

\emph{Let the Hermitian generators \(Q_1, \dots, Q_n\) be broken in the \(V \to \infty\) limit.
If \(\ev{\comm{Q_i}{Q_j}}{0} = 0\), then the number of Goldstone bosons is equal to the number of generators.}

In this formulation it is important to work in finite volume and think of low-momentum Goldstone bosons by acting with the \(Q_i\) on the vacuum \(\ket{0}\) (a more rigorous proof using currents can be found in~\cite{Brauner:2010wm}).
If there are less than \(n\) of them, there are complex numbers \(a_i \in \setC\), \(i = 1, \dots, n\) so that
\begin{equation}
  \sum_{i=1}^n a_i Q_i \ket{0} = 0.
\end{equation}
Some of the \(a_i\) cannot be real, otherwise we would find a linear combination of generators in the symmetry algebra annihilating the vacuum, contrary to the hypothesis.
Let
\begin{align}
  Q_a &= \sum_{i} \Re(a_i) Q_i, & Q_b &= \sum_{i} \Im(a_i) Q_i,
\end{align}
and define \(\ket{g} = Q_a \ket{0}\), and
\begin{align}
  (Q_a + i Q_b) \ket{0} &= 0, & Q_b \ket{0} &= i \ket{g}.
\end{align}
But then
\begin{equation}
  \ev{\comm{Q_a}{Q_b}}{0} = 2i \braket{g} \neq 0 ,
\end{equation}
contrary to assumption.
We learn that if the number of Goldstones is smaller than the number of broken generators, there must be Goldstone states coupling to more than one current.
There is \emph{de facto} a reduction in the number of \acp{dof}: the zero modes of two charges \(Q_a\), \(Q_b\) satisfying \(\ev{\comm{Q_a}{Q_b}}{0} \neq 0\) become canonically conjugate variables lowering the number of \ac{dof}~\cite{Nambu:2004yia,Das:2002cw,Das:2002ij}, and the kinetic term will be Schrödinger-like.
This will become clear in the effective Lagrangian approach of Leutwyler, Watanabe and Murayama that will be presented later in this section.

To conclude the operator approach to \ac{ssb} before we turn to the effective Lagrangian methods we present a very nice result that we call \emph{Brauner's theorem}~\cite{Brauner:2010wm}:

\emph{Every time we have two broken generators \(Q_a\), \(Q_b\), such that \(\ev{\comm{Q_a}{Q_b}}{0} \neq 0\), there is a single Goldstone boson coupling to both charges, and generically the dispersion relation is quadratic.}\footnote{We should really consider \(\ev{\comm{Q_a}{Q_b}}{0}/V\) or \(\ev{\comm{Q_a}{j^0_b(0)}}{0}\) to be precise.}

The starting point of the analysis is to use the available symmetries to write an ansatz for the matrix elements of the conserved currents:
\begin{equation}
  \mel{0}{j^\mu_a(0)}{n, \mathbf{k}} = i k^\mu_{0 n} F_{an}(k) + i \delta^{\mu0} G_{an}(k),
\end{equation}
where the \(k^\mu_{0n}\) are understood on-shell, \emph{i.e.} \(k^0_{0n} = E_{n,\mathbf{k}}\) is the correct dispersion relation for the state \(\ket{n, \mathbf{k}}\).
Current conservation implies
\begin{equation}
  \label{eq:Brauner-dispersion}
  \pqty{E^2_{n,\mathbf{k}} - \mathbf{k}^2} F_{a n}(\abs{\mathbf{k}}) + E_{n,\mathbf{k}} G_{an}(\abs{\mathbf{k}}) = 0 .
\end{equation}
If we try to understand the behavior of this equation, we encounter three cases:
\begin{enumerate}
\item if \(G_{an} = 0\) we are back to the Lorentz-invariant case;
\item if \(G_{an}/F_{an} = \order{\abs{\mathbf{k}}}\) as \(\mathbf{k} \to 0\), then the dispersion relation is linear in \(k\), \(E_{n, \mathbf{k}} = c \abs{\mathbf{k}} + \dots\), with \(c \neq 1\);
\item if \(G_{an}/F_{an} \neq 0\) for \(\mathbf{k} \to 0 \), then one of the roots of the quadratic equation must be \(E_{n, \mathbf{k}} = F_{an}/G_{an } \mathbf{k}^2 \).
\end{enumerate}
To study which case applies we consider in detail the two-point functions involving the currents:
\begin{align}
  \label{eq:Brauner-two-point}
  i D_{ab}^{\mu\nu}(x - y) &= \ev{T j_a^\mu(x) j^\nu_b(y)}{0}, \\
  \comm{j^0_a(t,x)}{j^0_b(t,y)} &= i \delta^d(x-y) c_{ab}(t,x),
\end{align}
where \(c_{ab}\) is related to the charge density in the ground state.
It is easy to show that
\begin{equation}
  \pdv{\alpha^\mu} i D_{ab}^{\mu0}(x-y) = \delta(x^0 - y^0) \ev{\comm{j^0_a(x^0, x)}{j^0_b(y^0,y)}}{0} .
\end{equation}
Next we compute this in momentum space using again the insertion of a complete set of states, together with the relation \(\Phi(x) = e^{i p x} \Phi(0) e^{-i px}\).
Using the standard Feynman prescription for time integrals,
\begin{equation}
  \int_{-\infty}^\infty \dd{x^0} \theta( \pm x^0) e^{\pm i x^0 (k^0- E_{n,\mathbf{k}} \pm i \epsilon)} = \pm \frac{i}{k^0 - E_{n,\mathbf{k}} \pm i \epsilon}
 ,
\end{equation}
we find:
\begin{multline}
	i k_\mu D_{ab}^{\mu 0}(k) = i \sum_n \left[ \frac{\pqty{k_\mu k^\mu_{0n} F_{an} + k^0 G_{an}} \pqty{k^0_{0n} F_{bn}^* + G_{bn}^*}}{k^0 - E_{n, \mathbf{k}} + i \epsilon}\right.\\
	\left. - \frac{\pqty{k^0_{0n} F_{bn} + G_{bn}} \pqty{k_\mu k^\mu_{0n} F_{an}^* + k^0 G_{an}^*}}{k^0 + E_{n, \mathbf{k}} - i \epsilon}\right].
\end{multline}
In the first term we expand \(k^0\) around \(E_{n,\mathbf{k}}\), then the coefficient of the pole enforces the dispersion relation in Eq.~(\ref{eq:Brauner-dispersion}), and similarly for the antiparticle contribution, expanding \(k^0\) around \(-E_{n, \mathbf{k}}\).
Adding up leads to
\begin{equation}
  i k_\mu D_{ab}^{\mu 0}(k) = -2 i \Im(\pqty{E_{n,\mathbf{k}} F_{an} + G_{an}}\pqty{E_{n,\mathbf{k}} F_{bn}^* + G_{bn}^*}).
\end{equation}
In the \(\mathbf{k} \to 0\) limit,
\begin{equation}
  \lim_{k \to 0} i k_\mu D_{ab}^{\mu 0}(k) = -2 i \Im(G_{an} G_{bn}^*).
\end{equation}
From Eq.~(\ref{eq:Brauner-two-point}) we recover
\begin{equation}
  2 \Im(G_{an}G_{bn}^*) = \ev{c_{ab}(0)}{0},
\end{equation}
as was to be shown.
When \(\ev{\comm{Q_a}{Q_b}}{0}/V \neq 0\), \(G_a\) and \(G_b\) have finite limits as \(\mathbf{k} \to 0\) and we generically have a quadratic dispersion relation.

Note that all the arguments presented apply to the case of a fixed charge sector, with the proviso that the energies \(E_{n, \mathbf{k}}\) are really measured with respect to the energy of the lowest-energy state with a given charge \(Q\).

Now we change gears and address the problem of \ac{ssb} in the non-relativistic domain using the theory of effective Lagrangians.
The conclusions will be quite similar, but we get a better view of the inner works of \ac{ssb} when the ground state contains charge densities.

\subsection{Non-linear realization of broken symmetries}
\label{sec:non-linear-realization}

At the end of the sixties two papers appeared, laying down the general structure of phenomenological Lagrangians~\cite{Coleman:1969sm,Callan:1969sn}.
They considered the general setup to describe theories where a global symmetry group \(G\) breaks into a subgroup \(H\).
The low-energy excitations live in the coset space \(G/H\) and the geometric properties of these homogeneous spaces can be used to describe the effective Lagrangians for the associated Goldstone bosons and their couplings to the other low-energy fields.

\medskip
We start with an observation which is the basis for constructing low-energy effective actions, especially for theories with spontaneously broken continuous symmetries.
It is a systematic way of representing all associated Ward identities.

\paragraph{Ward identities and the effective action}

This result is the basis for the construction of effective Lagrangians for theories with global symmetries (for a review see~\cite{Manohar:2018aog}).
It is important to note that in the arguments that follow, relativistic invariance plays no role.
We want a useful way of representing the properties of a field theory with a collection of conserved currents, \emph{i.e.} an effective way of summarizing the Ward identities for the associated symmetries of a field theory with a set of conserved currents
\begin{equation}\label{eq:consCurr}
  \del_\mu j^\mu_A = 0
\end{equation}
satisfying the standard canonical commutation relations
\begin{align}\label{eq:commRelCurr}
  \comm{j_A^0(t,x )}{ j_B^0(t,y)} &= f_{ABC} j_C^0(t,x) \delta(x-y), \\
  \comm{j_A^0(t,x )}{ j_B^m(t,y)} &= f_{ABC} j_C^m(t,x) \delta(x-y) +\text{Schwinger terms} .
\end{align}
Let \(\Psi_a(t,x)\) represent all the fields in the theory transforming in a given representation of the symmetry algebra,
\begin{equation}\label{eq:repSymmAlg}
  \comm{j_A^0(t,x)}{\Psi_a(t,y)} = (T_A)_{ab} \Psi_b(t,y) \delta(x- y) .
\end{equation}
The quantity of interest is the generating functional for the correlation functions of the \(\Psi_a\) and the currents \(j^\mu_A\):
\begin{equation}\label{eq:correlCurr}
  \ev*{T \Psi_a(x_1) \Psi_b(x_2) \dots j^{A_1}_{\mu_1}(y_1) j^{A_2}_{\mu_2} \dots}{0}
\end{equation}
written using the sources \(\eta\) and \(A_\mu\)
\begin{equation}\label{eq:genFun}
  Z[\eta, A_{\mu}]= \ev*{T \exp[i \int\dd{x} \eta_a \Psi_a + A_{\mu,A} j^{\mu,A}]}{0} = e^{iW[\eta, A_\mu]}.
\end{equation}
The next step is to compute the Legendre transform in \(\eta\) and define the standard effective action
\begin{equation}\label{eq:GammaW}
  \Gamma[\Psi, A_\mu] + W[\eta, A_\mu] = \int\dd{x} \eta_a \Psi_a.
\end{equation}
The Ward identities associated to the current algebra Eq.~\eqref{eq:commRelCurr},~\eqref{eq:repSymmAlg} are elegantly reproduced by the condition that \(\Gamma[\Psi, A_\mu]\) is gauge-invariant under~\cite{Zumino:1970tu} 
\begin{equation}
  \begin{cases}
  \delta \Psi_a = \epsilon^A (T_A)_{ab} \Psi_b, \\
  \delta A_\mu^A = D_\mu \epsilon^A,
\end{cases}
\end{equation}
where \(D_\mu\) is the covariant derivative with connection \(A_\mu\).
In case there are anomalies associated with the currents we need to include the relevant \ac{wzw} terms~\cite{Wess:1971yu,Witten:1983tw}.

\paragraph{Non-linear realization of symmetries}
Our general context is the study of \acp{nlsm}.
Given a Riemannian manifold \(M\), the fields are maps \(\phi: \setR^{d+1} \to M\), where \(\setR^{d+1}\) represents the spacetime (Euclidean or Minkowski), and the kinetic term has the form
\begin{equation}
  \label{eq:CWZ1}
  \Lag = \frac{1}{2} g_{a b}(\Phi)  \del_\mu \Phi^a \del^\mu \Phi^b ,
\end{equation}
where \(g_{ab}(\Phi)\) is a metric on \(M\).
We will later also consider terms with one derivative that appear in theories without Lorentz invariance.
The isometries of the metric \(g_{ab}\) correspond to the global symmetries of \(\Lag\).
Let \(X^a_A(\Phi)\), \(A, B = 1, \dots, \dim(G)\) be the vector fields generating the isometry group \(G\), \emph{i.e.} the Lie derivatives of \(g_{ab}\) with respect to \(X_A\) vanish,
\begin{equation}
  \label{eq:CWZ2}
  \Lie(X_A) g_{ab} = \nabla_a X_{A,b} + \nabla_b X_{A,a} = 0,
\end{equation}
and the symmetry transformations of the fields are
\begin{equation}
  \label{eq:CWZ3}
  \delta \Phi^a = \epsilon^A X_A^a(\Phi) .
\end{equation}
To implement the Ward identities of the global symmetry group it is convenient to gauge the symmetry.
For each generator \(X_A\) we introduce a gauge field \(A_\mu^A\).
The Killing vectors \(X_A\) satisfy the Lie algebra relations
\begin{equation}
  \label{eq:CWZ4}
  \comm{X_A}{X_B} = f_{ABC}X_C .
\end{equation}
For simplicity we take \(G\) to be compact and semisimple and normalize the generators so that the Cartan--Killing metric is proportional to the identity matrix, and \(f_{ABC}\) is totally antisymmetric.
It is easy to see that
\begin{align}
  \label{eq:CWZ5}
  \Lag &= \frac{1}{2} g_{ab} D_\mu\Phi^a D^\mu \Phi^b, & D_\mu \Phi^a &= \del_\mu \Phi^a - A_{\mu}^A X_A^a(\Phi)
\end{align}
is invariant under
\begin{equation}
  \label{eq:CWZ6}
  \begin{cases}
  \delta \Phi^a = \epsilon^A(x) X_A^a(\Phi),\\
  \delta A^A_\mu = D_\mu \epsilon^A = \del_\mu \epsilon^A + f^{ABC} A_\mu^B \epsilon^C .
\end{cases}
\end{equation}
For details on the properties of gauged \acp{nlsm} including \ac{wzw} terms and their role in string theory see for instance~\cite{Hull:1989jk,Jack:1989ne}.

\subsection{The CCWZ construction}
\label{sec:CCWZ}

We now review the seminal works~\cite{Coleman:1969sm,Callan:1969sn}.
Let us consider a theory with a global continuous symmetry \(G\) broken spontaneously to a subgroup \(H\).
Let \(\mathfrak{g}\) and \(\mathfrak{h}\) be the associated Lie algebras, so that \(\mathfrak{g} = \mathfrak{h} \oplus \mathfrak{k}\), where \(\mathfrak{k}\) is the tangent space at the origin of the coset space \(G/H\) and represents the broken directions.
We take
\begin{align}
  A,B, \dots &= 1, \dots, \dim(G), & i,j, \dots &= 1, \dots, \dim(H), \\ 
  a,b,\dots &= 1, \dots, \dim(G) - \dim(H), \\
  (T^A)^\dagger &= - T^A, & T^A &= \acomm{H^i}{X^a},
\end{align}
with Lie algebra relations
\begin{align}
  \comm{H^i}{H^j} &= f^{ijk} H^k, & \comm{H^i}{X^a} &= f^{iab} X^b, & \comm{X^a}{X^b} &= f^{abi} H^i + f^{abc} X^c.
\end{align}
Symmetric spaces are those equipped with a \(\setZ_2\) isometry,
\begin{align}
  R(H) &= H, & R(X) &=- X,
\end{align}
implying \(f^{abc} = 0\).
For simplicity, we will concentrate most of the time on such spaces.
Since the action of \(G\) is transitive on \(G/H\), the local geometric properties of this space can be obtained from those of the Lie algebra \(\mathfrak{g}\) at the origin of \(G/H\).
This is a remarkable simplification with respect to other manifolds with isometries.
We can consider the fibered space \( H \to G \to G/H\) with \(G/H\) the base and \(H\) the fiber.
Choosing a coset representative \(l(\varphi)\) is a local section of this fibered space,
\begin{equation}
\begin{tikzcd}
  H \arrow[r] &G \arrow[r] & G/H \arrow[l, "l(\varphi)"', bend right].
\end{tikzcd}
\end{equation}
Close to the origin it is common to choose \(l(\varphi) = e^{\varphi X}\), so that \(\varphi\) are the local coordinates close to the origin.
These are analogues of the pion fields in the chiral Lagrangian.

The left action of an element \(g \in G\) on the coset representative \(l(\varphi)\) normally requires a compensating right action of \(H\):
\begin{equation}
  \label{eq:CWZ-right-compensating}
  g l(\varphi) = l(\varphi') h(g, \varphi).
\end{equation}
For instance, if \(g = h \in H\) the compensating transformation is \(h\) itself.
For symmetric spaces we can derive a useful identity which makes the analysis independent of the precise compensator \(h\).
If we take the inverse of Eq.~\eqref{eq:CWZ-right-compensating}, \(l(\varphi)^{-1} g^{-1} = h^{-1}l(\varphi')^{-1}\), act with the isometry \(R\): \(l(\varphi) R(g^{-1}) = h^{-1} l(\varphi')\) and multiply with Eq.~\eqref{eq:CWZ-right-compensating} on the right term-by-term, we obtain
\begin{equation}
  g l^2(\varphi) R(g^{-1}) = l^2(\varphi'),
\end{equation}
or, in the exponential parametrization,
\begin{equation}
  g e^{2 \varphi X}R(g^{-1}) = e^{2 \varphi' X} .
\end{equation}
Once we have constructed the coset representation \(l(\varphi)\) we define the Maurer--Cartan one-forms
\begin{equation}
  \Omega = l^{-1} \dd{l} = \Omega_A T_A
\end{equation}
satisfying
\begin{align}
  \dd{\Omega} &= - \Omega^2 ; & \dd{\Omega_A} &= - \frac{1}{2} f_{ABC} \Omega_B \wedge \Omega_C .
\end{align}
We can split \(\Omega\) along \(H\) and \(G/H\):
\begin{equation}
  \Omega = e + \omega
\end{equation}
with \(\omega \in H\).
Under the left action of \(G\) on \(l(\varphi)\) in Eq.~\eqref{eq:CWZ-right-compensating},
\begin{equation}
  \Omega \to (g l h^{-1})^{-1} \dd{(g l h^{-1})} = h e h^{-1} + h\omega h^{-1} + h \dd{h^{-1}} .
\end{equation}
Since the broken generators (on \(G/H\)) belong to a representation of \(H\) we find
\begin{align}
  e & \to h e h^{-1} &\text{\(G/H\) vielbein}, \\
  \omega & \to h (\omega + \dd{}) h^{-1} &\text{\(H\) connection}.
\end{align}
We can use \(e\) to construct a \(G\)-invariant metric.
Furthermore, if we want to couple other fields to the \ac{nlsm} invariantly with respect to \(G\), we use the \(H\)-connection to define covariant derivatives.
Thus, using \(e\) and \(\omega\) we have the basic tools to construct phenomenological Lagrangians according to the prescription of~\cite{Coleman:1969sm,Callan:1969sn}.
It is also possible to discuss general properties of \ac{wzw} terms~\cite{Hull:1989jk,Jack:1989ne,Watanabe:2014fva}, but we will not pursue this subject here.

\subsection{The Leutwyler equations}
\label{sec:Leutwyler}

In~\cite{Leutwyler:1993gf,Leutwyler:1993iq}, Leutwyler extended the considerations in~\cite{Coleman:1969sm,Callan:1969sn} to the non-relativistic case.
He limited his study to theories that still maintain translation and rotational invariance as is the case for fixed charge.
He derived a set of equations that the lowest-order terms of the effective action should satisfy and worked out the example of ferromagnetic and antiferromagnetic systems to illustrate the fact that the count of Goldstone bosons is subtle in these systems, in agreement with~\cite{Nielsen:1975hm}.
We will see in detail how the effective-action point of view captures neatly the results obtained previously using the operator analysis of current commutators with order parameters.
The general solution to the Leutwyler equations was obtained in~\cite{Watanabe:2014fva} and will be presented below.

Leutwyler introduced a systematic way of obtaining the effective action in Eq.~\eqref{eq:GammaW} concentrating on the Goldstone excitations and writing the low-energy Lagrangian as an expansion in derivatives.
Unlike the number of Goldstone particles, which depends on the dispersion law, the number of fields needed to describe them is universal.
The effective theory involves \(\dim(G)-\dim(H)\) real fields.
As we have seen in the discussion of the Nielsen--Chadha theorem discussed in Subsection~\ref{sec:operator-approach}, the form of the dispersion relation \(\omega(k) = \gamma \abs{k}^n + \dots \) determines the number of independent one-particle states.
A more explicit count of such states appears in~\cite{Watanabe:2014fva}.
Since we do not have boost invariance, the Lagrangian in Eq.~\eqref{eq:GammaW} is a double sum,
\begin{equation}
  \Lag = \sum_{s,t} \Lag_{\text{eff}}^{(s,t)},
\end{equation}
where \(s\) and \(t\) are respectively the number of space and time derivatives in the effective action.
When the external gauge field is included we count \(V_0^A\) and \(\mathbf{V}^A\) as the same order as \(\del_0\) and \(\nabla\).
If we are interested in the counting of Goldstone bosons and their leading \ac{ir} properties it is sufficient to consider \(s,t = 0,1, 2\).

\paragraph{Global invariance}

Before including gauge fields, let us consider the structure of the lowest-derivative terms:
\begin{equation}
\begin{aligned}
  \label{eq:Leutwyler-lowest-terms}
  \Lag^{(0,1)} &= c_a(\varphi) \dot \varphi^a, \\
  \Lag^{(0,2)} &= \frac{1}{2} g_{ab}(\varphi) \dot \varphi^a \dot \varphi^b, \\
  \Lag^{(2,0)} &= \frac{1}{2} h_{ab}(\varphi) \nabla \varphi^a \cdot \nabla \varphi^b  .
\end{aligned}
\end{equation}
For \(\Lag^{(0,2)}\) and \(\Lag^{(2,0)}\), invariance under the global transformation Eq.~\eqref{eq:CWZ3} implies that \(g_{ab}\) and \(h_{ab}\) behave as Riemannian metrics Eq.~\eqref{eq:CWZ2}.
Depending on the properties of \(G/H\) there may be more than one second-rank invariant tensor that can be used for \(g \) or \(h\).
The conditions for the invariance of \(\Lag^{(0,1)}\), up to a total time derivative, lead to new and interesting geometric conditions.
The combination
\begin{equation}
  c_a(\varphi) \dot \varphi^a = c_a(\varphi) \dv{\varphi^a}{t},
\end{equation}
\emph{i.e.} \(c(\varphi) = c_{a}(\varphi) \dd{\varphi^a}\) is a one-form on the target manifold.
Using Eq.~(\ref{eq:CWZ-12}), the change of \(\Lag^{(0,1)}\) is related to the Lie derivative of \(c\):
\begin{equation}
  \Lie(X_A) c = \dd(i(X_A) c) + i (X_A) \dd{c}.
\end{equation}
Hence the condition of invariance of \(\Lag^{(0,1)}\) up to a total time derivative becomes the existence of a function \(e_A(\varphi)\) for each generator \(X_A\) such that
\begin{align}
  \label{eq:L01-invariance}
  i(X_A) \dd{c} &= \dd{e_A} ,\\
  \epsilon^A \delta_A \Lag^{(0,1)} &= \epsilon^A \pdv{t} (i(X_A) c + e_A).
\end{align}
Before further discussing the properties of \(c_a(\varphi)\) and \(e_A(\varphi)\), note that the conserved current associated to the symmetry \(\delta_A \varphi^a = X_A^a\) receives a contribution from the total derivative in Eq.~(\ref{eq:L01-invariance}).
Using the first two terms in Eq.~(\ref{eq:Leutwyler-lowest-terms}) we find that the charge density becomes
\begin{equation}
  j_A^0 = g_{ab} X_A^a \dot \varphi^b - e_A(\varphi) + \dots
\end{equation}
and we learn that
\begin{equation}
  \ev*{j_A^0} = - e_A(0) .
\end{equation}
This, together with Eq.~(\ref{eq:L01-invariance}) implies that the existence of a charge density in the ground state is at the origin of having a kinetic term for the \(\varphi^a\) fields linear in time derivatives.
Hence, at low energies, this term will dominate the dispersion relations.
We recover the conclusion that dispersion relations of the form \(\omega \sim k^2\) come generically from the existence of charge densities.

We work out some useful properties of \(c_a \dd{\varphi^a}\) and \(e_A(\varphi)\).
The first is that \(\dd{c}\) is an invariant two-form:
\begin{equation}
  \Lie(X_A) \dd{c} = \dd \Lie(X_A)c = \dd( \dd (i(X_A) c) + i(X_A) \dd{c}) = \dd^2(e_A) = 0.
\end{equation}
Thus, although the one-form is only defined locally, its exterior differential \(\dd{c}\) is a \(G\)-invariant closed two-form.
Once we know \(G\) and \(H\), it is not difficult to determine how many such two-forms are there.
As we show later, if the first homotopy group vanishes, \(\pi_1(G) = \varnothing \), then \(H^2(G/H) \simeq \pi_2(G/H)\), and we can use the long exact sequence of homotopy groups associated to the fibered space \(H \to G \to G/H\) to count the dimension of \(H^2(G/H)\).
This is the analysis in~\cite{Watanabe:2014fva}.
Before working out the details, we derive some useful properties of \(e_A(\varphi)\).
First note that
\begin{equation}
  \Lie(X_A) e_B = i(X_A) \dd{e_B} = i(X_A) i(X_B) \dd{c},
\end{equation}
hence\footnote{In fact, the 1-form \(c\) itself may be globally well-defined. Whether it actually is or is not depends on whether the charge density in the ground state belongs to a broken or an unbroken generator. The cohomology problem for \(\dd{c}\) classifies possible \ac{wzw} terms, corresponding to densities of the unbroken charge in the ground state.}
\begin{equation}
  \Lie(X_A) e_B + \Lie(X_B) e_A = 0 .
\end{equation}
Furthermore,
\begin{equation}
  \Lie(X_B) \dd{e_A} = \Lie(X_B) i(X_A) \dd{c} = i \comm{X_B}{X_A} \dd{c} =  f\indices{_B_A^C} i(X_C) \dd{c} = f\indices{_B_A^C} \dd{e_C}
\end{equation}
and since \(\comm{\dd}{\Lie(X_A)} = 0\) we learn that
\begin{equation}
  \dd(\Lie(X_B)e_A - f\indices{_B_A^C}e_C) = 0.
\end{equation}
In other words,
\begin{equation}
  \Lie(X_A)e_B = f\indices{_A_B^C}e_C + Z_{AB},
\end{equation}
where \(Z_{AB}\) are antisymmetric constants. They are related to the second Lie algebra cohomology of \(G\)~\cite{Watanabe:2014fva}.
If \(G\) is semisimple, \(H^2(G) = \varnothing\).
We will consider only this simpler case (see~\cite{Watanabe:2014fva} for the general case, which is \emph{e.g.} necessary to describe the free Schrödinger particle in Eq.~\eqref{eq:SchroedingerParticle}).
The Leutwyler equations are
\begin{equation}
  \label{eq:Leutwyler-equations}
  \begin{aligned}
    i(X_A) \dd{c} &= \dd{e_A}, \\
    \Lie(X_A) e_B &= f\indices{_A_B^C} e_C.
  \end{aligned}
\end{equation}
The latter equation implies that \(e_A\) belongs to the adjoint representation of the Lie algebra.
The complete solution to these equations was provided in~\cite{Watanabe:2014fva}.

\paragraph{Gauge invariance}

The transformation in Eq.~(\ref{eq:L01-invariance}) of \(\Lag^{(0,1)}\) implies that the gauging of \(\Lag^{(0,1)}\) does not follow from replacing \(\dot \varphi^a\) by \(D_0 \varphi^a\).
The gauging will have the form
\begin{equation}
  \label{eq:L01-gauging}
  \Lag^{(0,1)} = c_a(\varphi) \dot \varphi^a + f_A A^A_0 .
\end{equation}
For global transformations, \(A^A_0\) transforms like an adjoint field.
Hence \(f_A\) should also transform in the adjoint.
Then from Eq.~(\ref{eq:L01-invariance}) it is clear that the only possible choice is \(f_A = e_A(\varphi)\).
It is possible to show that under gauge transformations,
\begin{equation}
  \delta( c_a(\varphi) \dot \varphi_a + e_A(\varphi) V^A_0) = \pdv{t} \pqty{e^A (i(X_A)c + e_A)} .
\end{equation}
If this term cannot be cancelled by some local term in the action, then the physical wave functions get a change of phase under gauge transformations.
The expression in Eq.~(\ref{eq:L01-gauging}) again reinforces the identification of \(e_A(0)\) with the ground-state charge density: simply recall Eq.~(\ref{eq:CWZ4}) to Eq.~(\ref{eq:CWZ6}).
The other terms in the effective Lagrangian in Eq.~\eqref{eq:Leutwyler-lowest-terms} can be made gauge-invariant using the standard formulas in Eq.~(\ref{eq:CWZ5}) and Eq.~(\ref{eq:CWZ6}).

We now work out the solution to the Leutwyler equations~(\ref{eq:Leutwyler-equations}) following~\cite{Watanabe:2014fva}.
First recall the action of the symmetry group on the coset representative in Eq.~(\ref{eq:CWZ-right-compensating}).
If we make a specific choice for the coset representatives, then we can explicitly construct expressions for \(X_A^a\) and \(h(g, \varphi)\) close to the origin in terms of the Lie algebra structure constants.
Regardless of which parametrization we choose, at the origin on \(G/H\), the Killing vector fields associated to \(H\) should vanish:
\begin{equation}
  \ev*{X_i^a(\varphi)}_{\varphi = 0} = 0 .
\end{equation}
In fact, using the parametrization \(l(\varphi) = e^{\varphi X}\), it is possible to show that \(X_a = \pdv*{\varphi^a} + \dots \) and \(X_i = -f\indices{_i _a^b} \varphi^a \pdv*{\varphi^b}\).
This provides an initial condition for the Leutwyler equations~(\ref{eq:Leutwyler-equations}).
Choosing \(X_i\) in the unbroken group we learn that
\begin{equation}
  \label{eq:IC-for-e-C}
  f\indices{_i_B^C} e_C(0) = 0 .
\end{equation}
We will analyze this condition presently.
To solve the equations we use the Maurer--Cartan form \(\Omega_A\).
The infinitesimal form of Eq.~(\ref{eq:CWZ-right-compensating}) is
\begin{equation}
  l(\varphi + \epsilon^A X_A) = ( 1 - \epsilon^A T_A) l(\varphi) ( 1 + \epsilon^A k_A^i H_i) .
\end{equation}
Recalling that \(\Omega = l^{-1} \dd{l} = \Omega_A T_A\), we obtain
\begin{equation}
  i(X_A) \Omega = - l^{-1} T_A l + k_A,
\end{equation}
where \(k_A = k_A^i H_i\). 
Having chosen an orthonormal scalar product on the Lie algebra \(\ev*{T_A, T_B} = \delta_{AB}\) we can make the following ansatz for \(c(\varphi)\) and \(e_A(\varphi)\):
\begin{align}
  c(\varphi) &= \ev*{e(0), \Omega}, \\
  e_A(\varphi) &= \ev*{e(0), l^{-1}T_A l}.
\end{align}
Then we find
\begin{align}
  i(X_A) \dd{c} &= \dd \ev*{e(0), l^{-1}T_A l} - \ev*{e(0), \comm{k_A}{\Omega}}, \\
  \Lie(X_B)e_A &= i(X_B) \dd{e_A} = f\indices{_B_A^C} e_C + \ev*{e(0), \comm{l^{-1}T_A l}{k_B}},
\end{align}
where we have used the fact that \(\dd (l^{-1}T_A l) = \comm{l^{-1}T_A l}{\Omega}\).
Observing that for Lie algebras
\begin{equation}
  \ev*{X, \comm{Y}{Z}} = \ev*{Y, \comm{Z}{X}} = \ev*{Z, \comm{X}{Y}}, 
\end{equation}
the Leutwyler equations are satisfied if
\begin{equation}
  \ev*{e(0), \comm{l^{-1}T_A l}{k_B}} = \ev*{l^{-1} T_A l, \comm{e(0)}{k_B}} = 0 .
\end{equation}
But this is precisely equivalent to the initial condition for \(e_C(0)\) in Eq.~\eqref{eq:IC-for-e-C}.
Note that for generic \(g\) transformations, \(k_B\) ranges over the full Lie algebra of \(H\), and since \(l^{-1}T_A l = D_{AB}(l)T_B\) is the adjoint representation that is irreducible for simple groups.
In any case the condition is
\begin{equation}
  \comm{H_i}{e(0)} = 0 ,
\end{equation}
which is equivalent to Eq.~\eqref{eq:IC-for-e-C}.

\paragraph{Goldstone counting} Before developing a geometric understanding of the Watanabe--Murayama solution to the Leutwyler equations, one obtains as a spin-off  a precise way of counting type-I and type-II Goldstone bosons (which they call type A and type B).
It suffices to consider the quadratic expansion of the Lagrangian terms in Eq.~(\ref{eq:Leutwyler-lowest-terms}) around the origin in the parametrization \(l(\varphi) = e^{\varphi X}\).
The answer is
\begin{equation}
  \Lag_{\text{eff}} = \frac{1}{2} f\indices{_a_b^A}e_A(0) \dot \varphi^a \varphi^b + \frac{1}{2} g_{ab}(0) \dot \varphi^a \dot \varphi^b - \frac{1}{2} h_{ab}(0) \nabla \varphi^a \cdot \nabla \varphi^b .
\end{equation}
We know that the time component of the current associated to the symmetry \(\delta \varphi^a = X^a_A\) is
\begin{equation}
  j_A^0(x) = e_A(\varphi) - g_{ab}(\varphi) X_A^b(\varphi) \dot \varphi^b .
\end{equation}
The currents satisfy the standard commutation relations
\begin{equation}
  \comm{j_A^0(t,x)}{j_B^0(t,y)} = i f\indices{_A_B^C} j_C^0(t,x) \delta(x-y).
\end{equation}
The authors of~\cite{Watanabe:2014fva} define the antisymmetric matrix
\begin{equation}
  i \rho_{ab} = \ev*{\comm{Q_a}{j_b^0(x)}} .
\end{equation}
By translational invariance, \(\rho_{ab}\) is independent of \(x\), and it is related to the first term in the effective Lagrangian,
\begin{equation}
  \rho_{ab} = f\indices{_a_b^A} \ev*{j_A^0(0)} = - f\indices{_a_b^A}e_A(0) .
\end{equation}
This is an antisymmetric matrix, so it can always be skew-diagonalized and its rank is always even.
Let \(\rank(\rho) = 2m\).
Then we can perform an orthogonal rotation so that the first \(2m\) fields \(\varphi^a\) are associated in pairs with the skew-eigenvalues of \(\rho\). The effective Lagrangian takes the form
\begin{equation}
  \label{eq:skew-symmetrized-Lagrangian}
  \Lag = \sum_{j=1}^m \lambda_j \varphi_{2j -1} \dot \varphi_{2j} + g_{ab}(0) \dot \varphi^a \dot \varphi ^b + h_{ab} \nabla \varphi^a \cdot \nabla \varphi^b
\end{equation}
and we observe a reduction in the number of excitations.

As expected from the operator approach discussed in Subsection~\ref{sec:operator-approach}, when there are charge densities, the zero modes of the currents become each other's canonical conjugates, which cuts the number of Goldstone excitations in half. The Lagrangian in Eq.~(\ref{eq:skew-symmetrized-Lagrangian}) clearly shows this phenomenon.
Note that for the first \(2m\) fields, the symplectic (canonical) structure has changed and \(\varphi_{2j -1}\) and \(\varphi_{2j}\), \(j = 1, \dots, m\) become canonically conjugate pairs.
The dispersion relation is generically of the form \(E = \gamma k^2\), and we find
\begin{align}
	n_{II} &= 1/2 \rank(\rho), & n_I &= \dim(G/H) - \rank(\rho).
\end{align}
Hence,
\begin{equation}
  n_I + 2 n_{II} = \dim(G/H) 
\end{equation}
is the more precise counting rule for Goldstone bosons in the non-relativistic domain when rotation and translational invariance are preserved.\footnote{One can find similar rules relaxing these assumptions, but this requires taking into account other terms in the decomposition of the Lagrangian, see \emph{e.g.}~\cite{Watanabe:2014qla} for a discussion of the breaking of rotational invariance.}

\paragraph{Geometrical interpretation and classification}
Let us finally review the beautiful geometrical description of the Watanabe--Murayama solution to the Leutwyler equations.
In \(G/H\), \(\dd{c}\) is a \(G\)-invariant closed two-form.
If \(\det(\dd{c}_{ab}) \neq 0\) it defines a symplectic structure and in this case the theory describes only type-II Goldstone bosons.
The charge densities are represented by the matrix \(T^A e_A = e(0)\) and it represents a torus \(\mathbb{T}\) inside \(G\).
As we have seen, \(e(0)\) commutes with the generators \(H_i\), so the torus of charges commutes with \(H\).
One can consider the centralizer of \(\mathbb{T}\) in \(G\), \(U\).
A theorem by Borel~\cite{Borel} plays a central role. 

\emph{Let \(G\) be compact and semisimple, and \(U\) be the centralizer of a torus in \(G\). Then the homogeneous space \(G/U\) is homogeneous Kähler and algebraic.}

Thus, once \(e(0)\) is specified, we know the symplectic structure.
We have the fibration
\begin{equation}
  F \to G/H \to B= G/U
\end{equation}
and the fiber \(F = U/H\).
The invariant closed two-forms in \(B\) are pulled back to a two-form \(\dd{c}\) which is used to construct the effective Lagrangian terms in Eq.~(\ref{eq:Leutwyler-lowest-terms}).
Counting the number of two-forms is relatively straightforward in some circumstances.
If we consider the case where \(\pi_1(G) = \varnothing\), by a theorem of Hurewicz,
\begin{equation}
	H^2_{\text{dR}} = H_2(G/U) = \pi_2(G/U),
\end{equation}
and by the long exact sequence of homotopy groups for the fibration,
\begin{align}
	U  &\to G \to G/U, & \pi_2(G/U) &= \pi_1(U),
\end{align}
which corresponds to the number of \(U(1)\) factors in \(U\) (there are no \(U(1)\) factors in \(G\)).
\emph{We learn that the type-II Nambu--Goldstone bosons are described by the base manifold \(G/U\) and the type-I bosons by the fiber \(U/H\).}
This is one of the major results of~\cite{Watanabe:2014fva}.
They also provide a detailed classification when \(G\) is any of the classical groups.
We are interested in the \(SU(n)\) and \(SO(n)\) groups because they are related to the examples we will work out later.
For \(SU(n)\), the possible form of \(e_A(0) T^A\) is
\begin{equation}
  e(0) = \diag( \underbrace{\theta_1, \dots, \theta_1}_{n_1}, \underbrace{\theta_2, \dots, \theta_2}_{n_2}, \dots, \underbrace{\theta_k, \dots, \theta_k}_{n_k} )
\end{equation}
and the centralizer is
\begin{align}
  U &= U(1)^{k-1} \times \prod_k SU(n_k); & \sum_{i=1}^k n_i &= n , \sum_{i=1}^k n_i \theta_i = 0 .
\end{align}
For \(SO(n)\) groups, any element in the adjoint is an antisymmetric matrix. Thus:
\begin{equation}
  e(0) = \diag( \underbrace{0, \dots, 0}_m, \underbrace{\theta_1, \dots, \theta_1}_{n_1},  \dots, \underbrace{\theta_k, \dots, \theta_k}_{n_k} ) \otimes i \sigma_2 ,
\end{equation}
with centralizer
\begin{align}
  U &= SO(m) \otimes \prod_{i=1}^k U(n_i), & n = m + 2 \sum_{i=1}^k n_i.
\end{align}
Consider for instance the coset space \(SU(n+1)/U(1)^n\), which is a Kähler manifold according to Borel's theorem.
Hence \(G= SU(n+1)\) and \(H=U(1)^n\).
If we could charge all \(U(1)\)s with different charges, then \(U=U(1)^n\), and we would have no type-I Goldstones and \(\dim(SU(n+1)/U(1)^n)/2 = n(n+1)/2\) type-II.
Another example is to charge a single \(U(1)\), in which case \(U = SU(n) \times U(1)\), then \(\dim(G/H) = 2n\) and \(n_{II} = n\) and \(n_I = \dim(U/U(1)^n) = n(n-1)\).
Clearly there are plenty of intermediate cases whose realization depends on the details of the theory under consideration and the inequivalent ways of implementing charge densities.
Many more details and examples can be found in~\cite{Watanabe:2014fva}.

\section{The \(O(2)\) model}
\label{sec:original-paper}

After having worked out a number of general concepts needed for the large-charge expansion, we are ready to apply it to concrete cases. Let us first study the simplest example, namely the $O(2)$ model in $d+1$ dimensions.
In the \ac{uv}, it is the theory of a complex scalar field with the Lagrangian
\begin{equation}
	\Lag_{UV}= \del_\mu\varphi_{UV}^*\del^\mu\varphi_{UV} - r \varphi_{UV}^*\varphi_{UV} - 4 u(\varphi_{UV}^*\varphi_{UV})^2,
\end{equation}
where the coupling $r$ is fine-tuned such that in the \ac{ir}, the model flows to the strongly interacting \ac{wf} fixed point for $d<3$.\footnote{For $d>3$ a similar argument can be made for a flow to a conformal \ac{uv} fixed point~\cite{Fei:2014yja}.}
This strongly interacting \ac{cft} is what we will be focusing on in the following.
We will consider it on $\mathbb{R}\times \mani$, where for now $\mani$ is an unspecified compact homogeneous manifold with a characteristic length scale $L$. 

\subsection{Top-down approach: the RG flow}
\label{sec:top-down}

In this section we want to show -- using a top-down orthodox \ac{rg} flow analysis -- how the low-energy physics of a sector of fixed charge is described by an approximately scale-invariant action.\footnote{In the next section we will use a different approach to reach the same conclusion, following the old adage that ``\emph{conclusions based on the renormalization group arguments [\dots] are dangerous and must be viewed with due caution.}''~\cite{Bjorken:1965zz}.}
We follow the argument given in~\cite{Hellerman:2015nra}. While we present here the salient features of the \ac{rg} analysis, detailed calculations will follow in later sections.
For simplicity, we will consider the case of the \(O(2)\) model in \(d + 1\) dimensions. The main idea is to use the radial mode of the field to set the energy scale of the \ac{rg} flow once we are near the \ac{ir} conformal fixed point.
We will use a toy-model version of the \ac{rg} flow to illustrate this argument, where we consider only the flow of the \(\varphi^4\) coupling and neglect all other operators that can appear in the flow (in the spirit of~\cite{Polchinski:1992ed}).
In particular, we will assume that no singularities arise along the flow.
Separating two \acp{dof} into a radial mode \(a\) and an angular mode \(\chi\), the microscopic action becomes\footnote{This system cannot be quantized in $a=0$, but since we will consider it only in the context where $a$ has a \ac{vev}, this poses no issue.}
\begin{equation}
  \Lag_{UV} = \frac{1}{2} \del_\mu a \del^\mu a + \frac{1}{2} a^2 \del_\mu \chi \del^\mu \chi - u a^4.  
\end{equation}
In the flow towards the \ac{ir}, the first scale that we encounter is fixed by \(u\).
Once the cutoff scale \(\Lambda\) reaches \(\Lambda = \Lambda_{UV} = u^{1/(3-d)} \), the quantum effects become of the same order as the tree-level terms, and the renormalized coupling \(u(\Lambda)\) is completely fixed by the fixed point physics.
Below this point it has to satisfy
\begin{align}
  \frac{u(\Lambda)}{\Lambda^{3-d}} &= h, & \Lambda \ll \Lambda_{UV},
\end{align}
where \(h\) is a dimensionless parameter that depends only on the universality class.
Fixing the charge
\begin{equation}
  Q = \int \dd{V} \fdv{\Lag}{\del_0 \chi}
\end{equation}
results in a spontaneous symmetry breaking which gives rise to a \ac{vev} for both fields \(a\) and \(\del \chi\):
\begin{align}
  \ev{a} & = v, & \ev{\norm{\del \chi}} &= \mu .
\end{align}
There is a major difference between the two fields.
The fact that there is a potential term for \(a \) suggests that the spectrum of the fluctuations \(\hat a\) is gapped, and they are frozen below a certain energy.
This is to be contrasted with the \(\del \chi\) sector that, by virtue of Goldstone's theorem, has to remain free and massless in the \ac{ir}.
Keeping this in mind, we see that in the flow to the \ac{ir}, the next scale that we encounter is fixed by the \ac{vev} \(v\).
More precisely, expanding the potential, we see that the fluctuations of the \(a\) field will have a mass fixed by
\begin{equation}
  u ( v + \hat a)^4 \sim uv^4 + 4 uv^3\hat a + 6u v^2 \hat a^2 + \dots \ni \frac{1}{2} m_a^2 \hat a^2 . 
\end{equation}
Once the cutoff scale reaches \(\Lambda = m_a\), \(a \) is fixed to its \ac{vev} and the fluctuations will only give contributions controlled by \emph{positive powers} of \(\Lambda/m_a\).
At this point, the coupling \(u(\Lambda)\) is frozen and its value is fixed by the condition
\begin{equation}
  u(m_a) = h \eval{\Lambda^{3-d}}_{\Lambda = m_a} .
\end{equation}
Using that \(m_a^2 = 12 u v^2\), we get an equation for \(u(m_a)\) which yields:
\begin{equation}
   u(m_a) = \bar u = \pqty{ \pqty{12 v^2}^{3-d} h^2}^{1/(d-1)}
\end{equation}
whence it follows that \(m_a = (12 h v^2)^{1/(d-1)}\).
Plugging this back into our effective action we find that the entire $a$-dependence is encoded in the \ac{vev} $v$:
\begin{equation}
  \label{eq:action-a-chi-low-Lambda}
  \Lag_\Lambda[a, \del \chi] = \frac{v^2}{2} \del_\mu \chi \del^\mu \chi - \pqty{12^{3-d} h^2}^{1/(d-1)} v^{2(d+1)/(d-1)} + \order{\frac{\Lambda}{v^{2/(d-1)}} } %
\end{equation}
In this regime, the effective action is approximately scale-invariant with higher-order corrections controlled by \(\Lambda/v^{2/(d-1)}\).

Now that the dynamics of \(a\) is frozen, let us concentrate on \(\del \chi\).
For ease of exposition, it is convenient to introduce a new dynamical field \(B = \pqty{\del_\mu \chi \del^\mu \chi}^{1/2}\), which is the simplest scalar that can be constructed with the derivatives of \(\chi\) (remember that only the derivative of \(\chi\) is physical, since the field \(\chi\) is shifted by a constant by the action of the \(U(1)\) symmetry).
At these energies, the \ac{vev} \(\ev{B} = \mu\) is related to \(v\) by \(\mu^2 = \rho^2/v^4= 2(d+1)/(d-1) \pqty{12^{3-d} h^2}^{1/(d-1)} v^{4/(d-1)}\), which in turn we can use to express both \acp{vev} in terms of the charge density \(\rho = Q/ V\):
\begin{align}
  v &= 6^{(d-3)/(4d)} \pqty{\frac{d-1}{d+1}}^{(d-1)/(4d)} (2h)^{-1/(2d)} \rho^{(d-1)/(2d)} \\
  \mu &= 6^{(3-d)/(2d)} \pqty{\frac{d+1}{d-1}}^{(d-1)/(2d)} \pqty{2 h \rho}^{1/d} ,
\end{align}
which could have been guessed from dimensional analysis, since the system is at the fixed point where there are no dimensionful parameters, \(\rho\) has mass dimension \([\rho] = d\), \(\mu\) has mass dimension \([\mu] = 1\), and \(v \) is a canonical scalar and has dimension \([d] = (d-1)/2\).

In fact, there is no need anymore to refer to the frozen mode at all.
We can integrate \(a\) out and write an action for the \(B\) field alone.
Just as above, since we are expanding around a non-vanishing \ac{vev}, the effective action will contain a scale-invariant part, plus quantum corrections depending on the derivatives \(\del B\), controlled by positive powers of \(\Lambda/m_a \approx \Lambda/\ev{B} = \Lambda/\mu\):
\begin{align}
  \Lag_\Lambda[B] &= \Lag_{cl}[B] + \sum_{\Delta< d+1} \Lambda^{d+1- \Delta} \Lag_{q}^{(\Delta)}[B], & \Lambda \ll \mu,
\end{align}
where we have collected the quantum terms according to their classical dimension.
All in all we have the hierarchy
\begin{center}
  \begin{tikzpicture}[node distance=.7cm]
    \node [] (IR) {\(\Lambda\)};
    \node [right= of IR] (IRm) {\(\ll\)};
    \node [right=of IRm] (m) {\(m_a \sim v^{2/(d-1)} \sim \mu \sim \rho^{1/d}\)};
    \node [right= of m] (mUV) {\(\ll\)};
    \node [right=of mUV] (UV) {\(\Lambda_{UV} = u^{1/(3-d)}\).};

    \node [below=of IR] (IRexp) [align=left]{scale-invariant action\\for \(B = \del \chi\)};
    \node [below=of m] (mexp) {\(a\)-fluctuations are frozen};
    \node [below=of UV] (UVexp) {\(u\) is frozen};

    \draw [->] (IR) edge (IRexp);
    \draw [->] (m) edge (mexp);
    \draw [->] (UV) edge (UVexp);
  \end{tikzpicture}
\end{center}
The charge-fixing condition results in a new scale proportional to the charge density \(\rho\), \(\Lambda_Q \sim Q^{1/d}/L = \rho^{1/d}\) that controls the physics at energy \(\Lambda\).

What terms can enter in \(\Lag_{cl}[B]\)? As we have seen above, Goldstone's theorem tells us that at low energies, the system is described by a free massless field.
In the action in Eq.~\eqref{eq:action-a-chi-low-Lambda}, \(\del_\mu\chi \del^\mu \chi = B^2\) appears as a source for the \(a\) field that is then integrated out.
Generically this means that we will obtain terms with powers of \(B\) at the denominator, while its derivatives can only appear in the numerator, since we are in a regime \(\Lambda \ll \ev{B}\) where the fluctuations of \(B\) are small with respect to the \ac{vev}.
The \ac{rg} flow imposes strong constraints on the structure of the terms in \(\Lag_{\Lambda}[B]\)~\cite{Weinberg:1979327}.
The terms in \(\Lag_q\) are completely fixed by the renormalization of the scale-invariant part \(\Lag_{cl}\): at the fixed point we have
\begin{equation}
  \Lambda \fdv{\Lambda} \Lag_{cl} = \sum_{\Delta < d+1} (\Delta - d - 1) \Lambda^{d+1 - \Delta} \Lag_q^{(\Delta)}.
\end{equation}
Heuristically, if the renormalization of the scale-invariant part gives schematically
\begin{equation}
  \Lambda \fdv{\Lambda} \Lag_{cl}[B] = K_1 \Lambda^d + K_2 \frac{\Lambda^{d+2}}{B^2} + K_3 \Lambda^{d+2} \frac{(\del B^2)^2}{B^8} + \dots ,
\end{equation}
then, necessarily the corresponding coefficients of \(\Lambda\) in \(\Lag_q\) are fixed to be
\begin{equation}
  \Lag_q[B] = - \frac{1}{d} K_1 \Lambda^d - \frac{1}{d+2} K_2  \frac{\Lambda^{d+2}}{B^2} - \frac{1}{d+2}  K_3 \Lambda^{d+2} \frac{(\del B^2)^2}{B^8} + \dots .
\end{equation}
Equivalently, in terms of \(\del \chi\):
\begin{equation}
  \Lambda \fdv{\Lambda} \Lag_{cl}[\del \chi] = K_1 \Lambda^d + K_2 \frac{\Lambda^{d+2}}{\del_\mu\chi \del^\mu \chi} + K_3 \Lambda^{d+2} \frac{(\del (\del_\mu\chi \del^\mu \chi)^2)^2}{(\del_\mu\chi \del^\mu \chi)^4} + \dots ,
\end{equation}
corresponds to
\begin{equation}
  \Lag_q[\del \chi] = - \frac{1}{d} K_1 \Lambda^d - \frac{1}{d+2} K_2  \frac{\Lambda^{d+2}}{\del_\mu\chi \del^\mu \chi} - \frac{1}{d+2}  K_3 \Lambda^{d+2} \frac{(\del (\del_\mu\chi \del^\mu \chi)^2)^2}{(\del_\mu\chi \del^\mu \chi)^4} + \dots .
\end{equation}
A systematic derivation of the terms in the effective action, that makes use of Weyl invariance, will be presented in Section~\ref{sec:non-linear-sigma-model}.

\subsection{Bottom-up approach: the dilaton dressing}%
\label{sec:bottom-up-approach}

We have seen in the analysis above how the radial mode (or more precisely its \ac{vev}) plays the role of the \ac{rg}-flow parameter.
In this sense it plays the role of a dilaton, since the \ac{rg} transformation is a local scale transformation.

In this section we try to make this connection more precise, using a \emph{bottom-up} approach to write the \ac{eft} for the \ac{ir} physics of the \(O(2)\) model in a sector of fixed charge, \emph{i.e.} we start directly at the \ac{ir} fixed point.
The idea is that since the global symmetry is spontaneously broken, the system is naturally described by a free massless (Goldstone) field.
In order to realize the expected conformal symmetry at the fixed point, this Goldstone has to be supplemented by another field: the dilaton~\cite{Coleman:1988aos}. 
We will see that this is precisely the radial mode of the \ac{eft}~\cite{Orlando:2019skh}.
The role of the dilaton was originally investigated in Zumino's lectures~\cite{Zumino:1970tu} and independently in~\cite{Isham:1970gz,Isham:1970xz,Isham:1971dv}.
The aim is to transform any given local theory into one that is conformally invariant.
The prescription is simple and elegant: first, make the action generally covariant.
Second, make it Weyl-invariant by introducing an additional \ac{dof}.

We start with a two-derivative \ac{eft} for the prospective Goldstone of the type
\begin{equation}
  \Lag_2[\chi] = \frac{f_\pi^2}{2} \del_\mu \chi \del_\mu \chi -  C^{d+1},
\end{equation}
where \(f_\pi\) and \(C\) are dimensionful constants related to the underlying theory.
If we want to describe a (near) conformal theory,  we can introduce 
a new field \(\sigma\) -- the dilaton -- that realizes non-linearly conformal invariance and under dilatations \(x \to e^\alpha x\) transforms as \(\sigma(x) \to \sigma(e^\alpha x) - (d-1)\alpha/(2f)\), where \(f\) is a constant of dimension \([f] = -(d-1)/2\). 
Using this field we can turn any action into a non-linearly realized conformally-invariant one by dressing all the operators \(\Op_k\) of dimension \([\Op_k] = k\) as
\begin{equation}
  \Op_k(x) \to e^{2(k-d-1)/(d-1) f \sigma} \Op_k(x) .
\end{equation}
In our \(U(1)\) case we obtain
\begin{multline}
	\Lag_{CFT}[\chi,\sigma] = \frac{1}{2} g^{\mu\nu} f_\pi^{2} e^{-2\sigma f} \del_\mu \chi \del_\nu \chi - C^{d+1} e^{- 2(d+1)/(d-1) \sigma f} \\
	+ \frac{1}{2}e^{-2\sigma f}\pqty{g^{\mu\nu} \del_\mu\sigma \del_\nu \sigma- \frac{\xi R}{f^2}} +\order{R^2},
\end{multline}
where we have also added a kinetic term for the dilaton (including the curvature coupling that generates an improved energy-momentum tensor)\footnote{The dilaton here should not be confused with a modulus of the \ac{cft}. %
  It is a gapped Goldstone that realizes classical conformal invariance in the \ac{eft}. \acp{cft} with moduli space are typically supersymmetric and are described at large charge by a qualitatively different \ac{eft}~\cite{Hellerman:2017sur,Hellerman:2018xpi} that we will discuss later on in this review.}.
We now have obtained an effective action for the two Goldstones resulting from the breaking of the internal and of the conformal symmetry. 

The two fields can be combined into a complex one, akin to the string-theoretical axio-dilaton:
\begin{equation}
  \Sigma = \sigma + i f_\pi \chi.
\end{equation}
Now the action can be recast in the form
\begin{equation}
  \label{eq:phi-4}
  \Lag[\phi] = \del_\mu \phi^*\del^\mu\phi - \xi R \phi^* \phi - g \pqty{2 \phi^* \phi}^{(d+1)/(d-1)} + \dots ,
\end{equation}
where \(\phi = 1/(\sqrt{2}f) e^{-f \Sigma}\), which means the dilaton appears as the radial mode of $\phi$.
We are describing a \ac{cft}, which by definition has no dimensionful parameters.
The three dimensionful constants \(f_\pi\), \(C\) and \(f\) are combined into the two dimensionless quantities \(b = f f_\pi\) and \(g =(Cf^{2/(d-1)})^{d+1}\). 
The former controls the deficit angle for the field \(\phi\), which covers the whole complex plane only if \(b = 1\). Since this is not a microscopic, but an effective Lagrangian, an angle deficit may have arisen along the flow.
In odd dimensions, also a Weyl-anomaly term has to be added to the effective Lagrangian, however, 
it does not affect correlation functions of local operators~\cite{Simmons-Duffin:2016gjk}.

The description of the radial mode via the dilaton can also be harnessed to introduce a small mass term for $\sigma$ which explicitly breaks conformal invariance~\cite{Coleman:1988aos}.
This idea has been used to explore near-conformal dynamics at large charge~\cite{Orlando:2019skh,Orlando:2020yii}.
In general, the systematic use of the dilaton simplifies the arguments considerably, as we will see in the explicit examples.

\subsection{The linear sigma model}%
\label{sec:linear-sigma-model}

In the following, we will write down our treatment for general dimension $d+1$, to keep maximum generality but cover the same material that appeared first in~\cite{Hellerman:2015nra}.
We have argued above that we can describe the \ac{ir} physics in terms of a complex scalar $\phi$, (which is generically related to the microscopic field \(\varphi_{UV}\) by a complicated transformation).
As before, we parametrize it in terms of a radial and an angular \ac{dof}:
\begin{equation}
	\phi = \frac{a}{\sqrt 2} e^{ib\chi},
\end{equation}
where $a \in \mathbb{R}^+$, $\chi$ is $2\pi$ periodic, and $b$ encodes the possibility of a conical singularity in the \ac{eft}.
This choice also makes the global $O(2)$ or $U(1)$ symmetry manifest that lends its name to the model: the $U(1)$ acts by shifting $\chi$ by a constant, $\chi \to \chi + \alpha$.

We want to study the \ac{ir} fixed point in a sector of fixed and large global charge $Q$ and aim to write down an \ac{eft} in terms of the light \ac{dof} in this sector.
As discussed in Section~\ref{sec:Goldstone}, the ground state at fixed charge (which does not coincide with the ground state of the full theory) will give rise to a spontaneous symmetry breaking. This is the effect that will actually furnish us with the light \ac{dof}, namely the Goldstone fields, in terms of which we can write down an \ac{eft} for a strongly coupled system.

As discussed in the \ac{rg}-flow analysis, we will work in the range
\begin{equation}
  \frac{1}{L} \ll \Lambda \ll \frac{Q^{1/d}}{L} \ll u^{1/(3-d)},
\end{equation}
where $\Lambda$ is the cut-off of our effective theory. 
The \ac{lsm}, written in terms of $a$ and $\chi$ must be approximately scale-invariant and takes the form
\begin{align}%
  \label{eq:LSM}
  \Lag_{LSM} &= \frac{1}{2}\del_\mu a\del^\mu a +\frac{1}{2}b^2 a^2\del_\mu \chi \del^\mu \chi - \frac{\xi R}{2} a^2 - \frac{d-1}{2(d+1)}g a^{2(d+1)/(d-1)}\\
  & + \text{higher-derivative terms} + \text{higher-curvature terms},
\end{align} 
where $b$ and $g$ are dimensionless constants in the sense of Wilsonian couplings, $\xi =(d-1)/(4d)$ is the conformal coupling and $R$ the Ricci scalar of \mani. The form of the \ac{lsm} Lagrangian is determined by dimensional reasoning and scale invariance: the kinetic term for $a$ shows that $a$ has approximately dimension $(d-1)/2$, while $\chi$, being a pure phase, must be dimensionless. This in turn determines the kinetic term of $\chi$ to come with $a^2$. Scale invariance requires us to couple to the Ricci curvature with a power of $Ra^2$.
Lastly, a potential term proportional to $a^{\frac{2(d+1)}{d-1}}$ must be included on dimensional grounds.
In general there will also be higher-derivative terms and, depending on dimensionality, higher-curvature terms.
In writing the above Lagrangian, we have assumed a large \ac{vev} for $a$, 
\begin{equation}
  \Lambda \ll \ev{a}^{2/(d-1)} \ll u^{1/(3-d)},
\end{equation}
which we will justify in the following.

\paragraph{Ground state}
We now want to identify the classical minimal energy solution at fixed charge. To do so, we solve the Euler--Lagrange equation for the \ac{lsm}~\eqref{eq:LSM}. Since we start out with only two real \ac{dof} in the \ac{uv}, the naive counting of \ac{dof} suggests that there is only room for one Goldstone boson, corresponding to the broken global symmetry\footnote{This argument appears to apply to relativistic theories. In nonrelativistic models with a nonlocal interaction, a single complex scalar field can accommodate the spontaneous breaking of both the internal U(1) symmetry and spatial translations and rotations, see~\cite{Watanabe:2011dk}. We thank the referee for pointing this out to us.}. The rotational symmetry remains unbroken, leading to a homogeneous solution (\emph{i.e.} $\nabla\phi =0$). We write the \ac{eom} directly for this simpler case:
\begin{align}
  \fdv{\Lag_{LSM}}{\chi} - \del_t \fdv{\Lag_{LSM}}{(\del_t \chi)} &= b^2\del_t(a^2 \dot \chi) = 0,\\
  \fdv{\Lag_{LSM}}{a} - \del_t \fdv{\Lag_{LSM}}{(\del_t a)} &= b^2 a \dot\chi^2- \xi R a- g a^{\frac{2(d+1)}{d-1}-1} -\ddot a=0.
\end{align}
The first equation is nothing but the equation of charge conservation, $\del_t j^0=0$ and gives the charge density
\begin{equation}
  \rho = b^2 a^2 \dot \chi,
\end{equation}
or $Q=\rho\cdot \text{Vol}(\mani)$. The lowest-energy solution corresponds to $a=v = \text{const.}$ When plugged into the first equation, this results in the solution
\begin{align}\label{eq:GS}
  \chi &= \mu t, & \mu &=\frac{\rho}{b^2v^2}.
\end{align}
Put back into the \ac{eom} for $a$, and multiplying both sides by $v$ we find an implicit function definition for $v(\rho)$:
\begin{equation}
  \frac{\rho^2}{b^2v^2}-\xi R v^2 - g v^{2(d+1)/(d-1)}=0 .
\end{equation}
We see that via the \ac{vev}, the classical potential has acquired a centrifugal term $\propto v^{-2}$ and in consequence, its minimum is shifted away from the origin.
The explicit minimum solution for $v$ is easy to find for $R=0$, otherwise, we have to give an asymptotic expansion in the limit $\rho \gg 1$.
The relevant scale for this expansion, $\Lambda_Q$, is fixed by the charge.
This is precisely the scale that we have found in the \ac{rg}-flow analysis and marks the threshold of the \ac{eft}:
\begin{equation}
  \Lambda_Q = \rho^{1/d}.
\end{equation}
In terms of \(\Lambda_Q\), the \acp{vev} of \(a \) and \(\dot \chi\) are:
\begin{align}\label{eq:O2GroundState}
	v^2 & = (gb^2)^{(1-d)/(2d)} \Lambda_Q^{d-1} \left[ 1 - \frac{(d-1)b^2\xi R}{2d(gb^2)^{(d-1)/d}} \frac{1}{\Lambda_Q^2} + \order{\Lambda_Q^{-4}} \right], \\
	\mu &= \frac{1}{b^2}(gb^2)^{(d-1)/(2d)} \Lambda_Q \left[1 + \frac{(d-1)b^2\xi R}{2d(gb^2)^{(d-1)/d}} \frac{1}{\Lambda_Q^2}+ \order{\Lambda_Q^{-4}} \right].
\end{align}
At large charge, this result justifies our earlier assumption of $a$ having a large \ac{vev}.

We are now ready to calculate the energy of the ground state by performing the Legendre transform of the \ac{lsm} Lagrangian~\eqref{eq:LSM} and plugging in the lowest-energy solution~\eqref{eq:GS}. The resulting Hamiltonian is given by
\begin{align}\label{eq:HLSM}
	\mathcal{H}_{\ac{lsm}} &= \frac{1}{2}P_a^2 +\frac{1}{2}(\nabla a)^2 + \frac{1}{2b^2a^2}P_\chi^2 +\frac{1}{2}b^2a^2(\nabla \chi)^2 + \frac{\xi R}{2} a^2 + \frac{d-1}{2(d+1)}g a^{\frac{2(d+1)}{d-1}} + \dots,
\end{align}
where $P_a = \dot a$ and $P_\chi = b^2a^2 \dot \chi = \rho$.
The resulting energy density of the ground state is given by
\begin{equation}\label{eq:energy-density-GS}
  \frac{E_0}{\text{Vol}{\mani}} = \frac{d  g^{(d-1)/(2 d)} }{(d+1) b^{(d+1)/d}} \Lambda_Q^{d+1} \bqty{ 1  + \frac{(d+1) \xi  R b^{2/d}  }{2 d g^{(d-1)/d}} \frac{1}{\Lambda_Q^2} +  \order{\Lambda_Q^{-4}}} .
\end{equation}

\paragraph{Fluctuations}

The classical lowest-energy state at fixed and large charge we have found above serves as the vacuum around which we will now consider quantum fluctuations:
\begin{equation}
	\phi(t,x) = e^{i b \mu t}\left(\frac{v}{\sqrt{2}} + \pi(t,x) \right),
\end{equation}
where the fluctuations are parametrized by the complex field $\pi$. Expanding the \ac{lsm} Lagrangian on the vacuum solution~\eqref{eq:GS} to second order in the fields gives the following action for $\pi$ (where we did not write constant terms, which were studied in the vacuum solution):
\begin{equation}\label{eq:O2GoldstoneL}
	\Lag_\pi = D_0 \pi^* D_0 \pi - \nabla \pi \nabla \pi^* - b^2 \mu^2 \pi^*\pi - \frac{b^2 \mu^2 -\xi R}{d-1}(\pi +\pi^*)^2,
\end{equation}
where we have introduced the covariant derivative $D_0 = \del_0 -i b \mu$. 
From this expression, we can compute the inverse propagator for the real and imaginary part of \(\pi(t,x)\),
\begin{equation}
	D^{-1} = \begin{pmatrix}
		\frac{1}{2}(\nabla^2 - \del_0^2)-\frac{2(b^2 \mu^2- \xi R)}{d-1} & \mu \del_0\\
		-\mu \del_0 & \frac{1}{2}(\nabla^2 - \del_0^2)
	\end{pmatrix}.
\end{equation}
From its determinant, we find the dispersion relations:
\begin{align}
	\omega &= \frac{1}{\sqrt{d}} k \bqty{1 - \frac{(d-1) \xi R}{2d} b^{2/d} g^{1/d-1} \frac{1}{\Lambda_Q^2} +\order{\Lambda_Q^{-4}} } + \order{k^3},\label{eq:confGoldO2}\\
	\omega &=
       \begin{aligned}[t]
&\sqrt{\frac{4 d}{d-1} } \frac{g^{(d-1)/(2d)}}{b^{1/d}}  \Lambda_Q \bqty{ 1- \frac{d-2}{2d} \frac{\xi R b^{2/d}}{g^{1-1/d}} \frac{1}{\Lambda_Q^2} + \order{\Lambda_Q^{-4}}  }\\
	& + \frac{k^2}{4\Lambda_Q} \sqrt{1 - \frac{1}{d} } \frac{(2d-1) b^{1/d} }{g^{(d-1)/(2d)}} + \order{k^4} .
 \end{aligned}
\end{align}
The first observation is that, since the charge-fixing condition breaks Lorentz invariance, the modes are not simply relativistic.
We find a massless mode and a massive mode:
\begin{itemize}
\item The massless mode is, at leading order, the Goldstone associated to the spontaneous breaking of the global symmetry.
  It has a linear dispersion relation and its velocity is \(1/\sqrt{d}\). Note that this speed of sound is fully model-independent, determined solely by the underlying symmetries. It is simply the consequence of conformal invariance: the tracelessness of the stress tensor for a perfect fluid with energy density \(\rho\) and pressure \(p\) requires \(c^2 = \dv*{p}{\rho} = 1/d\).
\item The massive mode is, at leading order, the radial component of the field \(\phi\).
  As expected from the \ac{rg}-flow analysis, it has a mass of order \(\Lambda_Q\) and decouples from the low-energy theory when \(\Lambda \ll \Lambda_Q\).
\end{itemize}
\paragraph{Symmetry breaking pattern} Let us stop to see what we have learned about the symmetry breaking pattern.
We started with a conformal system having a global $O(2)$ symmetry.
The time-dependent ground state breaks boosts and time-translation invariance, as well as the global symmetry.
As discussed in Section~\ref{sec:bottom-up-approach}, the conformal invariance is non-linearly realized by the massive dilaton or radial mode, while the spontaneously broken global symmetry leads to the massless Goldstone boson with dispersion~\eqref{eq:confGoldO2}.

Starting from a system that lives in \(\setR^{d,1}\) and has conformal symmetry \(SO(d+1,2)\) times a global \(O(2)\), the breaking pattern (see also the discussion in the~\nameref{sec:introduction} and in~\cite{Monin:2016jmo}) is given by
\begin{equation}\label{eq:symmBreakingO2}
	SO(d+1,2) \times O(2) \to SO(d+1) \times D \times O(2) \leadsto SO(d+1) \times D',
\end{equation}
where $D$ is the generator of time translations and $D'=D+\mu H$, and $H$ is the generator of the global O(2) (see for example also~\cite{Nicolis:2011pv}). $D'$ is the helical symmetry of the ground state.
Here we have separated the symmetry breaking leading to massive and to massless Goldstones into two steps to emphasize the difference (a more detailed discussion will follow in Section~\ref{sec:O2n-Goldstones}). The massive mode has mass of $\order{\Lambda_Q}$ and is at the threshold of the \ac{eft}.

In the O(2), we only have two \ac{dof} to begin with. One turned out to be a massive mode which non-linearly realized the conformal invariance, and the other a massless mode corresponding to the broken global O(2) symmetry. We see that there is no room to break also the spatial $SO(d+1)$ symmetry, since this would require more \ac{dof}. In other words, the large-charge ground state of the O(2) model must be homogeneous.
In the following we will use the state-operator correspondence to compute the conformal dimensions of the lowest operators with fixed charge.
Then the \(SO(d+1)\) group that survives the first step is interpreted as the symmetry of the sphere preserved by the insertion of an operator at the origin in the cylinder frame (see Section~\ref{sec:observables}).

The above breaking pattern forms the basis for working out the \ac{nlsm}, where the effective theory is expressed in terms of the massless \ac{dof} only.

\subsection{The non-linear sigma model}%
\label{sec:non-linear-sigma-model}

As we have seen in Section~\ref{sec:linear-sigma-model}, the radial mode $a$ is massive and should thus be eliminated from the low-energy description of the theory. In principle, this is done by integrating out $a$ from Eq.~\eqref{eq:LSM}, resulting in a \ac{nlsm} action in terms of $\chi$ only. In practice, this is however not feasible. To leading order, we can use the saddle-point approximation and simply eliminate $a$ via its \ac{eom}.

To capture higher-order tree-level contributions we will constrain the form of the terms appearing in the \ac{nlsm} action via dimensional analysis and conformal invariance. This leads to an effective action in the sense of Wilson, containing infinitely many terms compatible with the symmetries of the model. Here, the power of working at large charge makes itself manifest: By examining the $Q$ scaling of the various terms, we are able to truncate the effective action after few terms by retaining only terms with a non-negative $Q$ scaling.

The leading term in the \ac{eft} for \(\chi\) is obtained using the \ac{eom} for \(a\) in the \ac{lsm}.
Up to a constant this must be
\begin{equation}%
  \label{eq:nlsm-lo}
  \Lag_{\textsc{nlsm}}[\chi] = k_0 (\del_\mu \chi \del^\mu \chi)^{(d+1)/2} + \dots,
\end{equation}
where \(k_0\) is a dimensionless constant.
This action is both scale and Weyl invariant.
The corresponding \ac{eom} is just the charge conservation for the \(O(2)\) symmetry that shifts \(\chi \to \chi + \alpha\):
\begin{equation}
  \rho = \fdv{\Lag[\chi]}{\del_0 \chi} = (d+1)k_0 \del_0 \chi (\del_\mu \chi \del^\mu \chi)^{(d-1)/2} = \text{const.}
\end{equation}
and admits the homogeneous solution \(\chi = \mu t\), where now
\begin{equation}\label{eq:mu-nlsm}
  \mu = \pqty{ \frac{\rho}{k_0 (d+1 )} }^{1/d} = \frac{\Lambda_Q}{(k_0 (d+1 ))^{1/d}} .
\end{equation}
The energy of the ground state is then
\begin{equation}
  \frac{E_0}{\text{Vol}{\mani}} = \mu \rho - \Lag[\chi] = \frac{d }{(k_0 (d+1)^{d+1})^{1/d}}  \Lambda_Q^{d+1} .
\end{equation}
It should be pointed out that the Lagrangian~\eqref{eq:nlsm-lo} is singular at the origin $\chi=0$ and cannot be used there. It must always be understood as an expansion around the fixed-charge ground state $\chi=\mu t$.
It is convenient to write the fluctuations over this homogeneous ground state by expanding the field as\footnote{Note that by abuse of notation, we again call the fluctuation $\pi$, but as opposed to the \ac{lsm} case, this is now a real field.}
\begin{equation}\label{eq:fluc-nlsm}
  \chi(t,x) = \mu t + \frac{\mu^{(1-d)/2}}{\sqrt{d(d+1)}} \pi(t,x) ,
\end{equation}
so that the second-order action for the fluctuations \(\pi\) is canonically normalized and reads
\begin{equation}\label{eq:quadrLagFluc}
  \Lag[\pi] = \frac{1}{2} \pqty{ (\del_0 \pi)^2 - \frac{1}{d} (\nabla \pi)^2  } .
\end{equation}
This is precisely the massless mode that we had found in the analysis of the fluctuations of the \ac{lsm}.
It is immediate to check that the corresponding energy-momentum tensor is traceless, \(T\indices{^\mu_\mu} = 0\).

\medskip
Now that we have found the leading term of the large-charge expansion of the effective Lagrangian, we can write the subleading terms in the \ac{eft}.
A generic term in the effective action has to obey conformal symmetry and parity.

To go beyond the leading order of the \ac{nlsm} Lagrangian, we will now invoke a number of arguments to constrain the form of the terms that can appear:
\begin{itemize}
  \item the \ac{eft} cannot have singularities, except at points where it breaks down;
	\item since we write the action in terms of the Goldstone boson $\chi$, only combinations of derivatives of $\chi$ can appear;
	\item the terms must all have energy dimension $d+1$;
	\item Lorentz invariance requires that the terms be Lorentz scalars, so $\del_\mu\chi$ must be fully contracted. This can be achieved either by contracting with the metric or (powers of) curvature tensors;
	\item the terms must obey parity invariance ($\chi \to -\chi$)\footnote{The phenomenology of generic parity-breaking systems is quite richer as shown in~\cite{Cuomo:2021qws}.};
	\item the terms must form fully Weyl-invariant combinations.
\end{itemize}
In the \ac{lsm} the combination \(\del_\mu \chi \del^\mu \chi \) appears as a source for the field \(a\) that we have integrated out; it follows that it can appear with any positive or negative power in the action.
On the other hand, to avoid singularities, there can only be an integer number of extra derivatives of \(\del \chi\) and they have to appear in the numerator.
Similarly, there can only be an integer number of curvature tensors, and they must appear at the numerator in order to have a well-defined flat-space limit.

It is possible to construct infinitely many terms fulfilling these requirements. Therefore, we now invoke the $Q$-scaling of the various ingredients and truncate all the terms in the effective Lagrangian that scale with a negative power of $Q$. Using the scaling of $\mu$ given in Eq.~\eqref{eq:mu-nlsm} and the scaling of the fluctuations in Eq.~\eqref{eq:fluc-nlsm}, we find that
\begin{align}
	\del_0 \chi &\sim Q^{1/d}, &  \del_i\chi &\sim Q^{(1-d)/(2d)}, & \del \dots \del \chi &\sim Q^{(1-d)/(2d)}.
\end{align}
We can write the most general term based on dimensional arguments and parity invariance and then consider its $Q$-scaling. Invariance under conformal symmetry will be checked in a second step. 
The most general term can schematically have the form
\begin{equation}\label{eq:general_op}
	\left( \prod_{k=1}^T(\del^{2p_k+1}\chi)_i \del^i \chi \right) (\del_\mu \chi \del^\mu \chi)^r \langle \mathcal{G}_{2s}^{(2l)}, (\del \chi)^{2s}\rangle,
\end{equation}
where the last term $\langle \mathcal{G}_{2s}^{(2l)}, (\del \chi)^{2s}\rangle$ refers to contractions of $\del \chi$ with geometric invariants $\mathcal{G}_{2s}^{(2l)}$, where the lower index refers to the number of indices and the upper to the dimension of the operator, \emph{e.g.} $\mathcal{G}_0^{(2)}=R,\ \mathcal{G}_2^{(2)}=R_{\mu\nu},\ \mathcal{G}_4^{(2)}=R_{\mu\nu\rho\sigma},\ \mathcal{G}_0^{(4)}=R_{\mu\nu\rho\sigma}R^{\mu\nu\rho\sigma}$, etc. The full geometry will always have the form $\mathbb{R} \times \mani$, which means that $\mathcal{G}_{2s}^{(2l)}$ has only non-vanishing spatial components, which have zero contraction with the homogeneous solution $\chi = \mu t$. 

The condition on the total dimension of this operator reads
\begin{equation}
	D = 2\sum_{k=1}^T p_k+2T + 2r +2l + 2s = d+1.
\end{equation}
The leading $Q$-scaling of~\eqref{eq:general_op} is given by
\begin{equation}
	D_Q=\frac{2r}{d}- (T+s)\left(\frac{d-1}{d}\right).
\end{equation}
The $Q$-scaling of the last term of Eq.~\eqref{eq:general_op}, $-s(1-1/d)$, is due to the fact that $\langle \mathcal{G}_{2s}^{(2l)}, (\del \chi)^{2s}\rangle$ contains only spatial derivatives.
Since \(d D_Q \ge 0\), adding \(D\) to both sides of the inequality yields 
the condition for a dimensionally correct term to have non-negative $Q$-scaling:
\begin{equation}
  D-d\cdot D_Q = (d+1)(T+s) + 2l + 2\sum_{k=1}^T p_k \leq d+1.
\end{equation}
If \(T \) or \(s\) are non-zero, so must be also \(p_k\) or \(l\), so the only solution that is admitted   is
\begin{align}
	T = s &=0, & l&\leq \frac{d+1}{2}.
\end{align}
This means that at fixed dimension \(d\) only a few scalar curvature invariants can be added.
The most general possible Lagrangian with non-negative $Q$ scaling has thus the form
\begin{equation}
  \label{eq:NLSM-U1}
  \Lag_{\ac{nlsm}} = \sqrt{\det(g)} \sum_{l=0}^{(d+1)/2} k_l \mathcal{G}_0^{(2l)} (\del_\mu  \chi \del^\mu \chi)^{(d+1)/2-l} .
\end{equation}
We will see in the following that in the large-\(N\) case these terms have a natural interpretation in terms of heat kernel coefficients for \(\mani\).
Each of these terms will appear in the Lagrangian with a coupling which is generically of order \(1\) and that depends only on the details of the fixed point \emph{i.e.} on the universality class of the system.

In general, the terms that we have found will not be Weyl-invariant and need a completion.
An efficient way of computing this completion is based on the observation that a Weyl transformation can always be compensated with a dilaton; and in our \ac{nlsm}, the role of the dilaton is played by $\log(\del_\mu\chi \del^\mu\chi)$, as this solves the classical \ac{eom} for $\sigma$ as given in Section~\ref{sec:bottom-up-approach}.
We can define the Weyl-invariant combination
\begin{equation}
  g'_{\mu \nu} = \pqty{ g^{\rho \sigma} \del_\rho \chi \del_\sigma \chi} g_{\mu \nu} = \norm{\dd{\chi}}^2 g_{\mu \nu}
\end{equation}
and the corresponding scalar curvature invariants will be automatically Weyl-invariant.
The Lagrangian becomes
\begin{equation}
  \begin{aligned}
	\Lag_{\ac{nlsm}} &= \sqrt{\det(g')} \sum_{l = 0}^{(d+1)/2} k_l {\mathcal{G}'}_0^{(2l)}\\
	& = \sqrt{\det(g')} \bqty{k_0  +   k_1 R' + k_2^{(1)}  (R')^2 + k_2^{(2)}  R'_{\mu \nu} {R'}^{\mu \nu} + k_2^{(3)}  {W'}^2 + \dots}.
 \end{aligned}
\end{equation}
In these terms, the \ac{nlsm} becomes a purely gravitational theory for the metric \(g'\) where \(\Lambda_Q\) plays the role of the Planck's length.%
\footnote{In string theory this would be the ``Einstein frame'' as opposed to the ``string frame'' in Eq.\eqref{eq:NLSM-U1}.}

Explicitly, the leading terms in the development read
\begin{align}
  \sqrt{\det(g')} &= \sqrt{\det(g)} \norm{\dd{\chi}}^{d+1}, \\
  \sqrt{\det(g')} R' &= \sqrt{\det(g)} \norm{\dd{\chi}}^{d-1} \pqty{  R + 2 d \frac{g^{\mu \nu} \del_{\mu \nu} \norm{\dd \chi}}{\norm{\dd{\chi}}}   + d ( d - 3 ) \frac{g^{\mu\nu} \del_\mu \norm{\dd{\chi}} \del_\nu \norm{\dd{\chi}}}{\norm{\dd{\chi}}^2} } , \\
  \sqrt{\det(g')} R'_{\mu\nu}{R'}^{\mu\nu} &= \sqrt{\det(g)} \norm{\dd{\chi}}^{d-3} \norm{ R_{\mu\nu} - (d-1) \frac{\del_{\mu\nu} \norm{\dd{\chi}}^{-1}}{\norm{\dd{\chi}}^{-1}} + \frac{1}{d-1} \frac{g^{\rho \sigma} \del_{\rho \sigma} \norm{\dd{\chi}}^{d-1}}{\norm{\dd{\chi}}^{d-1}} g_{\mu \nu}   }^2.
\end{align}
The first term is precisely the leading contribution to the \ac{nlsm} that is Weyl-invariant by itself.
The others complete the expansion in Eq.\eqref{eq:NLSM-U1} with contributions that have negative \(Q\)-scaling.
For concreteness, in \(d =2\) we have only the following two terms:
\begin{equation}
  \Lag_{\ac{nlsm}}^{d=2} = k_0 (\del_\mu \chi \del^\mu \chi)^{3/2} + k_1 R (\del_\mu \chi \del^\mu \chi)^{1/2}.
\end{equation}
In \(d = 3 \) we can have
\begin{equation}
  \Lag_{\ac{nlsm}}^{d=3} = k_0 (\del_\mu \chi \del^\mu \chi)^{2} + k_{1} R \del_\mu \chi \del^\mu \chi + \sum_{i}^3 k_2^{(i)} K_i ,
\end{equation}
where the \(k_i\) are constants that cannot be computed within the \ac{eft} and the \(K_i\) are the three quadratic invariants of \(\setR \times \mani\). A different way of arriving at the \ac{nlsm} is the \ac{ccwz} coset construction, as was done in~\cite{Monin:2016jmo}.

\medskip

Now we can repeat the analysis that we have restricted above to the leading terms and look for the fixed-charge minima of this action of the form \(\chi = \mu t\).
The ground state equations are
\begin{align}
  \rho &= \fdv{\Lag}{\del_0 \chi} = \sum_{l=0}^{(d+1)/2} k_l {\mathcal{G}'}_0^{(2l)} \pqty{d+1 - 2l} \mu^{d - 2l}, \\
  \frac{E_0}{V} &= \mu \rho - \Lag = \sum_{l=0}^{(d+1)/2} k_l {\mathcal{G}'}_0^{(2l)} \pqty{d - 2l} \mu^{d + 1 - 2l},
\end{align}
Solving for \(\mu \) in the former equation and substituting in the latter one, we  obtain an expansion for \(E_0\) as a function of the charge:
\begin{equation}
  E_0 = \sum_{l=0}^{(d+1)/2}c_{(d+1-2l)/d}  Q^{(d+1-2l)/d },
\end{equation}
where the coefficients \(c_k\) can be computed in terms of the coefficients in the Lagrangian and the geometric invariants order by order. 
From the general form we see that the \(Q\)-dimension of the terms starts at \(Q^{(d+1)/d}\) and decreases by powers of \(Q^{2/d}\).
It follows that in \(d \) dimensions we will have at most \(\floor{(d+1)/2} + 1 \) terms and that if \(d \) is even there is no classical term that scales like \(Q^0\).
Again, for concreteness, in \(d = 2\) we have only two terms that are not suppressed:
\begin{equation}
  E_0^{d=2} = c_{3/2} Q^{3/2} + c_{1/2}  Q^{1/2} + \order{Q^{-1/2}}.
\end{equation}
In \(d = 3\) there are only three terms,
\begin{equation}
  E_0^{d=3} = c_{4/3} Q^{4/3} + c_{2/3}  Q^{2/3} + c_0 + \order{Q^{-2/3}}.
\end{equation}
This is a crucial result: independently of the values of the couplings at cutoff (which enter the coefficients \(c_k\)), the energy of the ground state is under perturbative control if the charge is sufficiently large.

\medskip
Let us stop a moment to understand to which extent the \ac{lsm} and \ac{nlsm} descriptions are equivalent and useful. For one, the \ac{lsm} correctly gives the pattern of the spontaneous symmetry breaking (which for O(2) was easy to guess otherwise, but this will not be the case for larger global symmetry groups). 

From the \ac{nlsm} analysis, we have learned that $\floor{(d+3)/2}$ classical terms are not suppressed in the large-charge limit, so any effective description of the system must contain at least as many free parameters. So we see that with the terms retained explicitly in Eq.~\eqref{eq:LSM}, we can only reproduce the LO and NLO terms, which for $d=2$ is sufficient, but not in higher dimensions. 

\subsection{Quantum corrections: the conformal Goldstone}%
\label{sec:conformal-goldstone}

We have seen on general grounds in Section~\ref{sec:top-down}, the quantum corrections are controlled by inverse powers of the charge. Only one correction is not suppressed, namely the Casimir energy of the Goldstone which in general is given by the product of the speed of sound, which here is $1/\sqrt{d}$, and a geometrical factor.

Based on the quadratic action of the fluctuations Eq.~\eqref{eq:quadrLagFluc}, the Casimir energy of a scalar field on \(\mani\) can be evaluated for example with a zeta-function regularization.
Wick rotating $t\to i \tau$ and computing the determinant, we find
that the one-loop correction to the vacuum energy is given by the usual Coleman--Weinberg formula:
\begin{equation}\label{eq:ColemanWeinberg}
   E_{\text{Cas}}=\lim_{T\to\infty} \frac{1}{2T}  \log\det\left(-\partial_\tau^2-\frac{1}{d} \Laplacian_{\mani}\right)= \frac{1}{2 \sqrt{d}} \Tr(-\Laplacian_{\mani}^{1/2}) = \frac{1}{2 \sqrt{2}} \eval{ \zeta(s|\mani)}_{s= -1/2} .
\end{equation}
We will want to invoke the state-operator correspondence, so in practice, we will want to evaluate the zeta function on $\mani = S^d$.
It is convenient to distinguish the case \(d \) even from \(d \) odd~\cite{Cahn1975}.
\begin{itemize}
\item For even-dimensional spheres we have
  \begin{equation}\label{eq:zeta-evenSphere}
    \zeta(s | S^{2n}) = \frac{2}{(2n - 1)!} \sum_{j=0}^{n-1} \beta_{j,n} \sum_{k = 0}^\infty \binom{-s}{k} (-1)^k \pqty{n - \tfrac{1}{2} }^{2k} \zeta( 2s + 2k - 2j - 1, n + \tfrac{1}{2}),
  \end{equation}
  where the \(\beta_{j,n}\) are defined by
  \begin{equation}
    \sum_{j=0}^{n-1} \beta_{j,n} x^{2j} = \prod_{j=1/2}^{n- 3/2} (x^2 - j^2) ,
  \end{equation}
  and \(\zeta(s, q)\) is the Hurwitz zeta function
  \begin{equation}
    \zeta(s, q) = \sum_{n=0}^\infty \frac{1}{(q + n)^s} .
  \end{equation}
  The zeta function has simple pole when its argument is one, while the binomial coefficient has simple zeros for $s=0,\,-1,\,-2,\,\dots,1-k$. When the two coincide, they cancel each other, so the series defines a meromorphic function that has a simple poles for \( s = 1, 2, \dots, n\).
\item For odd-dimensional spheres we have
  \begin{equation}\label{eq:zeta-oddSphere}
    \zeta(s | S^{2n-1}) = \frac{2}{(2n - 2)!} \sum_{j=0}^{n-1} \alpha_{j,n} \sum_{k = 0}^\infty \binom{-s}{k} (-1)^k \pqty{n - 1 }^{2k} \zeta( 2s + 2k - 2j , n ),
  \end{equation}
  where the \(\alpha_{j,n}\) are defined by
  \begin{equation}
    \sum_{j=0}^{n-1} \alpha_{j,n} x^{2j} = \prod_{j=0}^{n- 2} (x^2 - j^2) .
  \end{equation}
  In this case, poles and zeros never coincide, so the series defines a meromorphic function that has simple poles for \( s = n - 1/2, n - 3/2, \dots \).
\end{itemize}
For \(d \) even, the zeta function is analytic in $s=-1/2$ and there is no term in the effective action that contributes at order $Q^0$, so this term can be evaluated unambiguously. For example, in the case of \(d = 2\) we obtain \(R_0 E_{\text{Cas}} = -0.0937\dots \) (for an independent numerical verification, see~\cite{delaFuente:2018qwv}).

For \(d\) odd, on the other hand, the Casimir energy corresponds to a pole of the zeta function that still needs to be regularized and the result is scheme-dependent. 
This is perfectly consistent with the observation made in Sec.~\ref{sec:linear-sigma-model} that 
when \(d = 2n -1\), there is a counterterm of $\order{Q^0}$ proportional to the $n$-th curvature invariants of $\setR \times S^{2n - 1}$. The Casimir energy depends on a Wilsonian parameter and is incalculable in the \ac{eft}.

\subsection{Calculating the conformal data}%
\label{sec:observables}

Now that we have constructed the effective theory of the low-energy \ac{dof} of our model at large charge, we can finally start calculating physical observables. In \acp{cft}, we are primarily interested in the so-called conformal data, namely the scaling dimensions of operators and the three-point function coefficients. This data, together with the operator product expansion are enough to construct any n-point correlation function in \ac{cft} thanks to the constraints that conformal invariance places on their form. 

\medskip
While so far, we have worked in Minkowski space, for the following we must work in Euclidean space, which means that we assume that an analytic continuation of time has taken place.

\paragraph{Two-point functions and the state-operator correspondence} It is easy to see that conformal invariance constrains two-point functions of scalar primaries to have the form
\begin{equation}\label{eq:twoptfn}
  \ev*{\mathcal{O}_i (x_i)\mathcal{O}_j (x_j)} = \begin{cases}
		\frac{C_{ij}}{\abs{x_i-x_j}^{2\Delta}}, & \Delta_1=\Delta_2=\Delta,\\
		0 & \text{otherwise}.
	\end{cases} 
\end{equation}
$C_{ij}$ is a constant, which by a suitable normalization of the fields is often chosen to be $C_{ij} = \delta_{ij}$, and $\Delta_i$ is the scaling or conformal dimension of the operator $\mathcal{O}_i$, in other words, under a scale transformation $x\to \lambda x$, it transforms as
\begin{equation}
	\mathcal{O}_i(x) \to \mathcal{O}'(\lambda x) = \lambda^{-\Delta_i} \mathcal{O}_i(x).
\end{equation}
Knowing the conformal dimension of the operators appearing in a two-point function thus gives us the full expression of the two-point function.

These conformal dimensions can be obtained without actually calculating the two-point function via operator insertions in the path integral by invoking another property specific to \acp{cft}, namely the state-operator correspondence. It is based on the fact that $\mathbb{R}^{d+1}$ (flat space) and $\mathbb{R} \times S^d(R_0)$ (cylinder) are related by a Weyl transformation and are therefore conformally equivalent:
\begin{equation}\label{eq:metric_flat_cyl}
  \dd{s^2}_{\text{flat}} = \dd{r}^2 + r^2 \dd{\Omega} = \frac{r^2}{R_0^2} \pqty{ \frac{R_0^2}{r^2} \dd{r}^2 + R_0^2 \dd{\Omega} } = e^{2 \tau/R_0} \pqty{ \dd{\tau}^2 + R_0^2 \dd{\Omega}} = e^{2 \tau/R_0} \dd{s^2}_{\text{cyl}} ,
\end{equation}
where \(\dd{\Omega}\) is the volume form of a unit 1-sphere and $R_0$ is the radius of the d-sphere.

Since the time direction \(\tau \) on the cylinder is mapped to the radial direction \(r\) in flat space, using radial quantization on $\mathbb{R} \times S^d$, the role of the Hamiltonian $H$ in the cylinder frame is played by the generator of dilatations $D$, and the role of the energy eigenvalue is taken by the scaling dimension.
The evolution operator takes now the form
\begin{align}
	U &= e^{iD \tau}, & \tau &= R_0 \log \frac{r}{R_0}.
\end{align}
The origin $r=0$ corresponds to $\tau= -\infty$ and \(R_0\) corresponds to \(\tau = 0\).
Inserting unity at the origin in the flat frame gives rise to the vacuum $\ket{0}$ on the cylinder, whose energy is \(0\). 
Inserting a primary operator $\mathcal{O}_\Delta$ of scaling dimension $\Delta$ at the origin generates the state $\ket{\Delta}$, which satisfies
\begin{equation}
	D\ket{\Delta} = -i\Delta \ket{\Delta}.
\end{equation}
The state-operator correspondence relates the scaling dimensions of operators inserted in flat space to  the energies of states on the sphere $S^{d}$ via
\begin{equation}\label{eq:state-opDeltaE}
  \Delta = R_0 E_{S^d}.
\end{equation}

Since we are discussing configurations of fixed charge, it is also convenient to map the charge density from one frame to the other.
Under Weyl rescaling, \(\rho\) has weight \(d\), so we have
\begin{equation}
  \rho_{\text{flat}}(r, \Omega) = \pqty{\frac{R_0}{r} }^d \rho_{\text{cyl}}(r, \Omega) = e^{-d \tau /R_0} \rho_{\text{cyl}}(\tau, \Omega) .
\end{equation}
We have seen that the lowest-energy state is homogeneous (in the cylinder frame), so we have \(\rho_{\text{cyl}} = \rho_0 = Q/(R_0^d V_d)\), where $V_d$ is the volume of the unit d-sphere, which is mapped to
\begin{equation}
  \rho_{\text{flat}}(r, \Omega) = \pqty{\frac{R_0}{r} }^d \rho_0 = \frac{Q}{r^d V_d} ,
\end{equation}
which has a singularity at \(r =0\), as expected from the insertion of a (primary) operator.
In fact, we could have used the reverse logic.
A primary field corresponds to an insertion at the origin in flat space.
Generically, this will lead to an isotropic charge configuration.
Mapping such a configuration to the cylinder results in a homogeneous density.
Generically, a homogeneous configuration will have lower energy than an inhomogeneous one, just like a primary field has lower scaling dimension than its descendants.

\medskip
In order to set the stage for computing higher correlation functions, let us spell out the calculation of the two-point function of the lowest operator $\mathcal{O}_{\Delta,Q}$ of charge $Q$ via insertions in the path integral (see also~\cite{Hellerman:2017sur,Badel:2019khk,Arias-Tamargo:2019xld,Orlando:2019skh}). In the low-energy description, this operator must be expressible in terms of the Goldstones. 
We have seen in Section~\ref{sec:bottom-up-approach} that the dilaton $\sigma$ is the (massive) field that non-linearly realizes scale invariance while $\chi$ is the field that realizes the global O(2) symmetry. Up to a normalization constant, an operator with scaling dimension $\Delta$ and charge $Q$ must have the form\footnote{Note that only in a weak-coupling limit, this operator takes the form $\mathcal{O}_{\Delta,Q} = \phi^Q$ and $\Delta \propto Q$.}
\begin{equation}
	\mathcal{O}_{\Delta,Q}(x) \propto e^{-2\Delta/(d-1)\sigma(x) f}e^{iQ\chi(x)}.
\end{equation}
Charge conservation and conformal invariance imply that only two-point functions of the form
\begin{equation}
  \ev*{\mathcal{O}_{\Delta,-Q}(x_0)\mathcal{O}_{\Delta,Q}(x_1)} = \int \DD{\chi} \DD{\sigma} e^{-2\Delta/(d-1)f(\sigma(x_0) + \sigma(x_1))} e^{-iQ(\chi(x_0)-\chi(x_1))}e^{-S_E[\chi,\sigma]}
\end{equation}
are non-vanishing, where $S_E[\chi,\sigma]$ is the Euclidean version of the \ac{lsm} action. In the limit of large charge, the path integral localizes around the saddle point, which we must however compute for the effective action including the contributions of the field insertions:
\begin{multline}
	S_{tot}[\chi, \sigma] = S_E[\chi,\sigma] + \int \dd[d+1]{x} \left[ iQ \chi(x)\left( \delta(x-x_0)- \delta(x-x_1) \right) \right.\\
	\left.+ \frac{2\Delta}{d-1}f \sigma(x) \left( \delta(x-x_0)+ \delta(x-x_1) \right)\right].
\end{multline}
To perform the calculation, we use cylindrical coordinates and go to the limit of large separation of the insertions, $x_0=0$, $x_1\to \infty$.
The flat-space metric takes the form in Eq.~\eqref{eq:metric_flat_cyl} and, imposing homogeneity, the action becomes
\begin{multline}
	S_{tot}[\chi, \sigma] = V \int_{-\infty}^\infty \dd{\tau} e^{\frac{(d+1)\tau}{R_0}}\left[\frac{1}{2}e^{-2\sigma f}(\dot\sigma^2+ f_\pi^2 \dot \chi^2) + C^{d+1} e^{-2\frac{d+1}{d-1}\sigma f} \right] +\\
	+ \lim_{\substack{\tau_0 \to -\infty\\ \tau_1 \to \infty}}  \int_{-\infty}^\infty \dd{\tau} \bqty{iQ\chi(\tau)\pqty{\delta(\tau-\tau_0)-\delta(\tau-\tau_1)}+
	\frac{2\Delta}{d-1}f\sigma(\tau)\pqty{\delta(\tau-\tau_0)+\delta(\tau-\tau_1)}} .
\end{multline}
where \(V = \vol(S^d)\). 
The resulting \ac{eom} are
\begin{equation}
  \begin{aligned}
	V e^{\frac{d+1}{R_0}\tau-2\sigma f} f_\pi^2 \pqty{\ddot{\chi} +\pqty{\frac{d+1}{R_0}-2\dot \sigma f}\dot \chi } -iQ\pqty{\delta(\tau-\tau_0)-\delta(\tau-\tau_1)} &=0,\\
	 V e^{\frac{d+1}{R_0}\tau-2\sigma f}\left( \ddot\sigma + f( f_\pi^2 \dot \chi^2- \dot \sigma^2) +\frac{d+1}{R_0}\dot \sigma \right)+ 2 V C^{d+1}\frac{d+1}{d-1}f e^{(d+1)\pqty{\frac{\tau}{R_0}-\frac{2\sigma f}{d-1}}}& \\
	 +\frac{2\Delta}{d-1}f\pqty{\delta(\tau-\tau_0)+\delta(\tau-\tau_1)} &=0.
  \end{aligned}
\end{equation}
In the interval $\tau_0<\tau<\tau_1$, we use the ansatz
\begin{align}
	\sigma &= \frac{d-1}{2fR_0}\tau -\frac{1}{f}\log (v), & v&=\text{const.},
\end{align}
which reduces the \ac{eom} to
\begin{equation}
  \begin{dcases}
	v^2 V f_\pi^2 \ddot \chi  - iQ\pqty{\delta(\tau-\tau_0)-\delta(\tau-\tau_1)} =0,\\
	\frac{v^2 V (d^2-1)}{4f R_0^2} + v^2 V f f_\pi^2 \dot \chi^2 + 2 V C^{d+1} \frac{d+1}{d-1}f v^{2\frac{d+1}{d-1}} + \frac{2\Delta}{d-1}f\pqty{\delta(\tau-\tau_0)+\delta(\tau-\tau_1)} =0.
      \end{dcases}
\end{equation} 
The first equation admits the solution
\begin{equation}
	\chi(\tau) = \frac{iQ}{2v^2V f_\pi^2}\pqty{|\tau-\tau_0| - |\tau -\tau_1|}, 
\end{equation}
so for $\tau_0<\tau<\tau_1$ we can write (note that the \ac{vev} of the field \(\chi\) is imaginary after the analytic continuation)
\begin{align}
  \chi(\tau) &=  i \mu \tau, & \mu &= \frac{Q}{2v^2 V f_\pi^2}.
\end{align}
The source part of the action evaluated at the saddle point gives
\begin{equation}
  \begin{aligned}
	S_{\text{Source}} & = iQ (\chi(\tau_0) - \chi(\tau_1)) + \frac{2\Delta}{d-1}f(\sigma(\tau_0)+\sigma(\tau_1))\\
	&= \mu Q (\tau_0 -\tau_1) +\frac{\Delta}{R_0}(\tau_0 +\tau_1) + \text{const.},
      \end{aligned}
\end{equation}
where we have separated the $\tau$-independent part which can be absorbed in the normalization of the field $\mathcal{O}_{\Delta,Q}$. Adding also the contribution of $S_E$, we find
\begin{align}
  S_{\text{tot}} = \pqty{V \Lag_E(v,\mu) -\mu Q} (\tau_1-\tau_0) + 
  \frac{\Delta}{R_0}(\tau_1+\tau_0),
\end{align}
where $V\Lag_E - \mu Q=E_0$ is precisely the energy of the large-charge ground state on a cylinder geometry.
Collecting all terms, we find for $x_0 \to 0,\, x_1\to \infty$
\begin{equation}
  \ev*{ \mathcal{O}_{\Delta,-Q}(x_0)\mathcal{O}_{\Delta,Q}(x_1)} = \frac{1}{\abs{x_0}^\Delta}\frac{1}{\abs{x_1}^\Delta} \frac{1}{\abs{x_0-x_1}^{R_0 E_0}}.
\end{equation} 
Using Eq.~\eqref{eq:state-opDeltaE}, we see that we recover the form Eq.~\eqref{eq:twoptfn} of the two-point function required by conformal invariance.
The corrections to the saddle-point result can be computed perturbatively.

\paragraph{Spectrum and scaling dimensions of states of charge $Q$} We revert here to using the state-operator correspondence in order to relate the energy of a state on the sphere to the scaling dimension of the corresponding operator. In Eq.~\eqref{eq:energy-density-GS}, we had calculated the energy density of the ground state in a sector of fixed charge $Q$. We have moreover seen that this classical result receives a quantum contribution from the Casimir energy of the Goldstone boson. Putting the classical result together with the Casimir energy Eq.~\eqref{eq:ColemanWeinberg} evaluated on $S^d$ results directly in the scaling dimension of the lowest operator with charge $Q$:
\begin{equation}
  \Delta_Q = R_0 \pqty{E_0+E_{\text{Cas}}} = \sum_{\ell=0}^{(d+1)/2}c_{(d+1-2\ell)/d}  Q^{(d+1-2\ell)/d } + R_0 E_{\text{Cas}}.
\end{equation}
In $d=2$, it is given by
\begin{equation}%
  \label{eq:EVLapSphere}
	\Delta_Q^{d=2} =c_{3/2} Q^{3/2} + c_{1/2}  Q^{1/2} -0.0937\dots + \order{Q^{-1/2}}.
 \end{equation}
 
Next we consider the spectrum of excited states. The leading-order Lagrangian Eq.~\eqref{eq:quadrLagFluc} is that of a free field, so all we need to do to find the dispersion relation of the excited states is remembering the eigenvalues of the Laplacian on the d-sphere:
\begin{equation}
	\omega_\ell = \frac{1}{R_0}\sqrt{\frac{E_{\Laplacian_{S^d}}}{d}} = \frac{1}{R_0}\sqrt{\frac{\ell(\ell+d-1)}{d}}.
\end{equation}
The energy of an excited state with $n_\ell$ modes of angular  momentum $\ell$ turned on is then given by
\begin{equation}
	E_Q^{(n_1,n_2,\dots)} = \frac{\Delta_Q}{R_0}+ \frac{1}{R_0}\sum_\ell n_\ell\omega_\ell.
\end{equation}
Let us examine a moment the first few excited states. Taking one quantum of $\ell=1$, we get the lowest excited state for which the operator dimension is increased by one. We see that this is a descendant state and that the conformal raising operator $P_\mu$ corresponds to the Goldstone with $\ell=1$.

\ytableausetup{centertableaux,boxsize=0.7em}
Excited states with $n_1=0$ are on the other hand conformal primaries, as none of the other oscillators increases the scaling dimension by an integer value. 
Acting with a single mode with $\ell=2$ on the ground state results in a spin 1 state with energy $1/R_0(\Delta_Q+\sqrt{2(d+1)/d})$.
To construct the lowest primary with a given spin we can use the standard Clebsch--Gordan decomposition.
For example, from the tensor product of two \(\ell = 2 \) oscillators we obtain the first excited scalar, since
\begin{equation}
	\ydiagram{2} \otimes \ydiagram{2} = \ydiagram{4} \oplus \ydiagram{2} \oplus \mathbf{1},
\end{equation}
which has dimension \((\Delta_Q + 2\sqrt{2(d+1)/d})\). From the product of one \(\ell =3 \) and one \(\ell = 2\) oscillator we obtain the lowest vector:
\begin{equation}
	\ydiagram{3} \otimes \ydiagram{2} = \ydiagram{5} \oplus \ydiagram{3} \oplus \ydiagram{1}.
\end{equation}
This state has scaling dimension \(\Delta_Q + \sqrt{2(d+1)/d} + \sqrt{3(d+2)/d}\).

States with higher spins have been discussed~\cite{Cuomo:2017vzg,Cuomo:2019ejv}, making use of the particle-vortex duality which has been first employed in the context of large charge in~\cite{Hellerman:2015nra}.

\paragraph{Three-point functions} Three-point functions of scalar primaries are constrained by conformal invariance to have the form
\begin{equation}
  \ev*{\mathcal{O}_{\Delta_i, Q_i}(x_i) \mathcal{O}_{\Delta_j, Q_j}(x_j)\mathcal{O}_{\Delta_k, Q_k}(x_k)} = \frac{C_{ijk}}{x_{ij}^{\Delta_i+\Delta_j-\Delta_k}x_{jk}^{\Delta_j+\Delta_k-\Delta_i}x_{ik}^{\Delta_k+\Delta_i-\Delta_j}},
\end{equation}
where $x_{ij} = \abs{x_i - x_j}$. The three-point function or fusion coefficient $C_{ijk}$ is not constrained and must be calculated independently.
Since in \acp{cft} we can use the \ac{ope} to reduce an $n$-point function to an $n-1$-point function, the knowledge of the scaling dimensions and fusion coefficients is sufficient to construct any correlation function.

The actual computation can be carried out in the same large-charge saddle-point limit discussed in the case of the two-point function above.
We have:
\begin{multline}
	\ev*{\mathcal{O}_{\Delta_i, Q_i}(x_i) \mathcal{O}_{\Delta_j, Q_j}(x_j)\mathcal{O}_{\Delta_k, Q_k}(x_k)} \propto \\
	\int \DD{\chi} \DD{\sigma}  e^{-2/(d-1) f ( \Delta_i \sigma(x_i) + \Delta_j \sigma(x_j) + \Delta_k \sigma(x_k))} e^{i Q_i \chi(x_i) + i Q_j \chi(x_j) +i Q_k \chi(x_k) }e^{-S_E[\chi,\sigma]},
\end{multline}
which, by charge conservation is non-vanishing only if \(Q_i + Q_j + Q_k = 0\).
The case \((Q_i, Q_j, Q_k) = (-Q - q, q, Q)\) with \(q \ll Q\) in the limit \(Q \gg 1\) has been studied in~\cite{Monin:2016jmo}, where the authors found that the fusion coefficient is given by
\begin{equation}
  C_{ijk} \propto C_q c_{3/2}^\delta  Q^{\delta/2}, 
\end{equation}
where \(C_q\) is a constant and \(\delta\) is the dimension of the operator of charge \(q\).%

\section{The $O(2n)$ vector model}\label{sec:O2n}

The most obvious generalization of the O(2) model that we discussed in detail in Section~\ref{sec:original-paper} is the O(2n) vector model~\cite{Alvarez-Gaume:2016vff}. Here, we have a vector with $2n$ entries and the global symmetry is O(2n).\footnote{We choose $2n$ for convenience, the case $2n+1$ can be treated analogously.} 
As we will see, the non-Abelian symmetry group will give rise to new and interesting effects. In particular, we will encounter type-II Goldstone bosons.

\subsection{Fixing the charge in non-Abelian symmetry groups}\label{sec:charge-fixing-O(2n)}

Before delving into the \ac{eft} description, let us stop a moment to think about what we mean by ``fixing a charge'' in the case of a non-Abelian symmetry group. 

The \ac{lsm} model Lagrangian of the \(O(2n)\) model is given by 
\begin{align}\label{eq:lsm_ON}
	\Lag &= \frac{1}{2}\del_\mu \phi^a \del^\mu \phi^a - \frac{1}{2} V(\phi^a\phi^a) + \text{higher-derivative and higher-curvature terms}, 
\end{align}
$a=1,\dots 2n$, living on $\mathbb{R} \times \mani$, where $\mani$ is a general d-dimensional compact homogeneous manifold, and we consider also the conformal coupling term to be part of the general potential $V$. The vector $\phi^a$ transforms as 
\begin{align}\label{eq:O2n-transf}
	\phi^a &\to M\indices{^a_b} \phi^b, & M^TM=\mathbbm{1}.
\end{align}
The Lagrangian~\eqref{eq:lsm_ON} is obviously invariant under this transformation.
The conserved current associated to the global \(O(2n)\) symmetry is now matrix-valued and has the form
\begin{equation}
  (j^\mu)^{ab} = (\phi^a\del^\mu \phi^b - \phi^b\del^\mu \phi^a).
\end{equation}
The current $j^\mu$ and with it the charge density $j^0$ is however not an invariant object, it transforms in the adjoint representation of the \(O(2n)\) symmetry as
\begin{align}
	j^0 &\to M j^0 M^{-1}, & M&\in O(2n).
\end{align}
It follows that we can decompose the current in terms of the generators of the algebra $T^A$,
\begin{align}
  j^\mu &= \sum_A j^\mu_A T^A, & T^A &\in so(2n).
\end{align}
Similarly for the conserved charge,
\begin{equation}
  Q = \int \dd{x} j^0 = \sum_{A} q_A T^A .
\end{equation}
Naively, we could assume that we can fix $\dim(O(2n))$ coefficients $q_A$. 
However, we can only fix the $\rank(O(2n))$ coefficients in the directions of the mutually commuting Cartan generators \(H^I\). This approach is one possible generalization of the one used for \(O(2)\).
We will use the notation \(\weight{q_1,q_2, \dots, q_n}\) to denote the projections of \(Q\) on the orthonormal basis of Cartan generators:
\begin{align}
  q_I &= \frac{1}{2} \ev*{Q H^I}, & \comm{H^I}{H^J} &= 0, & \ev*{H^I H^J} = 2 \delta^{IJ}.
\end{align}
The $q_I$ transform under the action of $O(2n)$, while the spectrum of the system is invariant.
The energy of a state of fixed charge $Q$ can only depend on the conjugacy class of \(Q\).
This means that the energy in general only depends on the eigenvalues of \(Q\).
Moreover, it is always possible to choose a representative of the conjugacy class that lies in the Cartan.
Equivalently, for each \(Q\) there is always an \(O(2n)\) transformation \(M\) such that
\begin{equation}\label{eq:blockdiag-q}
  MQM^{-1} = \sum_{I=1}^n  q_I H^I = \spmqty{ 0 &  q_1 \\ -  q_1 & 0  \\ & & 0 &  q_2 \\ & & - q_2 & 0 \\ & & & & \ddots  }.
\end{equation}

\subsection{Semiclassical linear sigma model description}\label{sec:O2n-lsm}

As in the O(2) model, we start with the classical analysis and identify the lowest-energy state at fixed charge.  
We fix the projections of the matrix $Q^{ab}$ on the generators\footnote{For ease of notation we fix the projections on all the Cartan generators. It is however possible to pick only part of them and leave a spectator subsector (see \emph{e.g.} the discussion in~\cite{Alvarez-Gaume:2016vff}).}:
\begin{align}
	Q^{2I,2I-1} &= \int \dd{x} \left( \dot\phi^{2I}\phi^{2I-1} - \phi^{2I}\dot\phi^{2I-1} \right) = q_I, & I&=1,\dots, n.
\end{align}
This form suggests introducing polar coordinates in each of the two-planes spanned by $(\phi^{2I-1},\phi^{2I})$:
\begin{align}
	\phi^{2I-1} &= a^I\cos \chi^I, & \phi^{2I} &= a^I\sin \chi^I,
\end{align}
so that 
\begin{equation}
	Q^{2I, 2I-1} = \int \dd{x} (a^I)^2\dot\chi^I = q_I
\end{equation}
are the canonically conjugate momenta corresponding to the $\chi^I$.
Guided by our results for O(2), we make an ansatz for a homogeneous solution $a^I(t,x)=a^I(t)$, $\chi^I(t,x)=\chi^I(t)$, which, if it exists, will have the lowest energy at fixed charge. 
The \ac{eom} under this ansatz are given by
\begin{align}
	\ddot{a}^I &= -a^I(\dot\chi^I)^2 + a^IV'\pqty{(a^1)^2+\dots+(a^n)^2},\\
	(a^I)^2\ddot{\chi}^I &=0.
\end{align}
They admit the solution
\begin{align}
	a^I &= A^I=\text{const.}, & \chi^I &= \mu t, 
\end{align}
with
\begin{align}
	\mu^2 &= {V'(v^2)},\\
	(A^I)^2 &= \frac{q_I}{\mathrm{Vol}(\mani)\mu} = \frac{q_I}{\mathrm{Vol}(\mani)\sqrt{V'(v^2)}},
\end{align}
where
\begin{equation}
  v^2 = A_1^2+\dots+A_n^2. 
\end{equation}
The $\mu$ are the same for all the $\chi^I$.
This ground state describes a circular motion in each of the two-planes. The angular velocity $\mu$ is the same in all planes, while the radius of the circle is proportional to $(q_I)^{1/2}$. The second equation above gives an implicit definition for the $A^I$. Summing over the $I$, we find
\begin{equation}\label{eq:vevO2n}
  v^2 = \frac{1}{\mathrm{Vol}(\mani)}\frac{\hat q}{\sqrt{V'(v^2)}},
\end{equation}
where
\begin{equation}
  \hat q = \sum_{I} q_I .
\end{equation}
We see that $\mu$ depends on the sum of the charges $q_I$. It is also interesting to evaluate the charge matrix on the ground state. We find that
\begin{align}
	Q^{2I, 2I'} &= Q^{2I+1, 2I'+1} =0, & Q^{2I, 2I'-1} &=\sqrt{q_Iq_{I'}}. 
\end{align}
Equivalently, we can write the charge matrix as the tensor product
\begin{equation}
  Q = \mathbf{q}^{1/2} \otimes \mathbf{q}^{1/2} \otimes J,
\end{equation}
where
\begin{align}
  \mathbf{q}^{1/2} &= \pmqty{ q_1^{1/2} & q_2^{1/2} & \dots & q_n^{1/2} }, & J &=
                                                                        \begin{pmatrix}
                                                                          0 & 1 \\ -1 & 0
                                                                        \end{pmatrix}.
\end{align}
This shows explicitly that the \(Q\) matrix is singular.
In fact, using an \(O(2n)\) transformation we can rotate the vector \(\mathbf{q}\) into \(\pmqty{\hat q & 0 & \dots}\), which shows that \(Q\) can always be written as the product
\begin{equation}
  Q = \mqty(\dmat{\hat q,0,\ddots}) \otimes J .
\end{equation}
To see this more explicitly, we observe that since the model is invariant under \(O(2n)\) transformations, we can map one solution to another via a transformation~\eqref{eq:O2n-transf}.
If we introduce complex fields $\varphi^I$,
\begin{equation}
  \varphi^I = \frac{1}{\sqrt{2}}(\phi^{2I-1} + i \phi^{2I}),
\end{equation}
and think of the $\varphi^I$ as coordinates in $\mathbb{C}^N$, $U(n) \subset O(2n)$ transformations act as a change of basis.
In our initial complex basis, the ground state is the vector 
\begin{align}\label{eq:GSO2n}
	\varphi &=\frac{1}{\sqrt{2}}(A^1,\dots, A^n) e^{i\mu t}, & \norm{\varphi}^2 = \frac{v^2}{2} .
\end{align}
We can instead pick a basis $(e^0,e^i)$ defined by
\begin{align}\label{eq:simpleframe}
  e^0 &=\frac{(A^1,\dots,A^n)}{v}, & e^0\cdot e^i&=0, & e^i\cdot e^j = \delta^{ij} ,
\end{align}
where any orthonormal choice for the $e^i$, $i=1,\dots, n-1$ will do. This makes manifest that the ground state preserves a $U(n-1)$ symmetry. 
Here, the ground state takes the form
\begin{equation}
	\varphi = \frac{1}{\sqrt{2}}(v,0,\dots, 0) e^{i\mu t},
\end{equation}
or expressed in real coordinates,\footnote{This solution can be rotated into $\phi=(v,0,\dots,0)$ with a time-dependent O(2n) transformation.}
\begin{equation}
	\phi = (v \cos \mu t, v \sin \mu t, 0, \dots, 0). 
\end{equation}
This is the basis in which the charge matrix $Q^{ab}$ lies in the Cartan and takes the block-diagonal form (see Eq.~\eqref{eq:blockdiag-q})
\begin{equation}\label{eq:transformedO2nCharge}
	Q^{ab} = \hat q H^1 = \spmqty{ 0 & \hat q \\ - \hat q & 0  \\ & & 0  \\ & & &  \ddots  }.%
\end{equation}
We see that a homogeneous minimum-energy solution at fixed charges $q_I$ can always be rotated into a configuration in which the charge is concentrated in a single direction. All homogeneous fixed-charge ground states with the same sum of charges $q$ have the same energy. 
As long as we deal with a homogeneous ground state, we can always work in the basis~\eqref{eq:simpleframe} and fall back directly to the techniques and results of the O(2) model derived in Section~\ref{sec:linear-sigma-model}.

For example the energy of the ground state is given by
\begin{equation}
	E_0 = \frac{1}{2} v^2 V'(v^2)+\frac{1}{2}V(v^2),
\end{equation}
where $v$ is implicitly defined by Eq.~\eqref{eq:vevO2n}.
So far, we have worked with general potential. The leading terms of $V$ are fixed by scale and conformal invariance to be
\begin{equation}\label{eq:scaleinvpotentialO2n}
	V(\phi^a\phi^a) = {\xi R}\phi^a\phi^a + \frac{d-1}{d+1}g (\phi^a\phi^a)^{(d+1)/(d-1)},
\end{equation}
which amounts to the same potential as for the O(2) model in Eq.~\eqref{eq:LSM}. Now the result for the energy density in the O(2n) model is simply given by Eq.~\eqref{eq:energy-density-GS}.

\subsection{Type I and type II Goldstones}\label{sec:O2n-Goldstones}

We next study the fluctuations around the homogeneous ground state, parametrized as 
\begin{equation}
	\varphi^I(t,x) = e^{i \mu t + i Y_I/v} \left(\frac{1}{\sqrt{2}} A^I + X^I\right)
\end{equation}
which we can collect in \(n\) complex fields \(\pi^I = \frac{1}{\sqrt{2}}e^{i Y_I/v} X^I\).
We are working in the frame where all the fields rotate with the same angular momentum.\footnote{We could have also written the fluctuations as $\phi^I(t,x) = e^{i\mu t}A^I/\sqrt{2}+\pi^I(t,x)$, but this would have introduced a time-dependence in the potential.} 

We expand the \ac{lsm} Lagrangian up to quadratic order around the ground state and find
\begin{equation}\label{eq:LpiO2n}
  \begin{aligned}
  \Lag_\pi &= \overline{D_0 \pi^I}D^0\pi^I - \overline{\nabla_j \pi^I}\nabla^j \pi^I- \frac{1}{2}  V\left(2 \left(\frac{A^2}{2} + \frac{A^I}{\sqrt{2}}(\pi^I + \bar \pi^I) + \bar \pi^I \pi^I\right)\right) + \dots ,\\
  &= \overline{D_0 \pi^I}D^0\pi^I - \overline{\nabla_j \pi^I}\nabla^j \pi^I - \mu^2 \bar \pi^I\pi^I-\frac{1}{2}V''(A^2 ) (A^I(\pi^I+\bar\pi^I))^2 +\dots,
\end{aligned}
\end{equation}
where
\begin{equation}
	D_0\pi^I =(\del_0 + i\mu)\pi^I.
\end{equation}
If we work instead in the basis~\eqref{eq:simpleframe}, the fluctuations in the directions $e^0$ and $e^i$ decouple. The action then becomes
\begin{equation}
	\Lag_\pi = \Lag[\pi^0] + \sum_i \Lag[\pi^i],
\end{equation}
where
\begin{align}
  \label{eq:O2n-pi0-fluctuation-Lagrangian}
	\Lag[\pi^0] &= \overline{D_0 \pi^0}D^0\pi^0 - \overline{\nabla_j \pi^0}\nabla^j \pi^0 - \mu^2\bar\pi^0\pi^0 - \frac{v^2}{2}V''(v^2)(\pi^0+\bar\pi^0)^2,\\
	\Lag[\pi^i] &= \overline{D_0 \pi^i}D^0\pi^i - \overline{\nabla_j \pi^i}\nabla^j \pi^i - \mu^2\bar\pi^i\pi^i.
\end{align}
We see that $\Lag[\pi^0]$ describes directly the type I Goldstone boson found in the O(2) model~\eqref{eq:O2GoldstoneL} and discussed in Section~\ref{sec:Goldstone}. Using the potential~\eqref{eq:scaleinvpotentialO2n}, we find again a speed of sound of $1/\sqrt{d}$. As before, the Casimir energy of this Goldstone gives a quantum correction of $\order{Q^0}$ to the classical ground state energy.

 $\Lag[\pi^i]$ is a new contribution we need to analyze.
We can now arrange the Goldstones into representations of the unbroken $U(n-1)$ symmetry. The $\pi^0$ are a singlet, while the $\pi^i$ are in the vector representation of $U(n-1)$. 
Due to this, we have a single dispersion relation for the whole vector multiplet. 
The inverse propagator for the real and imaginary part of \(\pi^i(t,x)\) is
\begin{equation}
	D^{-1} = \begin{pmatrix}
		\frac{1}{2}\del_\mu\del^\mu & \mu \del_0\\
		-\mu \del_0 & \frac{1}{2}\del_\mu\del^\mu
	\end{pmatrix}.
\end{equation}
From its determinant, we find the dispersion relations:
\begin{align}
	\omega &= \pm \omega_p \pm \mu, & \omega_p = \sqrt{p^2 + \mu^2} .
\end{align}
Ordering them,
\begin{align}
  -\mu &- \omega_p, &  \mu &- \omega_p, & \omega_p &- \mu, & \omega_p &+ \mu,
\end{align}
we see that they can be described by a pair of creation and annihilation operators associated to \((\omega_p - \mu, -\omega_p + \mu)\) and to \(-\mu -\omega_p, \mu + \omega_p\).
For large \(\mu\) and small \(p\), the first pair corresponds to type-II Goldstone bosons:
\begin{equation}\label{eq:wrongTypeIIdisp}
  \omega = \frac{p^2}{2\mu} +\order{\mu^{-3}}.
\end{equation}
The other pair is a massive excitation with \(m = 2 \mu\).

To summarize, we have found the following fluctuations around the homogeneous ground state of the O(2n) model:
\begin{itemize}
	\item a universal sector already present for $n=1$ governed by a type I Goldstone with the "conformal" speed of sound of $1/\sqrt{d}$. We have seen in the O(2) model that this massless mode comes together with the massive radial mode we had associated to the dilaton which non-linearly realizes the conformal symmetry. 
	\item a sector of $n-1$ massless modes with the quadratic dispersion relation typical of the type II Goldstone. They are paired with $n-1$ massive modes. 
\end{itemize}
In the limit of $\mu \to \infty$, the massive mode is non-propagating and the Goldstones are described by the action for a Schrödinger particle. To see this, we observe that if $\mu$ is much larger than the typical time variation of the field, $\mu \gg \del_0$, the Lagrangian for $\pi^i$ takes the form of Eq.~\eqref{eq:SchroedingerParticle}.

While we have obtained the right qualitative picture, at the order we are working (\emph{i.e.} non-negative $Q$-scaling), we cannot make any quantitative predictions at $\order{\mu^{-1}}$ appearing in the leading term of Eq.~\eqref{eq:wrongTypeIIdisp}. We will see in the explicit example of the O(4) model discussed in Section~\ref{sec:NLSM-O2n} that there is a higher-order operator that changes the dispersion relation by a coefficient of $\order{1}$.

\paragraph{Canonical quantization}

If we want to proceed to a canonical quantization, we need to find the wave functions associated to the zero of the kinetic term.
The wave equation is
\begin{equation}
  \pqty{ D_0^2 + p^2 + \mu^2} \pi_p(t) e^{i p x} = 0,
\end{equation}
and as it is standard in \ac{qft},
\begin{equation}
  \pi_p(t) = a(p) e^{-i (\omega_p + \mu) t} + b^\dagger(-p) e^{i(\omega_p - \mu )t}.
\end{equation}
Hence,
\begin{equation}
  \pi(t,x) = \int [\dd{p}] \bqty{ a(p) e^{-i (\omega_p + \mu)t} + b^\dagger(-p) e^{i (\omega_p - \mu)t}} e^{i p x}
\end{equation}
for some measure \([\dd{p}]\) to be determined.
The canonically conjugate momentum is
\begin{equation}
  \Pi_\pi  = D_0 \pi = -i \int [\dd{p}] \omega_p \bqty{ a(p) e^{-i (\omega_p + \mu)t} - b^\dagger(-p) e^{i (\omega_p - \mu)t}} e^{i p x}.
\end{equation}
Imposing the canonical commutation relations, we obtain
\begin{align}
  \comm{a(p)}{a^\dagger(p')} &= \delta(p-p'), & \comm{b(p)}{b^\dagger(p')} &= \delta(p-p'), \\ \comm{a(p)}{b(p')} &= 0, & 
 [\dd{p}] &= \frac{\dd[d]{p}}{(2\pi)^{d/2} \sqrt{2 \omega_p}},
\end{align}
and the free Hamiltonian becomes
\begin{equation}
  H = \int \dd[d]{p} \pqty{ (\omega_p + \mu) a^\dagger(p) a(p) + (\omega_p - \mu) b^\dagger(p) b(p)} + \text{zero-point energy}.
\end{equation}
The oscillators \(b(p)\) and \(b^\dagger(p)\) represent the type-II Goldstone bosons for large \(\mu\), and there is one for each \(\pi^I\), \(I = 1, \dots, n-1\). We have obtained a redefinition of the problem in terms of separate particles and antiparticles that need not have the same energy as Lorentz invariance is broken. The same result could have been obtained using the classical Foldy--Wouthuysen transformation~\cite{Foldy:1950}.

\medskip
For completeness, we present also the canonical quantization of \(\pi^0\), which turns out to be surprisingly involved.
We can write the Lagrangian for \(\pi^0\) in Eq.~(\ref{eq:O2n-pi0-fluctuation-Lagrangian}) in terms of the real and imaginary parts of \(\pi^0 = (X + i Y)/\sqrt{2}\):
\begin{equation}
  \Lag[\pi^0] = \frac{1}{2} \del_\mu X \del^\mu X + \frac{1}{2} \del_\mu Y \del^\mu Y + \mu \pqty{X \del_0Y - \del_0 X Y} - \frac{1}{2} m^2 X^2 ,
\end{equation}
where \(m^2 = 2 v^2 V''(v^2)\).
The analysis is very similar and the free wave equation defined by the inverse propagator is
\begin{equation}
  \begin{pmatrix}
    \frac{1}{2}(\del_\mu \del^\mu - m^2) &  \mu \del_0 \\
    - \mu \del_0 &  \frac{1}{2}\del_\mu \del^\mu
  \end{pmatrix}
  \begin{pmatrix}
    X \\ Y 
  \end{pmatrix} = 0.
\end{equation}
Acting on plane waves \(e^{-i \omega t + i p x}\), the operator becomes
\begin{equation}
  \begin{pmatrix}
  \frac{1}{2}(\omega^2 - p^2 -m^2) & - i \mu \omega \\
   i \mu \omega & \frac{1}{2}(\omega^2 - p^2)
\end{pmatrix}.
\end{equation}
There are four zeros to the determinant \(\Delta(\omega, p)\) of this matrix:
\begin{equation}
  \omega^2 = \omega_\pm^2,         
\end{equation}
where
\begin{align}
  \omega_\pm^2 & = p^2 + 2 \mu^2 \pqty{1 + \frac{m^2}{4\mu^2} } \pm 2 \mu \Omega_p, \\
  \Omega_p^2 &= p^2 + \mu^2 \pqty{1 + \frac{m^2}{4\mu^2} }^2 .
\end{align}
The field operators and the Hamiltonian are determined as in the case of the type-II fluctuations, but the algebra is substantially more complicated.
The standard procedure is as follows:
\begin{enumerate}
\item Assign annihilation (resp. creation) operators to the positive (resp. negative) energy solutions of the wave equation \emph{i.e.} the corresponding zero-eigenvalues of \(\eval{\Delta(\omega,p)}_{\pm\omega_{\pm}}\).
\item Expand the vector \(
  \begin{psmallmatrix}
    X & Y
  \end{psmallmatrix}
  \) in terms of the zero modes of \(\Delta(\omega,p)\) with the relevant \(a\), \(a^\dagger\), \(b\) and \(b^\dagger\) coefficients.
\item Impose the canonical commutation relations to determine the quantum properties and the measure in momentum space.
\end{enumerate}
After some algebra the results are
\begin{align}
  X(t,x) &=
  \begin{aligned}[t]
{}& \int \frac{\dd[d]{p}}{(2\pi)^{d/2} \sqrt{4 \Omega_p}} \left[ i \sqrt{\frac{\omega_+}{\mu - \frac{m^2}{4\mu} + \Omega_p} } \pqty{a_p e^{-ip_+ x} - a_p^\dagger e^{i p_+ x} } \right. \\ 
  & \left.- i \sqrt{\frac{\omega_-}{-\mu + \frac{m^2}{4\mu} + \Omega_p} } \pqty{b_p e^{-ip_- x} - b_p^\dagger e^{i p_- x} } \right],
\end{aligned}
\\
  Y(t,x) &=
           \begin{aligned}[t]
{}& \int \frac{\dd[d]{p}}{(2\pi)^{d/2} \sqrt{4 \Omega_p}} \left[ i \sqrt{\frac{\mu - \frac{m^2}{4\mu} + \Omega_p}{\omega_+} } \pqty{a_p e^{-ip_+ x} + a_p^\dagger e^{i p_+ x} }\right.\\
  & \left. + i \sqrt{\frac{-\mu + \frac{m^2}{4\mu} + \Omega_p}{\omega_-} } \pqty{b_p e^{-ip_- x} + b_p^\dagger e^{i p_- x} } \right] ,
\end{aligned}
\end{align}
and the Hamiltonian is, up to the zero-point energy,
\begin{equation}
  H = \int\dd{p} \pqty{ \omega_+(p) a^\dagger(p) a(p) + \omega_-(p) b^\dagger(p) b(p)}.
\end{equation}
The explicit expression becomes rather simple in the large-\(\mu\) limit, and at leading order the field \(Y(t,x)\) is identified with the Goldstone~\cite{Alvarez-Gaume:2016vff}.

\paragraph{Casimir energy} Like before, we can compute the Casimir energy. Due to the form of the dispersion relation, this will result in a correction of $\order{1/\mu}$ to the energy of the ground state.
 In zeta-function regularization, we find
 \begin{equation}\label{eq:CasimirTypeII}
  E_{\text{CasII}} = \frac{1}{2}\frac{n-1}{2 \mu} \Tr(\Laplacian_{\mani}) = \frac{n-1}{4 \mu} \eval{ \zeta(s|\mani)}_{s= -1} .
\end{equation}
 It is well-known that the quantum Hamiltonian for a Schrödinger particle on flat space annihilates the vacuum, so there is no Casimir energy.  
This is consistent with the fact that the zeta function on a square torus of side $L$ vanishes at $s=-1$:
\begin{equation}
	\zeta(-1,T^d) = \frac{8 \pi^2}{L^2} d \zeta(-2) = 0.
\end{equation}
This is however not true in curved space.
For $\mani=S^d$, we can use the expansions in equations~\eqref{eq:zeta-evenSphere} and~\eqref{eq:zeta-oddSphere}.
Both are well-defined in $s=-1$.
For example, for \(S^2\), \(\zeta(-1| S^2) = -1/15\).
To make a quantitative statement, we need however to add higher-order terms to the effective action.
Two things will happen: the dispersion relation will be rescaled by a Wilsonian coefficient (see \emph{e.g.} Eq.\eqref{eq:O(4)-typeIIGoldstone-dispersion}), and further contributions at $\order{1/\mu}$ will appear (see \emph{e.g.} the four-derivative term in Eq.~\eqref{eq:four-derivative-term-in-O(n)}).
This is very different from the situation for the type I Goldstone which gives a universal contribution, independent of the details of the effective action.

For now, we will content ourselves with a qualitative result, which we will improve upon in the discussion of the non-linear sigma model.

\paragraph{Symmetry-breaking pattern}
In the case of the O(2n) model $(n>1)$, the symmetry breaking pattern is more interesting than for O(2). In fact, there are two ways of looking at it. One is to understand the breaking as a consequence of the helical ground state solution, the other is to see the breaking as a two-step process involving an explicit breaking due to charge fixing and a spontaneous breaking due to the ground state solution itself.
\begin{itemize}
	\item The ground state is a classical solution with helical symmetry $D'=D+\mu H$, \emph{i.e.}, a symmetry under a combined time translation and a global symmetry. It spontaneously breaks the global symmetry of the model. This gives rise to both massive and massless Goldstone bosons which realize the symmetry non-linearly. Those Goldstones commuting with the generators of the helical symmetry are massless, the others are parametrically massive in the limit of large charge and are invisible to the low-energy theory (see also the discussion in~\cite{Hellerman:2018sjf}).\footnote{Some properties of the massive Goldstones are still controlled by the symmetries alone and can be computed also in the strong-coupling regime, see the recent discussion in~\cite{Cuomo:2020gyl}.}
	\item Fixing the charge restricts us to a slice of Hilbert space, in which not the full symmetry of the system is linearly realized (as observed in Section~\ref{sec:Goldstone}, the lowest energy state is not annihilated by the charge operator \(Q\ket{0} \neq 0\)).
     We can understand this in terms of a chemical potential for a global symmetry $H$. Only the symmetries that commute with the chemical potential are linearly realized, while the others correspond to massive modes. Since these modes are parametrically heavy and do not enter the effective low-energy description, we can equivalently think of the chemical potential as an explicit symmetry-breaking term. The ground state will in general spontaneously break some of the remaining symmetries, thus leading to massless Goldstone bosons. In this sense we can think of the breaking pattern in terms of a two-step process.
\end{itemize}
The two descriptions are equivalent. 
The analysis of the case of the homogeneous ground state above provides an explicit example of this equivalence: expanding around the helical solution in Eq.~\eqref{eq:GSO2n} we obtain the action~\eqref{eq:LpiO2n} which contains a chemical potential \(\mu\) for the helical symmetry.
The generator of the chemical potential corresponds to the sum of the Cartan generators, $H=\sum_I H^I$. The system has $SO(d+1,2) \times O(2n)$ symmetry, of which $SO(d+1) \times D\times U(n)$ is linearly realized (see the discussion in the~\nameref{sec:introduction}).
The ground state breaks the latter spontaneously (with massless Goldstones) to $SO(d+1) \times D'\times U(n-1)$,
\begin{equation}
	SO(d+1,2) \times O(2n) \to SO(d+1) \times D \times U(n)  \leadsto SO(d+1)\times D' \times U(n-1),
\end{equation}
where $D$ is again the generator of time translations and $D' = D+\mu H$.
Based on this symmetry-breaking pattern, we expect $\dim\left(U(n)/U(n-1)\right) = 2n-1$ massless Goldstone degrees of freedom.
This is consistent with one type I scalar and one type II vector: since type II Goldstones account for two \ac{dof}, the counting gives precisely $1 + 2 \times (n-1) = 2n -1$ \ac{dof}.\footnote{In the first picture, the breaking takes the form $SO(d+1,2)\times O(2n) \to SO(d+1)\times D'\times O(2n-1)$.}
For $n>1$, there are now however enough \ac{dof} to admit different symmetry-breaking patterns, for example including a breaking of the spatial $SO(d+1)$ symmetry.
This allows for the possibility of an inhomogeneous ground state, which will be discussed in Section~\ref{sec:inhomGS}.

\paragraph{Scaling dimensions for the lowest operator of charge $Q$} 
Up to $\order{Q^0}$, the result for the scaling dimension of the lowest operator of charge $Q$ remains the same as for the O(2) model, as the classical contribution of the ground state can be brought to the same form as for O(2), and in both cases only one conformal Goldstone contributes to the Casimir energy:
\begin{equation}
	\Delta_Q = \sum_{l=0}^{(d+1)/2}c_{(d+1-2l)/d}  Q^{(d+1-2l)/d } + E_{\text{Cas}}.
\end{equation}
While the coefficients $c_{(d+1-2l)/d}$ depend on $n$, $E_{Cas}$ is universal for any $n$.
In $d=2$, the scaling dimension is again given by
\begin{equation}\label{eq:Delta_d2_On}
	\Delta_Q^{d=2} =c_{3/2}(n) Q^{3/2} + c_{1/2}(n)  Q^{1/2} -0.0937\dots + \order{Q^{-1/2}}.
\end{equation}
This result has been verified to high precision for O(2) and O(4) in lattice studies~\cite{Banerjee:2017fcx,Banerjee:2019jpw}.
Moreover, the universality of the $Q^0$ contribution from the conformal Goldstone has been verified independently in~\cite{delaFuente:2018qwv} by numerical means.

\subsection{Algebraic interpretation}
\label{sec:algebraic-homo-On}

Let us now interpret our results in algebraic terms.
We have fixed the charges \(\weight{q_1,q_2, \dots, q_n}\) in the directions of the Cartan and have found the state of minimal energy \(\phi_0\).
This state is homogeneous and its corresponding charge \(Q_0 = Q(\phi_0)\) belongs to the conjugacy class \(\mathcal{C}_{\hat Q}\) of the Cartan element \(\hat Q = \hat q H^1\), where \(\hat q = q_1 + q_2 + \dots + q_n\):
\begin{equation}
  \exists M \in O(2n) : M Q_0 M^{-1} = \hat Q = \spmqty{ 0 & \hat q \\ - \hat q & 0  \\ & & 0 \\ & & & \ddots  }.
\end{equation}
It follows in particular that the energy of the lowest state only depends on the sum of the charges.

We can reformulate this cylinder-frame description in terms of states inserted on the plane.
To each state we associate an operator \(\Op\) with the same charges, \emph{i.e.} weights, \(\weight{q_1,q_2, \dots, q_n}\) that belongs to some unspecified representation \(R\).
States with charges in the same conjugacy class correspond to operators in the same representation. The representative in the Cartan subalgebra corresponds to the operator with the highest weight.
Identifying the solution of lowest energy of the classical \ac{eom}, we find
	  that among all the operators \(\Op_{\weight{q_1,q_2, \dots, q_n}}^R\), the one of lowest dimension sits in the representation \(R\) with highest weight  \(\hat q = \weight{q,0, \dots, 0}\).
In the Dynkin basis (dual of the simple coroot basis) this is
\begin{equation}
  \hat q =
  \begin{cases}
    \omegaweight{q,q} & \text{for \(D_2\)} \\
    \omegaweight{q, 0, \dots, 0} & \text{for \(D_n\), \(n \ge 3\).}
  \end{cases}
\end{equation}
the lowest-dimensional operator always sits in the totally symmetric representation.\footnote{We would like to thank the authors of~\cite{Antipin:2020abu} for discussions about this point.}

\bigskip

We can ask whether we can find other solutions to the \ac{eom} with a charge which belongs to a different conjugacy class.
This will require relaxing the homogeneity hypothesis.
Such solutions correspond to operators in different representations.%
\footnote{In generalizing the $O(2)$ model we have chosen to fix the charges \weight{q_1, \dots, q_n}. The operator picture suggests yet a different approach: we could instead look for saddles in the path integral restricted to a given representation of the symmetry group.}
From this point of view the Abelian case discussed in Section~\ref{sec:original-paper} is special.
The irreps of an Abelian group have dimension one and there is no difference between weight and highest weight, so that fixing the charge is the same as fixing the representation.
This is reflected in the fact that all solutions to the \ac{eom} are necessarily homogeneous, as we have observed above.%
\subsection{Non-linear sigma model}\label{sec:NLSM-O2n}

Having identified the symmetry-breaking pattern we can write a \ac{nlsm} describing the low-energy \ac{dof}.
As discussed above, there are two equivalent ways of understanding the breaking.
In either case the coset manifold encoding the broken generators is an odd-dimensional sphere
\begin{equation}
  S^{2n-1} = \frac{O(2n)}{O(2n-1)}  = \frac{U(n)}{U(n-1)}, 
\end{equation}
which can be realized via \(2n\) fields \(\psi^a\) that satisfy the constraint
\begin{equation}
  \sum_{a=1}^{2n} \psi^a \psi^a = 1.
\end{equation}
Before writing the most general \ac{nlsm}, let us use this description to find a physical interpretation of the modes \(\pi^I\) that we have introduced in Section~\ref{sec:O2n-Goldstones}.
We repeat the same dilaton-dressing construction as for the \(O(2)\) model in Section~\ref{sec:bottom-up-approach}.
We start from a quadratic action
\begin{equation}
  \Lag_2[\psi] = \frac{f_\pi^2}{2} \sum_{a=1}^{2n} \del_\mu \psi^a \del_\mu \psi^a -  C^{d+1},
\end{equation}
and add a dilaton dressing for each operator of dimension \(k\):
\begin{equation}
  \Op_k \to e^{2(k-d-1)/(d-1) f \sigma} \Op_k .
\end{equation}
The resulting action can be recast in the form of the \ac{lsm} in Eq.~\eqref{eq:lsm_ON} if we identify the exponential of the dilaton with the overall radial mode $e^{-\sigma f}$ of the \ac{lsm} fields \(\phi^a\) and the two constants satisfy \(f f_\pi = 1\):
\begin{equation}
  \phi^a = \frac{1}{f} e^{-\sigma f} \psi^a .
\end{equation}
Apart from reproducing our previous result, this identification is interesting also for a different reason.
Expanding the radial mode around the ground state we find:
\begin{equation}
  \frac{1}{f} e^{-\sigma f} = \sqrt{\sum_{a=1}^{2n}(\phi^a \phi^a)} = \sqrt{v^2 + A^I (\pi^I + \bar \pi^I) + \pi^I \bar \pi^I} \approx v + \frac{1}{2} (\pi^0 + \bar \pi^0) + \dots  ,
\end{equation}
where we have expressed the fluctuations in the basis of Eq.~\eqref{eq:simpleframe}.
In Section~\ref{sec:O2n-Goldstones} we have seen that the universal sector \(\pi^0\) of the fluctuations describes the conformal Goldstone (with dispersion \(\omega \sim p/\sqrt{d}\)) and a massive mode.
Now we see that the real part of \(\pi^0\) describes the fluctuations of the overall radial mode, which is in turn identified with the dilaton.
For all \(O(2n)\) models the dilaton is the massive partner of the conformal Goldstone, and the respective excitations are collected into the complex field \(\pi^0\).

Now we can move on to writing the \ac{nlsm}.
The explicit construction of the terms follows the one in Section~\ref{sec:non-linear-sigma-model}.
Using the observation that the role of the dilaton is played by \(\log(\del_\mu \psi^a \del^\mu \psi^a)\), we can introduce the Weyl-invariant  combination
\begin{equation}
  g'_{\mu \nu} = \pqty{ g^{\rho \sigma} \sum_{a=1}^{2n} \del_\rho \psi^a \del_\sigma \psi^a } g_{\mu \nu} = \norm{\dd{\psi}}^2 g_{\mu \nu}
\end{equation}
and write an expansion in terms of the corresponding scalar curvature invariants:
\begin{multline}
	\Lag_{\ac{nlsm}}[g'] = \sqrt{\det(g')} \sum_{l = 0}^{(d+1)/2} k_l {\mathcal{G}'}_0^{(2l)} = \sqrt{\det(g')} \left[k_0  +   k_1 R' + k_2^{(1)}  (R')^2 \right.\\
	\left.+ k_2^{(2)}  R'_{\mu \nu} {R'}^{\mu \nu} + k_2^{(3)}  {W'}^2 + \dots\right],
\end{multline}
or equivalently,
\begin{equation}
  \label{eq:O(2n)-NLSM}
  \Lag_{\ac{nlsm}}[\psi] = \sqrt{\det(g)} \pqty{ k_0 \norm{\dd{\psi}}^{d+1} + k_1 R \norm{\dd{\psi}}^{d-1} + \dots } .
\end{equation}
The corresponding \(O(2n)\) conserved current is then
\begin{equation}
  J^{ab}_\mu = \frac{1}{\norm{\dd{\psi}}} \fdv{\Lag_{\ac{nlsm}}[\psi]}{\norm{\dd{\psi}}} \pqty{ \psi^a \del_\mu \psi^b - \psi^b \del_\mu \psi^a}.
\end{equation}
In the non-Abelian case, though, this is not enough.
The combination \(\norm{\dd{\psi}}\) is not the only Lorentz and \(O(2N)\) invariant quantity.
For example, we can consider the four-derivative term\footnote{We would like to thank Simeon Hellerman for pointing this out to us.}
\begin{equation}
  \label{eq:four-derivative-term-in-O(n)}
  K^2 = \norm{ \del_\mu \psi^a \del_\nu \psi^b - \del_\mu \psi^b \del_\nu \psi^a  }^2
\end{equation}
and add scale-invariant combinations of \(\norm{\dd{\psi}}\) and \(K\) (together with their Weyl completions) to the \ac{nlsm} actions.
We will not do this here, but we will limit ourselves to observing that if we add to the action the combination
\begin{equation}
  \Delta \Lag = \frac{3}{8} \pqty{\lambda -1} K^2 \norm{\dd{\psi}}^{d-3} ,
\end{equation}
this will change the overall coefficient of the dispersion relation for the type-II Goldstones in Eq.~\eqref{eq:wrongTypeIIdisp}.
While the overall structure and qualitative properties are protected by the symmetries, the dispersion relation becomes \(\omega = \lambda p^2/(2 \mu) \), and the vacuum energy is multiplied by a new Wilsonian parameter $\lambda$ which cannot be computed in the \ac{eft}.

\paragraph{Non-linear sigma model for O(4)} To simplify the following considerations, we will specialize our discussion to the simplest non-Abelian case, namely the O(4) vector model in $2+1$ dimensions on \(\setR \times S^2\). 

Our starting point is the \ac{nlsm}
\begin{equation}
  \Lag = \sqrt{\det(g)} \pqty{ \frac{2c_1}{3 } \norm{\dd{\psi}}^3 - 2 c_2 R \norm{\dd{\psi}} + \frac{c_1}{4 } \pqty{\lambda - 1} \frac{K^2}{\norm{\dd{\psi}}}   }.
\end{equation}
We can solve the constraint \(\psi^a \psi^a = 1\) choosing three fields \(\alpha, \beta, \gamma\):
\begin{equation}
  \label{eq:S3-parametrization}
  \begin{cases}
    \psi^1 = \sin(\gamma) \cos(\alpha), \\
    \psi^2 = \sin(\gamma) \sin(\alpha), \\
    \psi^3 = \cos(\gamma) \cos(\beta),\\
    \psi^4 = \cos(\gamma) \sin(\beta). \\
  \end{cases}
\end{equation}
This coordinate system is chosen to diagonalize the action of the two Cartans \(H_1\) and \(H_2\), which act as
\begin{align}
  \label{eq:O(4)-Cartan-action}
  H_1 &: \alpha \mapsto \alpha + \epsilon_1, & H_2 &: \beta \mapsto \beta + \epsilon_2.
\end{align}
The standard homogeneous ground state with charges \([q_1,q_2]\) is obtained for \(\gamma = \bar \gamma\), \(\alpha = \mu t\) and \(\beta = \mu t\).
The corresponding charge matrix is
\begin{equation}
  \begin{aligned}
  Q^{ab} &= 8\pi R_0^2 \pqty{c_1 \mu^2 - \frac{2}{R_0^2}  c_2}
  \begin{psmallmatrix}
    0 & \sin(\bar \gamma)^2 & 0 & \cos(\bar \gamma) \sin(\bar \gamma) \\
    -\sin(\bar \gamma)^2 & 0 & -\cos(\bar \gamma) \sin(\bar \gamma) & 0\\
    0 & \cos(\bar \gamma) \sin(\bar \gamma) & 0 & \cos(\bar \gamma)^2 \\
    -\cos(\bar \gamma) \sin(\bar \gamma) & 0 & -\cos(\bar \gamma)^2 & 0 
  \end{psmallmatrix}  \\
  &= 8\pi R_0^2 \pqty{c_1 \mu^2 - \frac{2}{R_0^2}  c_2} \pmqty{ \sin(\bar \gamma) \\ \cos(\bar \gamma)} \otimes \pmqty{ \sin(\bar \gamma) & \cos(\bar \gamma)}  \otimes J
\end{aligned}
\end{equation}
which fixes \(\mu \) and \(\bar \gamma\) to be the solutions of the equations
\begin{align}
 8 \pi R_0^2 \pqty{c_1 \mu^2 - \frac{2}{R_0^2}  c_2} \sin(\bar \gamma)^2 &= q_1, &
 8 \pi R_0^2 \pqty{c_1 \mu^2 - \frac{2}{R_0^2}  c_2} \cos(\bar \gamma)^2 &= q_2.
\end{align}
Note that the Wilsonian coefficient \(\lambda\) does not enter these expressions.
As expected and shown by the explicit form of the charge matrix, \(Q^{ab}\) can be rotated with an \(O(4)\) transformation into
\begin{equation}
  Q = \pqty{q_1 + q_2}
  \begin{pmatrix}
    0 & 1 & 0 & 0 \\
    -1 & 0 & 0 & 0 \\
    0 & 0 & 0 & 0 \\
    0 & 0 & 0 & 0 \\
  \end{pmatrix}
\end{equation}
and only depends on the projection of \(Q^{ab}\) on the Cartan generator \(H^1\), while the other projection vanishes:
\begin{align}
    \hat q_1 &= \frac{1}{2} \ev*{Q^{ab} H^1} = q_1 + q_2= q,  & \hat q_2 &= \frac{1}{2} \ev*{Q^{ab} H^2} = 0 .
\end{align}
The energy of the ground state is
\begin{equation}
  E = c_{3/2} q^{3/2} + c_{1/2} q^{1/2} + \order{q^{-1/2}},
\end{equation}
where
\begin{align}
	c_{3/2}&=\frac{1}{3\sqrt{2\pi c_1}}, & c_{1/2}&= 4c_2\sqrt{\frac{2\pi}{c_1}}.
\end{align}

The first correction to this expression comes from the Casimir energy of the quantum fluctuations.
For simplicity, we set \(q_1 = q_2 = q/2\) and the fluctuations over the ground state can be parametrized by three fields \(\pi_1\), \(\pi_2\), \(\pi_3\) defined by
\begin{equation}
  \begin{cases}
  	\alpha = \mu t + \frac{1}{2} \pi_1 - \frac{1}{\sqrt{2}} \pi_2, \\  
    \beta = \mu t + \frac{1}{2} \pi_1 + \frac{1}{\sqrt{2}} \pi_2, \\  
    \gamma = \frac{\pi}{4}  + \frac{1}{\sqrt{2}} \pi_3.  
  \end{cases}
\end{equation}
The corresponding quadratic action is
\begin{equation}
  \Lag[\pi_i] = \frac{\sqrt{\det(g)} \mu c_1}{2} \pqty{ \dot \pi_1^2 - \frac{1}{2} (\nabla \pi_1)^2 + \dot \pi_2^2 - \lambda (\nabla \pi_2)^2 + \dot \pi_3^2 - \lambda (\nabla \pi_3)^2 - 4 \mu \pi_3 \dot \pi_2 }.
\end{equation}
We recognize the expected conformal Goldstone \(\pi_1\), together with a type-II Goldstone. Note however that the parameter \(\lambda\) appears in front of the space derivatives, so that the dispersion relation of the massless mode is given by
\begin{equation}
  \label{eq:O(4)-typeIIGoldstone-dispersion}
  \omega = \abs{\lambda} \frac{\ell (\ell +1)}{2 \mu}, 
\end{equation}
where we used again the spectrum of the Laplacian on the 2-sphere as in Eq.~\eqref{eq:EVLapSphere}.

\subsection{Inhomogeneous ground states}\label{sec:inhomGS}

So far, we have only realized charge configurations in which all the charge is aligned with one Cartan generator. To access a generic charge configuration, we are forced to relax the assumption of homogeneity of the ground state.\footnote{Strictly speaking, this would require a suitable extension of the general discussion in Section~\ref{sec:Goldstone}.}
This is possible because now there are enough \ac{dof} to realize a symmetry-breaking pattern which also breaks spatial symmetries. We will explore this possibility using again the simplest non-trivial example, namely the O(4) vector model in 2+1 dimensions using the \ac{nlsm} we set up in the last subsection.

Let us now make a perturbative ansatz for an inhomogeneous solution:
\begin{align}\label{eq:inhomsol}
	\alpha & = \mu_1 t, & \beta &= \mu_2 t, & \gamma= \epsilon p(\theta, \phi) ,
\end{align}
where $\theta, \phi$ parametrize the angles of the 2-sphere.
In the limit of \(\mu_2 - \mu_1 \ll \mu_2 + \mu_1\), the \ac{eom} reduce to the form
\begin{equation}
  \label{eq:O(4)-sineGordon}
 \lambda \Delta^2_{S^2} p(\theta, \phi) + \frac{\mu_2^2 - \mu_1^2}{2 \epsilon}  \sin(\epsilon p(\theta, \phi))=0,
\end{equation}
where
\begin{equation}
  \Delta^2_{S^2}  = \pdv[2]{\theta} + \frac{\cos\theta}{\sin \theta} \pdv{\theta} + \frac{1}{\sin^2 \theta}\pdv[2]{\phi} 
\end{equation}
is the Laplacian on the sphere.
The system can be studied for any value of \(\epsilon\), but it is instructive and qualitatively equivalent to consider the limit \(\epsilon \ll 1\).
With this ansatz, the \ac{eom} simplifies to the Laplace equation on the two-sphere:
\begin{equation}
  \lambda \Delta^2_{S^2} p(\theta, \phi) + (\mu_2^2 - \mu_1^2)p(\theta, \phi)=0.
\end{equation}
If we impose smoothness of the solution we find
\begin{align}
  \label{eq:no-homo-O4-spherical-harmonic}
  p(\theta, \phi)& = Y^m_\ell (\theta, \phi), & \lambda \ell \pqty{\ell + 1} &= \mu_2^2-\mu_1^2,
\end{align}
where $Y^m_\ell$ are the usual spherical harmonics.
The further requirement that $p$ be real selects $m=0$.
Then \(p(\theta)\) is a Legendre polynomial:
\begin{equation}
  p(\theta) = \sqrt{\frac{2 \ell + 1}{4\pi}} P_\ell (\cos(\theta)).
\end{equation}
The inhomogeneity only depends on one of the angles of the sphere, replicating the symmetry breaking pattern found for the inhomogeneous ground state of the O(4) model on the torus in~\cite{Hellerman:2017efx, Hellerman:2018sjf}.
Note that in order for \(p\) to be smooth, the admissible values for \(\mu_2^2 - \mu_1^2\) are quantized.
This condition comes from the linear approximation that we are using and can be relaxed.
In particular, \(\ell = 0\) corresponds to the homogeneous solution \(\mu_2 = \mu_1\), but we have already seen that this leads necessarily to \(\hat q_2 = 0\).

The lowest-energy solution with $\hat q_2\neq 0$ corresponds to $\ell=1$ and the corresponding operator is not a Lorentz scalar, but a spin 1 field. The scale of the inhomogeneity is of the size of the sphere, which means that the \ac{eft} description is indeed applicable. The inhomogeneity of the lowest-energy solution however indicates that it cannot be the ground state of a local action. Here, we are no longer in the representation \(R\) with highest weight  \(\hat q = \weight{q,0, \dots, 0}\) as in the homogeneous case, see Section~\ref{sec:algebraic-homo-On}. To access a different representation, we need to fix the values of the higher Casimir operators, which are non-local quantities. This means that, even in principle, we cannot use a Lagrange multiplier. 

Using the normalization
\begin{equation}
	\int_{S^2} \dd{\Omega} Y_\ell^m Y_{\ell'}^{m'} = \delta_{\ell \ell'} \delta_{mm'},
\end{equation}
we find that for general values of \(\ell\) the charge matrix is in the Cartan subalgebra and has the form
\begin{equation}
  Q^{ab} =   \begin{pmatrix}
    0 & \hat q_1 & 0 & 0 \\
    -\hat q_1 & 0 & 0 & 0 \\
    0 & 0 & 0 & \hat q_2 \\
    0 & 0 & -\hat q_2 & 0 \\
  \end{pmatrix},
\end{equation}
where
\begin{align}
  \hat q_1 &= \frac{c_1 (4 \pi - \epsilon^2)  \lambda \ell (\ell + 1) \mu_1 }{\mu_2 - \mu_1} - 8 \pi c_2, \\
  \hat q_2 &= c_1  \lambda \ell (\ell + 1) \pqty{ \frac{\epsilon^2 \mu_1}{\mu_2 - \mu_1}  +  \epsilon^2 - 2 \pi } - 8 \pi c_2 .
\end{align}
The first observation is that using these expressions we can reframe the validity of our approximation in terms of charges.
The sum of the charges and the ratio \(\hat q_2/(\hat q_1 + \hat q_2)\) are given by
\begin{align}
  \label{eq:no-homo-O4-total-charge}
  \hat q_1 + \hat q_2 &= 2 \pi c_1 \ell (\ell + 1) \lambda \frac{\mu_1 + \mu_2}{\mu_2 - \mu_1} + \dots, & \frac{\hat q_2}{\hat q_1 + \hat q_2} &= \frac{\epsilon^2}{4 \pi} + \dots  .
\end{align}
From the first expression we see that the first approximation \(\mu_2 - \mu_1 \ll \mu_2 + \mu_1\) is consistent when the sum of the charges is large, \(\hat q_1 + \hat q_2 \gg 1\), while from the second expression we see that the linearization of \(\sin(p(\theta, \phi))\) obtained for \(\epsilon \ll 1\) is consistent when most of the charge is in one direction.

From the algebraic point of view, we see that these inhomogeneous solutions correspond to operators that sit in new representations of \(O(4)\), with highest weight \(\weight{\hat q_1, \hat q_2}\), that were not accessible without breaking rotational invariance.

The energy of the ground state is now
\begin{equation}
  E= c_{3/2} (\hat q_1 + \hat q_2)^{3/2} + c_{1/2} (\hat q_1 + \hat q_2)^{1/2} + \frac{\lambda \ell(\ell + 1)}{3 c_{3/2}} \frac{\hat q_2}{\sqrt{\hat q_1 + \hat q_2}} + \dots,
\end{equation}
which shows that, as expected, the lowest inhomogeneous solution is the one with \(\ell = 1\).
The energy has the form of the homogeneous case plus a correction proportional to the charge \(\hat q_2\), that vanishes in the homogeneous case.
We can identify this correction with the energy of one quantum of the type-II Goldstone fluctuation in Eq.~\eqref{eq:O(4)-typeIIGoldstone-dispersion}.
In fact, combining~Eq.~\eqref{eq:no-homo-O4-spherical-harmonic} and Eq.~\eqref{eq:no-homo-O4-total-charge}, at leading order, where \(\mu_1 \sim \mu_2 \sim \mu\), we have precisely
\begin{equation}
  \frac{\lambda \ell(\ell + 1)}{3 c_{3/2}} \frac{1}{\sqrt{\hat q_1 + \hat q_2}} = \frac{\lambda \ell (\ell + 1)}{2 \mu} + \dots 
\end{equation}
The type-II Goldstone is charged under the unbroken \(U(1)\) symmetry of the homogeneous solution: adding a quantum to the homogeneous solution of weight \(\weight{\hat q, 0}\) one obtains a state of weight \(\weight{ \hat q, 1}\) which must coincide with the linear approximation of the inhomogeneous solution that we have found here.

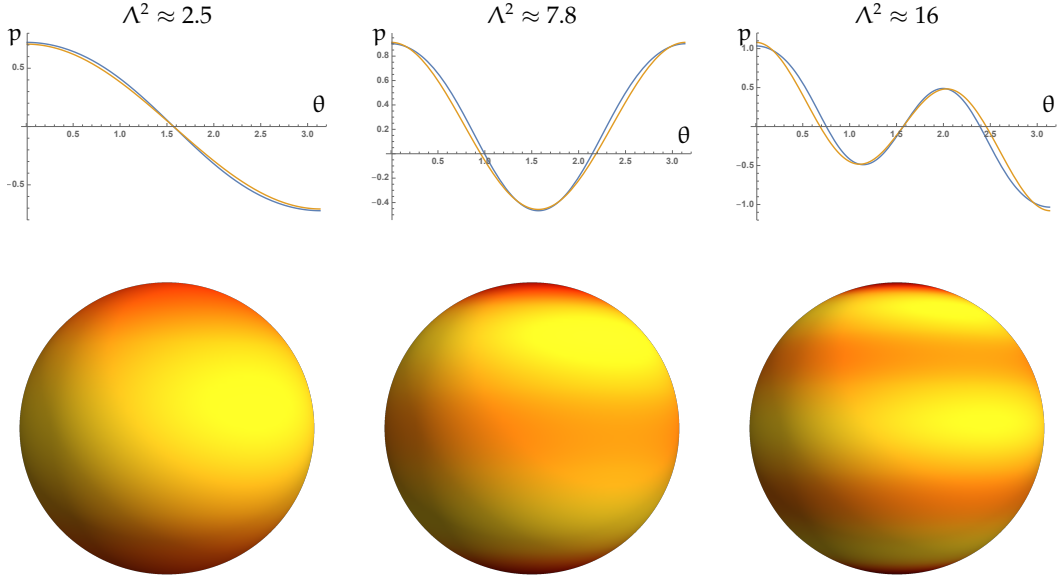
\begin{figure}
  \centering
  \begin{footnotesize}
    \begin{tikzpicture}
      \node at (-4.8,0) {\includegraphics[width=.3\textwidth]{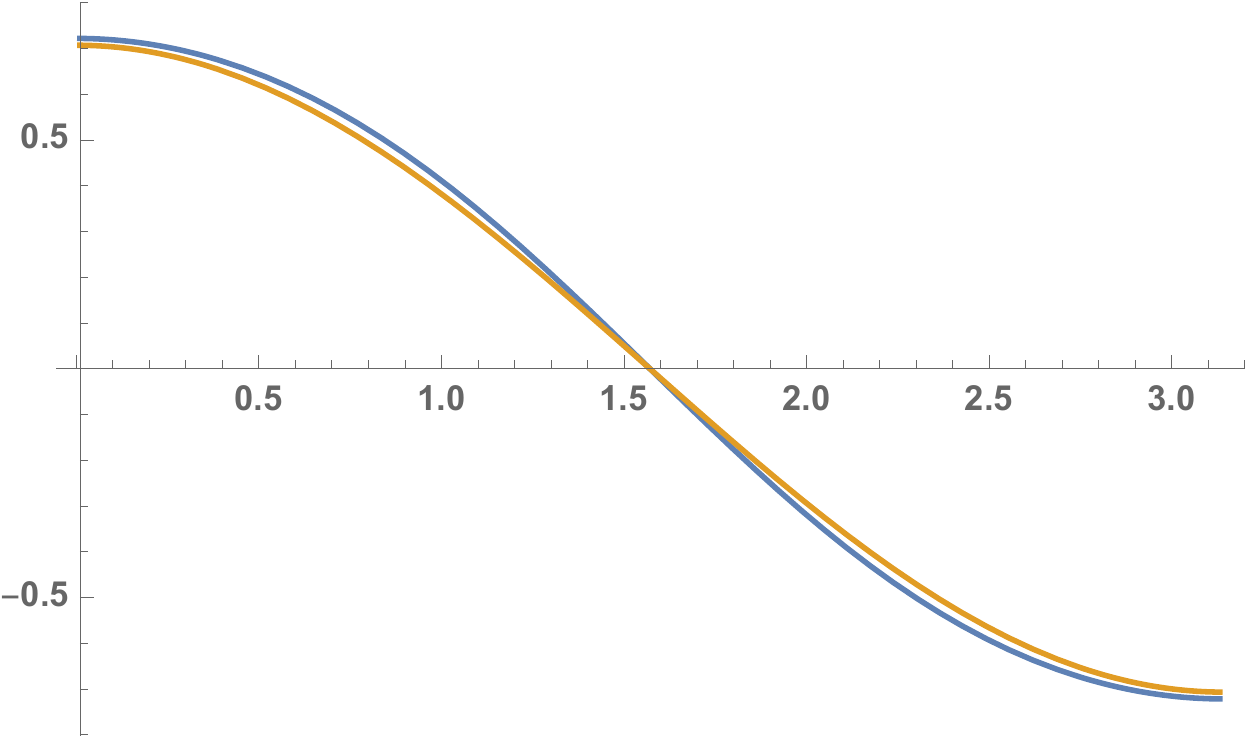}};
      \node at (0,0) {\includegraphics[width=.3\textwidth]{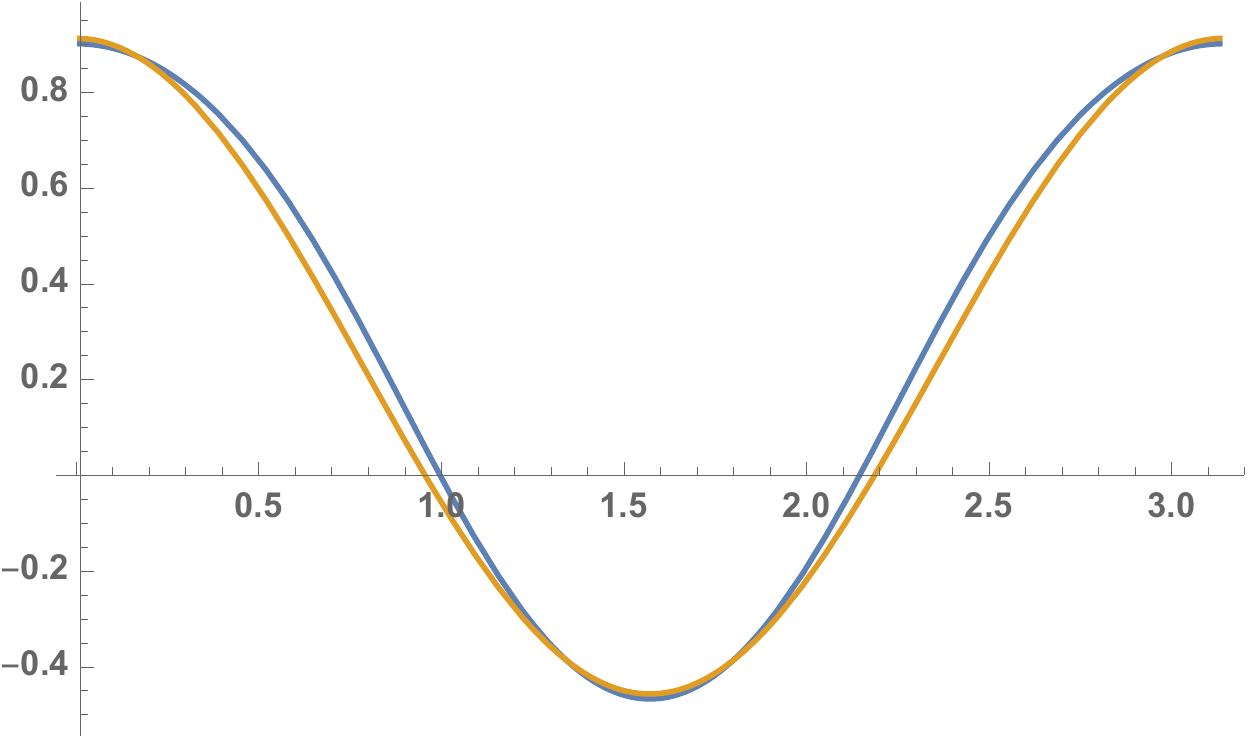}};
      \node at (4.8,0) {\includegraphics[width=.3\textwidth]{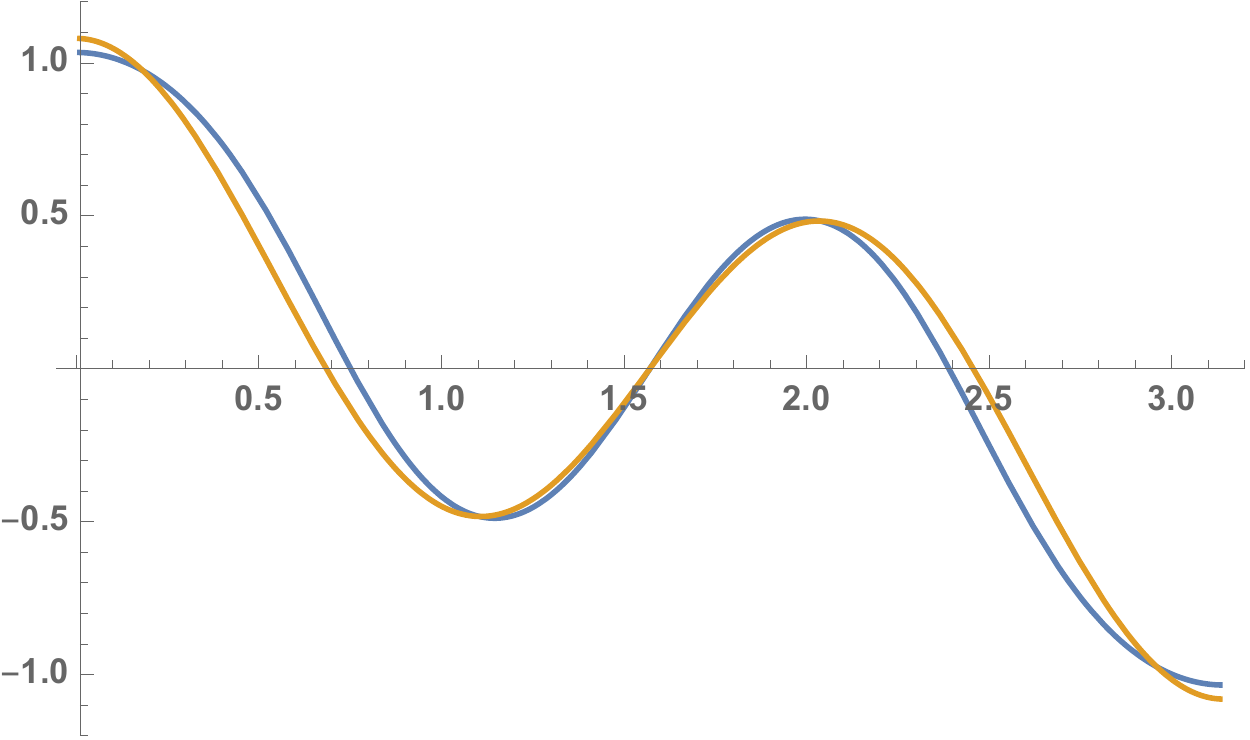}};
      \node at (-4.8,1.5) {\(\Lambda^2 \approx 2.5\)};
      \node at (0,1.5) {\(\Lambda^2 \approx 7.8\)};
      \node at (4.8,1.5) {\(\Lambda^2 \approx 16\)};

      \node at (2,-.1) {\(\theta\)};
      \node at (6.8,.3) {\(\theta\)}; 
      \node at (-2.8,.3) {\(\theta\)};

      \node at (2.8,1.2) {\(p\)};
      \node at (-2,1.2) {\(p\)};
      \node at (-6.8,1.2) {\(p\)};

      \node at (-4.8,-4) {\includegraphics[width=.3\textwidth]{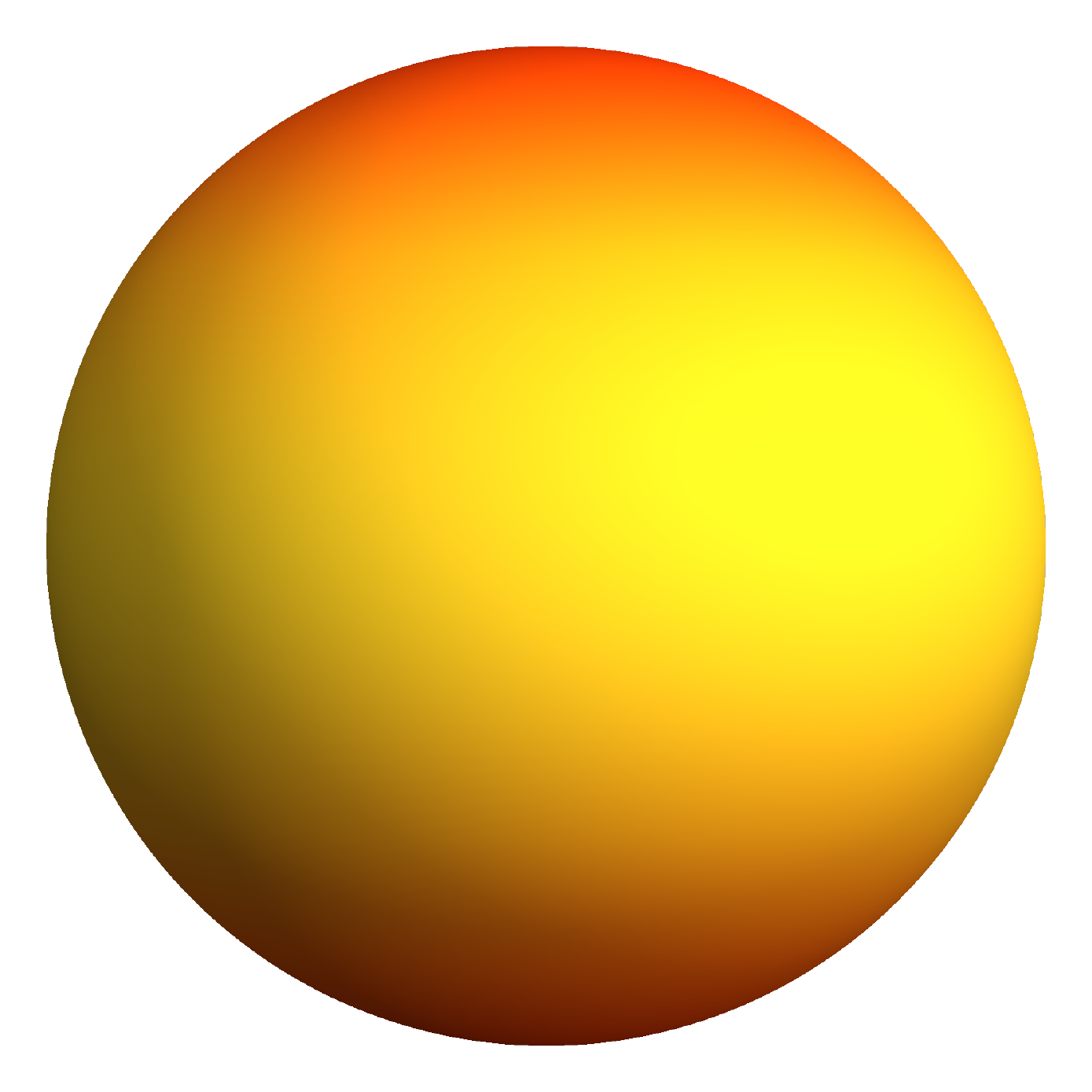}};
      \node at (0,-4) {\includegraphics[width=.3\textwidth]{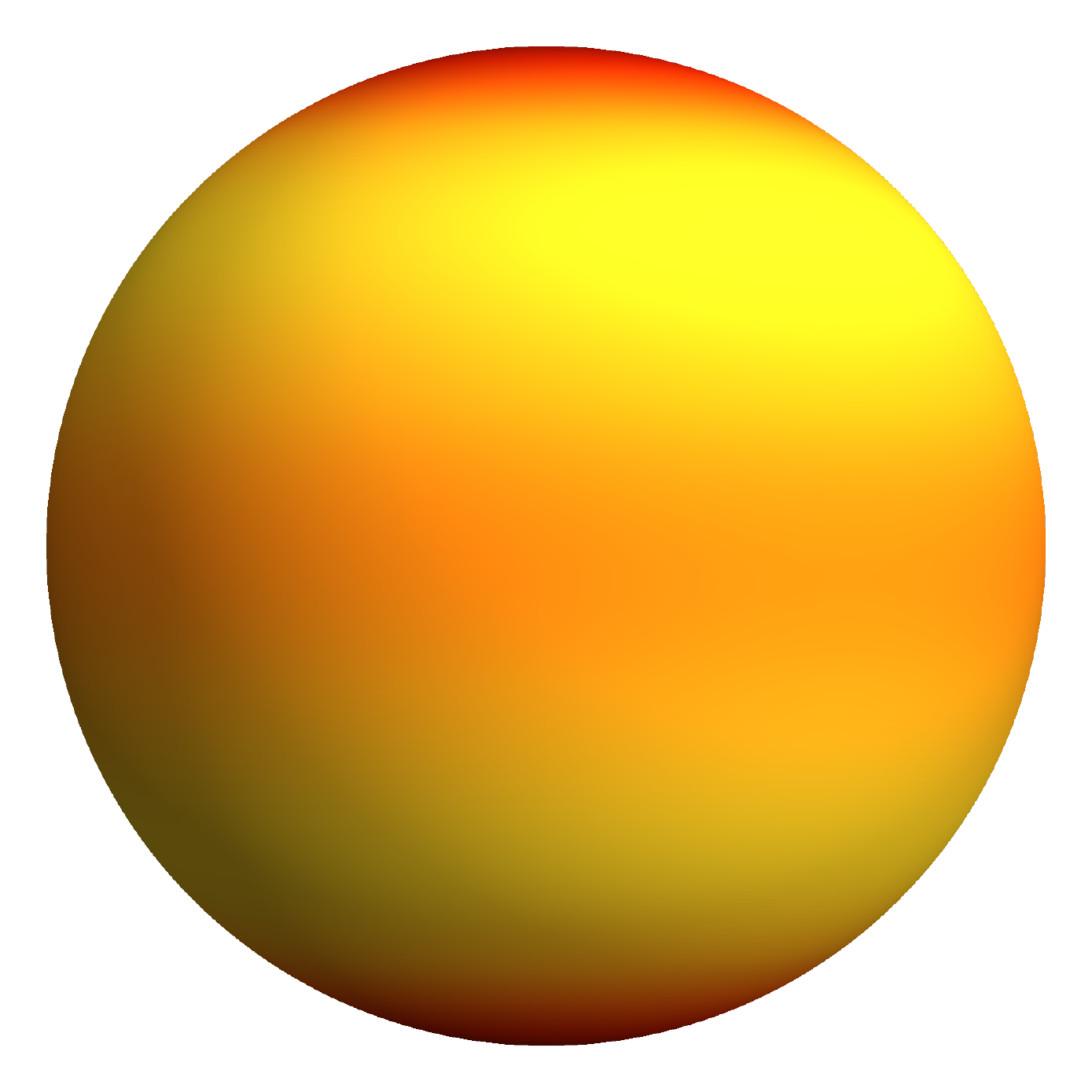}};
      \node at (4.8,-4) {\includegraphics[width=.3\textwidth]{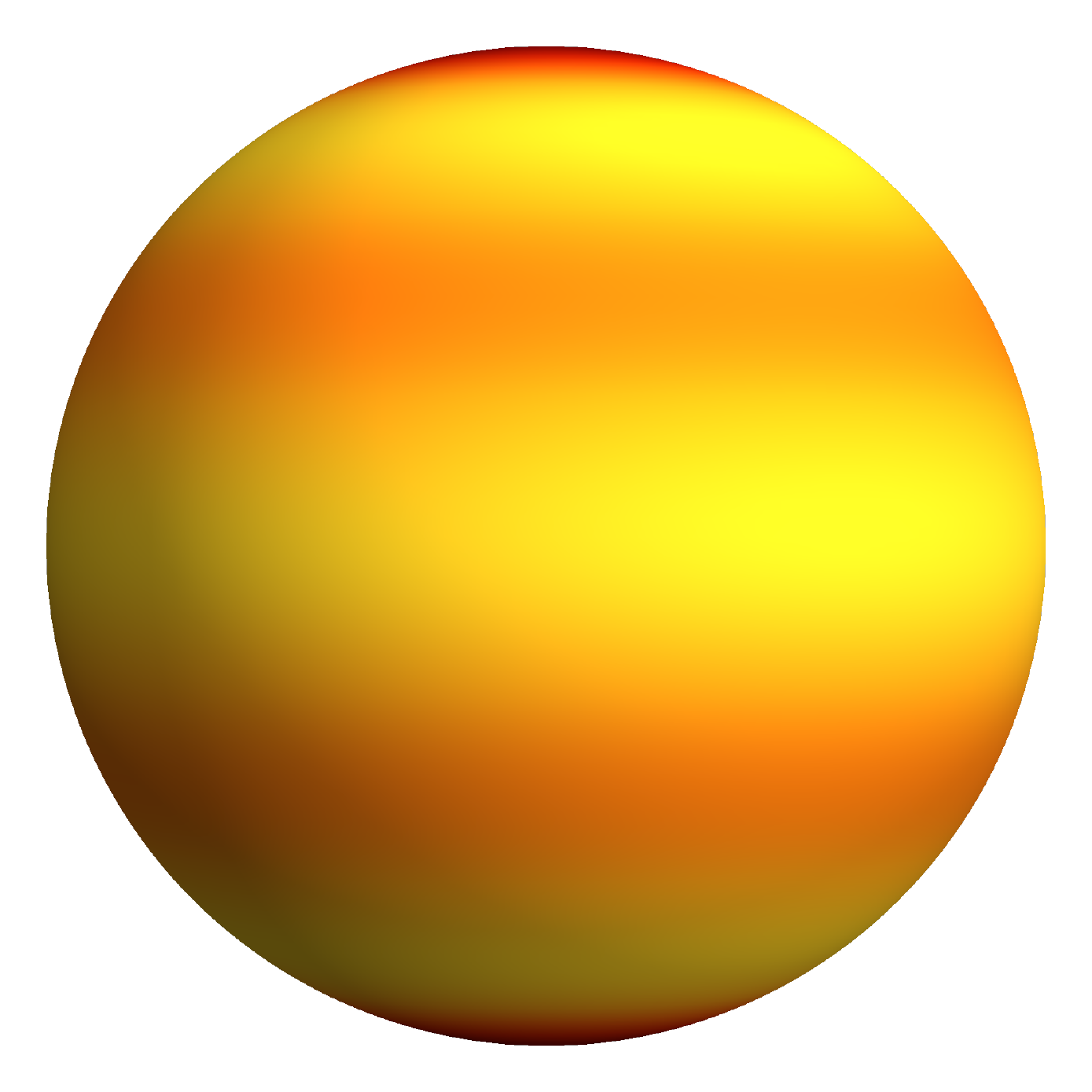}};

    \end{tikzpicture}
  \end{footnotesize}
  \caption{Solutions to the sine--Gordon equation and charge distributions on the sphere for \(\hat q_2/(\hat q_1 +  \hat q_2) = 1/6\) and \(\Lambda^2 = \mu_2^2 - \mu_1^2 \approx 2.5, 7.8, 16\) and Legendre polynomials \(P_\ell(\cos(\theta))\) for \(\ell = 1,2,3\).}
  \label{fig:p-and-legengre}
\end{figure}

The linear approximation can be relaxed (see~\cite{Banerjee:2019jpw}) and one can study (numerically) the full sine-Gordon problem in Eq.~\eqref{eq:O(4)-sineGordon}.
Since the associated linear problem admits smooth solutions for \(\Lambda^2 = \mu_2^2 - \mu_1^2 = \ell (\ell + 1)\), we expect a discrete spectrum starting around these values when \(\hat q_2\) is very small.
In the linear case, the allowed values of \(\Lambda\) do not depend on the normalization of the field \(p\).
Here on the other hand, different normalizations of \(p\) (which are fixed by the charges) will correspond to different values of \(\Lambda\).
Larger values of \(\hat q_2\) lead to larger values of \(\Lambda^2\).
The solution is qualitatively similar to the solution of the linearized problem, \emph{i.e.} the Legendre polynomial \(q(\theta ) = P_\ell(\cos(\theta))\).
Figure~\ref{fig:p-and-legengre} shows the qualitative behavior of \(p(\theta)\) for \(\hat q_2/(\hat q_1 + \hat q_2) = 1/6\) for three different allowed values of \(\Lambda^2 \) and we contrast it with the behavior of the Legendre polynomials with the same integral normalization.

The advantage of this non-linear equation is that it extends the validity of our approximation, which in this case is consistent for \(\hat q_1 + \hat q_2 \gg 1\), independently of the charge distribution between the two Cartan generators.

\paragraph{Symmetry-breaking pattern and number of Goldstone bosons}
We have seen in the previous discussion that the inhomogeneous ground state depends only on the angle $\theta$. 
Fixing the charge explicitly breaks the symmetry as follows:
\begin{equation}
  SO(3, 2) \times O(4) \to SO(3) \times D \times O(2) \times O(2), 
\end{equation}
where $D$ is the time translation $\del_t$ and the \(O(2)\)s are generated by the two Cartans \(H^1\) and \(H^2\).
The inhomogeneous ground state leads to the spontaneous symmetry breaking pattern
\begin{equation}
  SO(3) \times D \times O(2) \times O(2)  \to  SO(2) \times D',
\end{equation}
where $D' = D - (\mu_1 H^1 + \mu_2 H^2)$. We can convince ourselves of this by observing that the ground state \(\ev*{\psi}\) is annihilated by the infinitesimal generator of \(D'\),
\begin{equation}
  \pqty{ \del_t - \mu_1 \del_\alpha - \mu_2 \del_\beta } \ev*{\psi} = 0.
\end{equation}
The breaking of the global symmetries results in two Goldstone modes. 
The novelty of this solution is that it breaks the space symmetries and preserves only an SO(2).
More precisely, the generators of SO(3) are given by
\begin{align}
	T^1 &= \del_\phi,\\
	T^2 &= -\cos \phi \del_\theta + \cot \theta \sin \phi \del_\phi,\\
	T^3 &= \sin \phi \del_\theta + \cot \theta \cos \phi \del_\phi.
\end{align}
$T^1$ is unbroken since \(\ev{U}\) is function of \(t \) and \(\theta\).
In the case of spontaneously broken spacetime symmetries, there are in general fewer Goldstone bosons than broken generators as the broken generators may not be independent.
The breaking of the spatial symmetry can result in maximally two Goldstone bosons, minus the number of nontrivial solutions of the equation~\cite{Low:2001bw}
\begin{equation}\label{eq:invHiggs}
	\pqty{c_2(\phi) T^2 + c_3(\phi) T^3} \ev*{\varphi}  =0.
\end{equation}
Plugging in the generators, we find that equation~\eqref{eq:invHiggs} has one solution, $\{ c_2= \sin \phi,\ c_3 = \cos \phi\}$, so the breaking of the spatial SO(3) produces only one Goldstone \ac{dof}.
The explicit study of the fluctuations shows that the three \ac{dof} are again arranged into a type-I (conformal) Goldstone and a type-II Goldstone field (see~\cite{Banerjee:2019jpw}).

\subsection{Large n}\label{sec:O2n-largeN}

A way of going beyond the \ac{eft} treatment done so far, is to study the O(2n) model in a regime of $n \to \infty$. We can adapt standard large-$n$ methods~\cite{Zinn-Justin:572813} to working at large charge, see~\cite{Alvarez-Gaume:2019biu}. This extra control parameter enables us to verify the predictions of our \ac{eft} and study the interplay between the parameters.
While within the \ac{eft}, the model-dependent Wilsonian coefficients are not accessible, we can compute them from first principles in the path-integral formulation we adopt here.

\paragraph{The action}%
\label{sec:action}

We start with the Landau--Ginzburg model for \(2n\) real scalar fields in the vector representation of \(O(2n)\) in \((1+2) \) dimensions with Euclidean signature on \(S^1_\beta \times \Sigma\), where $\Sigma$ is a Riemann surface.\footnote{As usual, we will mostly work on $\Sigma=S^2$.}
Keeping all the terms up to mass dimension three, the \ac{uv} Lagrangian is
\begin{equation}\label{eq:UV-Hamiltonian}
  S_\theta[\varphi_I] = \sum_{i=1}^n  \int \dd{t} \dd{\Sigma} \bqty{g^{\mu\nu} \pqty{\del_\mu \varphi_I}^* \pqty{\del_\nu \varphi_I} + r \varphi_I^* \varphi_I  + \frac{u}{2}  \pqty{\varphi_I^* \varphi_I}^2 + \frac{v}{4}  \pqty{\varphi_I^* \varphi_I}^3},
\end{equation}
where \(\varphi_I\) are complex fields.

We are interested in the canonical partition function at fixed charge, where we fix the charges \(Q_I\) that act as rotations on  the complex fields \( \varphi_I \), \emph{i.e.} the Noether charges
\begin{equation}
  \hat Q_I = \int \dd{\Sigma} j_I^0 =  i \int \dd{\Sigma} \bqty{  \dot \varphi_I^* \varphi_I - \varphi_I^* \dot \varphi_I} .
\end{equation}
The partition function takes the form
\begin{equation}%
  \label{eq:canonicalPartitionFn}
  \begin{aligned}
  Z(Q_1, \dots, Q_n) &= \Tr[ e^{-\beta H} \prod_{I=1}^n \delta(\hat Q_I - Q_I)]\\
  & = \int_{-\pi}^{\pi} \prod_{I=1}^n \frac{\dd{\theta_I}}{2 \pi} \prod_{I=1}^n e^{i \theta_I Q_I} \Tr[ e^{-\beta H} \prod_{I=1}^n e^{- i \theta_I \hat Q_I}] .
\end{aligned}
\end{equation}
The integral over \(\theta_I\) can be solved in terms of an asymptotic expansion in \(1/Q_I\).
In general such an expansion will receive contributions from the end points of the integration \(\theta_I = \pm \pi\ \) and from the saddle points of the integrand.
In our case, however, the integrand is a \((2\pi)\)-periodic function of \(\theta_I\), so the contributions from the end points cancel each other and the leading contribution comes from the saddle point.

The trace describes the grand-canonical partition function for a theory with  imaginary chemical potentials \(\mu_I = i \theta_I/\beta \) associated to the currents \(j_I^0\).
The \(\theta\)-dependent terms break the original \(O(2n)\) symmetry to the \(U(n)\) that acts linearly on the complex fields \(\varphi_I\).
The saddle-point equation for \(\theta_I\) is $iQ_I + \del_{\theta_I} F_{gc}(i\theta/\beta) = 0$, where \(F_{\text{gc}}\) is the corresponding free energy.
  We will be discussing compact manifolds, so we expect \(F_{\text{gc}}\) to be smooth and the derivatives to be well-defined.  
If the theory is $CP$-invariant, \(F_{\text{gc}}\) is necessarily an even function of $\mu$.
The reason is that under $T$, the chemical potential transforms as $\mu \to - \mu$ and if the theory is $CP$ invariant, then $\mu \to  - \mu$ has to be a symmetry. %
It follows that at the saddle point $\theta$ has to be necessarily imaginary, since the derivative of $F_{gc}(\mu)$ is an odd real function:
taking the complex conjugate of the equation one finds that at the minimum $\theta_I^* = - \theta_I$, and equivalently, the \(\mu_I\) are real.

\bigskip

Since the current \(j^0\) in the trace depends on the momenta, the sum over the momenta is non-trivial.
The result can be understood in two equivalent ways: as imposing a non-trivial \ac{bc} on the fields \(\varphi_I\) around the thermal circle or as introducing a constant background field in the time direction for the gauged \(U(1)\) symmetry and keeping periodic \ac{bc}:
\begin{equation}
    \Tr[e^{-\beta H -i \theta \cdot \hat{Q}}] = \int\displaylimits_{\varphi_I(\beta, x) = \varphi_I(0,x)} \DD{\varphi_I}  e^{-S_\theta[\varphi_I]},
  \end{equation}
  where the gauged Euclidean action is
  \begin{equation}
    S_\theta[\varphi_I] = \sum_{i=1}^n  \int \dd{t} \dd{\Sigma} \bqty{g^{\mu\nu} \pqty{D_\mu^i \varphi_I}^* \pqty{D^i_\nu \varphi_I} + r \varphi_I^* \varphi_I  + \frac{u}{2}  \pqty{\varphi_I^* \varphi_I}^2 + \frac{v}{4}  \pqty{\varphi_I^* \varphi_I}^3}
  \end{equation}
  and the covariant derivative is
  \begin{equation}
    D^i_\mu \varphi_I = \begin{cases}
      \pqty{\del_0 + i \frac{\theta_I}{\beta} } \varphi_I & \text{if \(\mu = 0\)} \\
      \del_i \varphi & \text{otherwise.}
    \end{cases}
  \end{equation}
We are interested in the \ac{ir} behavior of the system at the conformal point where \(r\) is fine-tuned to remain of order \(\order{1}\), \(u\) diverges and \(v\) is generically of order \(\order{1}\) since it is dimensionless.
Since we are working at fixed charge, our problem has two intrinsic scales: the charge density \(\rho_I = Q_I/ V\) and the scale \(u\) that has mass dimension \(1\). We fix \(Q_I\) such that the hierarchy 
\begin{equation}
  \frac{1}{L} \ll \Lambda \ll \rho_I^{1/2} \ll u 
\end{equation}
is satisfied, where \(L\) is the typical scale of the surface \(\Sigma\), and \(\Lambda \) is the energy scale for the physics that we want to study.
Employing the standard approach to the vector model~\cite{Moshe:2003xn,ZinnJustin:2007zz} we use a Hubbard--Stratonovich transformation~\cite{stratonovich1957method,Hubbard:1959} and in this regime we can safely approximate the action as
\begin{equation}
  S_Q = \sum_{i=1}^n \bqty{-i   \theta_I Q_I + \int \dd{t} \dd{\Sigma} \bqty{ \pqty{D^i_\mu \varphi_I}^* \pqty{D^i_\mu \varphi_I} + \pqty{r + \lambda } \varphi_I^* \varphi_I  }} 
\end{equation}
with $D_0^i= \del_0 +i \theta_I /\beta $, and \(\lambda\) which has been promoted to a collective field from a Lagrange multiplier.

It is convenient to separate the zero modes:
\begin{align}
  \varphi_I &= \frac{1}{\sqrt{2}} A_I + \pi_I , & \ev{\pi_I} &= 0. \\
  \lambda &= \pqty{m^2 - r} + \hat \lambda , & \ev*{\hat \lambda} &= 0,
\end{align}
and perform the quadratic path integral over the \(\pi_I\), expanding the functional determinant as a formal series:
\begin{multline}
  \label{eq:expanded-tr-log-original}
  S_{\theta}[\hat \lambda] =  \sum_{i=1}^n \Bigg[ V \beta \pqty{\frac{\theta_I^2}{\beta^2} + m^2 } \frac{A_I^2}{2} +  \Tr[\log(-D_\mu^i D_\mu^i + m^2)] - \frac{A_I^2 }{2} \Tr(\hat \lambda \Delta^i \hat \lambda) \\
  - \sum_{n=2}^\infty \frac{(-1)^n}{n } \Tr(\Delta^i \hat \lambda)^n \Bigg] ,
\end{multline}
where \(\Delta\hat \lambda\) are non-local terms written in terms of the propagator as
\begin{align}
  \Tr(\Delta^i \hat \lambda)^n &= \int \dd{r_1} \dots \dd{r_n}  \hat \lambda(r_1) \dots  \hat \lambda(r_n) \prod_{i } \Delta^i(r_I - r_{i-1}), & r_0 &= r_n .
\end{align}
Our final result is an effective action for the fluctuations \(\hat \lambda\) which depends on the parameters \(m^2\), \(A_I\) and \(\theta_I\).
We have redefined the quantum structure of the problem: instead of our original fields $\varphi_I$, we now have an action in terms of $\hat\lambda$. The information about the fixed point and the symmetry-breaking pattern is contained in the zero-mode $m^2$, which acts as a \ac{rg} flow parameter.

\paragraph{The saddle point}%
\label{sec:saddle}

If we neglect the fluctuations  \(\hat \lambda \), we can identify the functional determinant with the grand-canonical (fixed chemical potential) free energy:
\begin{equation}
  \label{eq:grandcanonical-saddle}
  F^{\saddle}_{gc}(i \theta) = \sum_{i=1}^n \bqty{ V \pqty{\frac{\theta_I^2}{\beta^2} + m^2 } \frac{A_I^2}{2} +  \frac{1}{\beta}\Tr[\log(-D_\mu^i D_\mu^i + m^2)]}.
\end{equation}
The canonical (fixed charge) free energy is then
\begin{equation}
  F^{\saddle}_{c} (Q) = \eval{-i \sum_{i=1}^n \frac{\theta_I}{\beta} Q_I + F^{\saddle}_{gc}(i \theta)}_{Q_I = -i\beta \pdv{\theta_I} F_{gc}(\theta)} = \eval{- \sum_{i=1}^n \mu_I Q_I + F^{\saddle}_{gc}(\mu)}_{Q_I = \pdv{ F_{gc}}{\mu_I}} ,
\end{equation}
where the value of \(\theta\) is fixed by the saddle-point condition.
This is a non-trivial consistency check of our construction: at the saddle point we reproduce the usual Legendre transform that relates the two thermodynamic potentials.

The functional determinant that appears in Eq.~(\ref{eq:grandcanonical-saddle}) needs to be regularized.
In view of using the state/operator correspondence, we are interested in the theory living on a two-sphere. It is therefore convenient to use the zeta-function regularization as it does not affect the compactification manifold (see~\cite{Elizalde:2012zza,kirsten2001spectral} for an introduction).
In this scheme we introduce the sum over the eigenvalues of the operator \(-D_\mu D^\mu + m^2 \),
\begin{equation}
  \zeta(s | \theta, \Sigma, m) = \sum_{n \in \setZ} \sum_{p}  \pqty{ \pqty{\frac{2 \pi n}{\beta} + \frac{\theta}{\beta} }^2 + E(p)^2 + m^2}^{-s} .
\end{equation}
The \(E^2(p)\) are the eigenvalues of the Laplacian on \(\Sigma\),
\begin{equation}
  \Laplacian f_p(x) + E(p)^2 f_p (x) = 0 ,
\end{equation}
and we have used the fact that the \(\theta\)-terms shift the Matsubara frequencies on the thermal circle from \(2 \pi n/\beta\) to  \(2\pi n /\beta + \theta/ \beta\).
The functional determinant is written using the identity \(\log(x) = - \eval{\dv{x^{-s}}{s}}_{s= 0}\):
\begin{multline}
  \Tr[\log(- D_\mu D^\mu + m^2)] = - \eval{ \dv{s} \sum_{n \in \setZ} \sum_{p} \pqty{\pqty{\frac{2 \pi n}{\beta} + \frac{\theta}{\beta} }^2 + E(p)^2 + m^2 }^{-s} }_{s=0} \\ = - \eval{\dv{s} \zeta(s| \theta, \Sigma, m)}_{s=0}.
\end{multline}
For \(\beta \to \infty\), the zeta function \(\zeta(s| \theta, \Sigma, m)\) does not depend on \(\theta\), and is proportional to the zeta function on the compactification manifold \(\Sigma\):
\begin{multline}
  \label{eq:zeta-beta-infinity}
  \lim_{\beta \to \infty } \frac{1}{\beta}  \zeta(s | \theta, \Sigma, m) =  \frac{1}{2\pi} \int_{-\infty}^{\infty} \dd{\omega } \sum_p \pqty{ \pqty{\omega + i \mu  }^2 + E(p)^2 + m^2}^{-s} \\
  = \frac{\Gamma(s-\frac{1}{2})}{2 \sqrt{\pi} \Gamma(s)} \sum_p \pqty{E(p)^2 + m^2}^{1/2 - s} =\frac{\Gamma(s-\frac{1}{2})}{2 \sqrt{\pi} \Gamma(s)}  \zeta(s-\tfrac{1}{2}| \Sigma, m).
\end{multline}

The saddle-point equations are obtained by deriving the effective action with respect to  \(m^2\), \(A_I\) and \(\theta_I\) at \(\hat \lambda = 0\).  Using standard methods, we find:
\begin{equation}
  \label{eq:saddle-finite-Q-no-T}
  \begin{cases}
    \frac{1}{2} \beta V v^2 + \frac{n \beta}{2}  \zeta(\tfrac{1}{2}| \Sigma, m) = 0, \\
    i Q_I - \frac{\theta_I}{\beta} V A_I^2 = 0, & i = 1, \dots, n,\\
    2 V \beta \pqty{m^2 + \frac{\theta_I^2}{\beta^2}} A_I = 0, & i = 1, \dots, n,
  \end{cases}
\end{equation}
where %
we have introduced
\begin{equation}
  v^2 = \sum_{i=1}^n A_I^2. 
\end{equation}
The zero modes take the values
\begin{equation}
  A_I^2 = \frac{Q_I}{m V} 
\end{equation}
and the remaining equations are
\begin{equation}
  \begin{cases}
    m \zeta(\tfrac{1}{2} | \Sigma, m) = - \frac{Q}{n}, \\
    v^2 = \frac{Q}{m V},
  \end{cases}
\end{equation}
where
\begin{equation}
  Q = \sum_{i = 1}^n Q_I .
\end{equation}
As we have seen in Section~\ref{sec:O2n-lsm}, the saddle point depends only on the sum of the charges \(Q_I\). %

If we neglect the fluctuations \(\hat \lambda\), the corresponding free energy is then given by\footnote{The contribution from the zero modes to the energy vanishes at the saddle point.}
\begin{equation}
  F^{\saddle}(Q) = - \frac{1}{\beta}  \sum_{i =1}^n \bqty{ i \theta_I Q_I - \beta \zeta(-\tfrac{1}{2}| \Sigma, m)} = m Q + n \zeta(-\tfrac{1}{2}| \Sigma, m),
\end{equation}
where  \(m\) is the saddle-point value.

Our final result for the free energy at fixed charge \(Q\) at leading order in the fluctuations of \(\lambda\) on \(\setR \times \Sigma\) is
\begin{equation}
  \label{eq:result-finite-Q-zero-T-zeta}
  \begin{cases}
    F^{\saddle}_\Sigma(Q) = m Q + n \zeta(-\tfrac{1}{2}| \Sigma, m) , \\
    m \zeta(\tfrac{1}{2}| \Sigma, m) = - \frac{Q}{n} .
  \end{cases}
\end{equation}
The natural parameter that appears is \(Q/n\) which we hold fixed.
In the following we will study the systems in the limit of \(Q/n \gg 1\) in order to express the zeta functions as asymptotic expansions.

\bigskip

We use Mellin's representation of the zeta function,
\begin{equation}
  \zeta(s |  \Sigma, m) = \frac{1}{\Gamma(s)} \int_0^\infty \frac{\dd{t}}{t} t^s e^{-m^2 t } \Tr[e^{\Laplacian_\Sigma{} t}].
\end{equation}
In the limit of large \(Q/n\), we expect \(m\) to be parametrically large, so we can use the corresponding asymptotic expansion where the integral localizes around \(t \to 0\).
The trace over the eigenvalues of the Laplacian is expressed using Weyl's asymptotic formula in terms of heat kernel coefficients:
\begin{equation}
  \Tr(e^{\Laplacian_\Sigma{} t}) = \sum_{n=0}^{\infty} K_n t^{n/2 - 1} .
\end{equation}
The heat kernel coefficients \(K_n\) can be computed using geometric invariants of the surface \(\Sigma\) (see~\cite{rosenberg_1997,Vassilevich:2003xt} for a detailed introduction) and one finds that if \(\Sigma\) has no boundary, all the odd coefficients vanish, \(K_{2n+1} = 0\).
The leading coefficients are given in terms of the volume \(V\) and the scalar curvature \(R\) of \(\Sigma\),
\begin{align}
  K_0 &= \frac{V}{4 \pi}, & K_2 &= \frac{V R}{24 \pi} . 
\end{align}
Each consecutive order in the large-charge expansion is obtained by taking the next term in the heat kernel expansion.
This gives a precise geometrical interpretation of our expansion large-charge expansion:
the leading order is fixed by the volume, the first correction by the scalar curvature and so on.
This structure is of course completely consistent with the general form of the \ac{nlsm} found in Eq.~(\ref{eq:NLSM-U1}).
In the case of the unit sphere we can use the explicit integral representation of the heat kernel~\cite{McKean:1967xf} and express \(m\) and \(F\) as asymptotic expansions in \(1/Q\):
\begin{align}
  m_{S^2}(Q) &= \pqty{\frac{Q}{2n} }^{1/2} + \frac{1}{12} \pqty{\frac{Q}{2n} }^{-1/2} + \frac{7}{1440} \pqty{\frac{Q}{2n} }^{-3/2} + \frac{71}{120960} \pqty{\frac{Q}{2n} }^{-5/2} + \dots \\
  \label{eq:conformal-dimensions-large-Q}
  \frac{F^{\saddle}_{S^2}(Q)}{2n} &= \frac{2}{3} \pqty{\frac{Q}{2n} }^{3/2} + \frac{1}{6} \pqty{\frac{Q}{2n} }^{1/2} - \frac{7}{720} \pqty{\frac{Q}{2n} }^{-1/2} - \frac{71}{181440} \pqty{\frac{Q}{2n} }^{-3/2} +\dots
\end{align}
We find precisely the expected large-charge expansion with the powers of \(Q\) we found in the \ac{eft}, see Eq.~\eqref{eq:Delta_d2_On}. The big difference with respect to the \ac{eft}-treatment is that the extra control-parameter $n$ allows us now to compute the coefficients $c_{3/2}(n)$, $c_{1/2}(n)$ which are incalculable within the \ac{eft}:
\begin{align}
	c_{3/2}(n) &= \frac{2}{3}(2n)^{-1/2}, & c_{1/2}&=\frac{1}{6}(2n)^{1/2}, & c_{-1/2}&=-\frac{7}{120}(2n)^{3/2}, &\dots
\end{align}

The saddle-point equations~\eqref{eq:result-finite-Q-zero-T-zeta} are valid for any value of the charge.
So we can also use them in the opposite limit of \(Q/n \ll 1\). 
Using the appropriate expression for the zeta function and expanding at \ac{nlo} we find that \(m\) receives a \(Q/n\) correction to the conformal coupling value
\begin{equation}
  m = \frac{1}{2} + \frac{8}{\pi^2} \frac{Q}{2n} + \order{\pqty{\frac{Q}{2n} }^2},
\end{equation}
and the corresponding conformal dimension is
\begin{equation}
  \label{eq:conformal-dimension-small-charge}
  \Delta(Q) = n \pqty{\frac{Q}{2n}  + \frac{8}{\pi^2} \pqty{\frac{Q}{2n} }^2 + \order{\pqty{\frac{Q}{2n} }^3} }.
\end{equation}
The leading term was to be expected because in this limit we can identify the lowest operator of charge \(Q\) with \(\varphi^Q\), which has engineering dimension \(Q/2\).

The technology used here can also be used to study the case of small but finite temperature, see~\cite{Alvarez-Gaume:2019biu}.

\paragraph{The Goldstones}%
\label{sec:goldstones}

Up to this point, \(n\) has been a generic parameter.
We have derived the saddle-point equations and then computed the free energy assuming that the fluctuations \(\hat \lambda\) could be neglected.
This can be made more precise starting from the action~\eqref{eq:expanded-tr-log-original}.
In the standard treatment of the large-$n$ limit~\cite{PhysRevD.10.2491,ZinnJustin:2007zz}, a natural rescaling of the quantum fluctuations is introduced that results in a self-consistent $1/n$ expansion. In our case, we rescale the fluctuations as \(\hat \lambda \to \hat \lambda/n^{1/2}\).
In this way we introduce a hierarchy among the terms in the effective action. The results of the previous subsection are now understood as the leading effects in \(1/n\) and we can study the system perturbatively.
From now on we will take the limit
\begin{equation}
  n \gg 1 .
\end{equation}
We have however observed that the zeta functions have a natural expansion in terms of the parameter \(Q/n \gg 1\).
This means that we are effectively working in the hierarchy
\begin{equation}
  1 \ll n \ll Q \ll n^2.
\end{equation}
Equivalently, we have two large numbers, \(n\) and \(Q/n\), with \(n \gg Q/n\) controlling the splitting between tree-level and quantum effects in the theory and \(Q/n\) giving an expansion of the physical observables at each fixed order in \(n\).

We have seen in Section~\ref{sec:O2n-Goldstones} that one of the large-charge predictions for the O(2n) vector model~\cite{Hellerman:2015nra,Alvarez-Gaume:2016vff} is that in the large-charge expansion of the ground state energy there is a universal \(Q^0\) term due to the Casimir energy of the conformal Goldstone, see Eq.~\eqref{eq:Delta_d2_On}.
Since this term is \(n\)-independent, it has to appear as part of the first \(1/n\) correction to the (order \(n\)) results of before.

Consider the action for the \(\pi^I\) and use the fact that at the saddle point all the thetas are equal, \(\theta^I = i m \beta\):
\begin{equation}
  \begin{aligned}
    S_Q ={}& \eval{-i \theta Q + \pqty{\frac{\theta^2}{\beta^2} + m^2 } \frac{v^2}{2}V\beta}_{\theta = i m \beta } \\
    &+ \sum_{I=1}^n \int \dd{t}\dd{\Sigma} \bqty{ \pqty{ D_\mu \pi^I}^* \pqty{D_\mu \pi^I} + m^2 \abs{\pi^I}^2 + \frac{A^I}{\sqrt{2}} \hat \lambda \pqty{\pi^I + (\pi^I)^*} + \hat \lambda \abs{\pi^I}^2},
  \end{aligned}
\end{equation}
If we use the same orthonormal basis of \(\setC^n\) as in Eq.~(\ref{eq:simpleframe})
and project the fields \(\pi^I\) on it, the Hamiltonian reads
\begin{equation}
  \begin{aligned}
    S_Q ={}& m \beta Q + \int \dd{t} \dd{\Sigma} \bqty{\pqty{D_\mu \pi^0}^* \pqty{D_\mu \pi^0} + m^2 \abs{\pi^0}^2  + \frac{v}{\sqrt{2}} \hat \lambda \pqty{\pi^0 + (\pi^0)^*} + \hat \lambda \abs{\pi^0}^2} \\
    &+ \sum_{i = 1}^{n-1} \int \dd{t} \dd{\Sigma} \bqty{ \pqty{D_\mu \pi^i}^* \pqty{D_\mu \pi^i} + m^2 \abs{\pi^i}^2 + \hat \lambda \abs{\pi^i}^2 }.
\end{aligned}
\end{equation}
At the saddle point \(m^2 = - \theta^2/\beta^2\) and we recognize the fields \(\pi^i\) as the type-II Goldstones discussed in Section~\ref{sec:O2n-Goldstones}.

The conformal Goldstone mode appears as a combination of \(\pi^0\) (from now on, for ease of notation \(\pi^0 = \pi\)) and \(\hat \lambda\).
In order to see it, one can integrate out the type-II Goldstones, which at order \(\order{n^0}\) give a mass term for \(\hat \lambda\). 
The resulting one-loop effective action at quadratic order is
\begin{multline}
  \label{eq:action-sigma-lambda-zero-T}
  S^{(2)}[\pi, \hat \lambda] = \int \dd{t} \dd{\Sigma}  \Bigg[ \del_\mu \pi^* \del^\mu \pi + m \pqty{\pi^* \del_0 \pi - \pi \del_0 \pi^*} + \sqrt{-\frac{\zeta(\tfrac{1}{2}|\Sigma,m)}{2 V} } \hat \lambda  \pqty{\pi + \pi^*} \\- \frac{\zeta(\tfrac{3}{2}|\Sigma,m)}{8V}\hat \lambda^2 \Bigg].
\end{multline}
There is no kinetic term for \(\hat \lambda\) so we can integrate it out:
\begin{equation}
  S^{(2)}[\pi] = \int \dd{t} \dd{\Sigma}  \bqty{\del_\mu \pi^* \del^\mu \pi + m \pqty{\pi^* \del_0 \pi - \pi \del_0 \pi^*} - \frac{\zeta(\tfrac{1}{2}|\Sigma,m)}{ \zeta(\tfrac{3}{2}|\Sigma,m)} \pqty{\pi + \pi^*}^2} .
\end{equation}
The leading-order contribution in the \(1/Q\) expansion of this term comes from the leading term in the heat kernel expansion of the zeta function,
\begin{equation}
  \zeta(s| \Sigma, m) = \frac{K_0}{s-1} m^{2-2s} + \dots
\end{equation}
and we find
\begin{equation}
  S^{(2)}[\pi ] = \int \dd{t} \dd{\Sigma}  \bqty{\del_\mu \pi^* \del^\mu \pi + m \pqty{\pi^* \del_0 \pi - \pi \del_0 \pi^*} + m^2 \pqty{\pi + \pi^*}^2}.
\end{equation}
This is precisely the action for the conformal Goldstone mode and its massive partner in Eq.~(\ref{eq:O2GoldstoneL}).

\medskip
In summary, having another control parameter, we were able to confirm our earlier \ac{eft} predictions from first principles. Moreover, we were able to compute the parameters $c_i$ which are incalculable within the \ac{eft}.

\section{Matrix models}\label{sec:matrixModels}

The subject of matrix models at large charge is much richer than the vector models discussed so far, but is also far less developed. While a number of works have appeared which discussed theories of matrix-valued fields in sectors of large charge~\cite{Loukas:2017hic,Loukas:2017lof,Banerjee:2019jpw,Orlando:2019hte}, their general properties have not yet been developed in an exhaustive fashion. We will therefore confine ourselves to present those examples that have been understood so far.

\subsection{\(SU(2) \times SU(2)\)}\label{sec:SU2SU2}

As a first (and so far best-understood) example we study a model in which the order parameter is a unitary \(2 \times 2\) matrix field \(\Psi\).
We assume the Lagrangian to be invariant under the action of \(SU(2) \times SU(2)\) that transforms \(\Psi \) as
\begin{equation}
  \Psi(x) \mapsto V_L \Psi(x) V_R^{-1} .
\end{equation}
Following the approach to the \(O(2n)\) vector model in Section~\ref{sec:NLSM-O2n} we write a \ac{nlsm} based on a scalar function of \(\Psi\) which is Lorentz-invariant and invariant under the global symmetry.
The simplest such function is
\begin{equation}
  \norm{\dd{\Psi}}^2 = \ev*{\del_\mu \Psi^\dagger \del^\mu \Psi}.
\end{equation}
The Lagrangian takes the same form as in Eq.~\eqref{eq:O(2n)-NLSM}:
\begin{equation}
  \Lag_{\ac{nlsm}}[\Psi] = \sqrt{\det(g)} \pqty{ k_0 \norm{\dd{\Psi}}^{d+1} + k_1 R \norm{\dd{\Psi}}^{d-1} + \dots } .
\end{equation}
Unlike in the cases studied so far, we now have a left and a right current, which are represented by \(2 \times 2\) matrices that transform in the adjoint of the respective global groups:
\begin{align}
  J_L &= \frac{i}{2 \norm{\dd{\Psi}}} \fdv{\Lag}{\norm{\Psi}} \pqty{ \dd{\Psi} \Psi^\dagger - \Psi \dd{\Psi^\dagger}}, & J_R &= \frac{i}{2 \norm{\dd{\Psi}}} \fdv{\Lag}{\norm{\Psi}} \pqty{ \dd{\Psi^\dagger} \Psi - \Psi^\dagger \dd{\Psi}}, \\
  J_L &\mapsto V_L J_L V_L^{-1}, & J_R &\mapsto V_R J_R V_R^{-1} .
\end{align}
This is precisely the same situation that we had encountered for the \(O(2n)\) model and the only invariant objects are the eigenvalues of these matrices.
The conserved charges are obtained as usual,
\begin{align}
  Q_L &= \int \dd{x} J_{L,0}, & Q_R &= \int \dd{x} J_{R,0},
\end{align}
and we can fix \(2 \rank(SU(2)) = 2\) charges, one on the left and one on the right, corresponding to the projections of the charge matrices on the Cartan generator, which we can choose to be the Pauli matrix \(H_L = H_R = \sigma_3\):
\begin{align}
  q_L &= \frac{1}{2} \ev*{Q_L H_L }, & q_R &= \frac{1}{2} \ev*{Q_R H_R}.
\end{align}
The energy of a given configuration will depend in general only on the eigenvalues of \(Q_L \) and \(Q_R\), which can always be diagonalized with a \(SU(2) \times SU(2)\) transformation,
\begin{align}
  Q_L &= j_L \sigma_3,  & Q_R &= j_R \sigma_3 .
\end{align}
In this particular case, the parallel with the \(O(4)\) model is explained by the fact that \(SU(2)\times SU(2)\) is a double cover of \(O(4)\):
\begin{equation}
  O(4) = \frac{SU(2) \times SU(2)}{\set{(1,1), (-1,-1)}} .
\end{equation}
Explicitly, if we parametrize \(\Psi \in SU(2)\) as
\begin{equation}
  \Psi =
  \begin{pmatrix}
    e^{i \beta} \cos(\gamma) & e^{i \alpha} \sin(\gamma)\\
    -e^{-i \alpha} \sin(\gamma) & e^{-i \beta} \cos(\gamma)
  \end{pmatrix},
\end{equation}
we find that the invariant function \(\norm{\dd{\Psi}}\) coincides with the invariant function \(\norm{\dd{\psi}}\) defined for the \(O(4)\) model with the parametrization in Eq.~\eqref{eq:S3-parametrization}:
\begin{equation}
  \norm{\dd{\Psi}}^2 = 2 \norm{\dd{\psi}}^2 .
\end{equation}
The Cartans \(H_L\) and \(H_R\) in \(SU(2) \times SU(2)\) act as
\begin{align}
  H_L &:
  \begin{cases}
     \alpha \mapsto \alpha + \frac{\epsilon_L}{2}\\
     \beta \mapsto \beta + \frac{\epsilon_L}{2}
  \end{cases} &
  H_R &:
  \begin{cases}
    \alpha \mapsto \alpha - \frac{\epsilon_L}{2} \\
    \beta \mapsto \beta + \frac{\epsilon_L}{2} 
  \end{cases} .
\end{align}
Comparing with the action of \(H_1\) and \(H_2 \) in \(O(4)\) found in Eq.~\eqref{eq:O(4)-Cartan-action} we see that the actions are related by
\begin{align}
  H_1 &= H_L - H_R, &  H_2 = H_L + H_R .
\end{align}
With these identifications we can translate all the results of Section~\ref{sec:NLSM-O2n} into results for this matrix model (see also~\cite{Banerjee:2019jpw}).
\begin{itemize}
\item If we fix the projections \(q_L\) and \(q_R\) and impose homogeneity, the lowest-energy solution is such that the charge matrices can be rotated into the form
  \begin{align}
    Q_L &= j \sigma_3, &     Q_R &= j \sigma_3 ,
  \end{align}
  where \(j = q_L + q_R\). The corresponding state lives in the representation \((j,j)\) of \(SU(2) \times SU(2)\);
\item states that live in generic representations \((j_L, j_R)\), with \(j_L \neq j_R\) correspond to inhomogeneous solutions~\cite{Banerjee:2019jpw}. The only difference with respect to the \(O(4)\) case is that here, \(j_L\) and \(j_R\) are generically integers or half integers, while for \(O(4)\) they are constrained by \(j_L + j_R \in \setZ \);
\item the dimension of the lowest operator in the representation \((j_L, j_R)\) in \(2 + 1\) dimensions is given by
  \begin{equation}
    \Delta(j_L, j_R) = c_{3/2} j^{3/2} + c_{1/2} j^{1/2} + \lambda^2 \frac{\abs{j_L - j_R}}{j^{1/2}} + \dots,
  \end{equation}
  where \(j = \max(j_L, j_R)\), and \(c_{3/2}\), \(c_{1/2}\) and \(\lambda^2\) are Wilsonian parameters;
\item the fluctuations over the ground state are encoded by a (conformal) type-I Goldstone and a type-II Goldstone.
\end{itemize}

\subsection{SU(N) matrix models}
\label{sec:SU(N)matrix}

Another option is studying an SU(N) matrix model in 2+1 dimensions at large charge, where the field $\Phi \in su(N)$ and the global SU(N) symmetry acts on it via the adjoint map~\cite{Loukas:2017lof,Loukas:2017hic}.
These models display a richer phenomenology than what we have seen in the vector models.
For example, it is possible to \emph{fix more than one charge independently}, while still obtaining a homogeneous ground state.
These charge assignments will correspond in general to different symmetry-breaking patterns leading to distinct fixed points in the \ac{ir}.

\medskip

We study the \ac{lsm} description of the model. Conformal symmetry requires scale invariance of the action and fixes the potential to be a polynomial of order six in $\Phi$ plus the conformal coupling of $\Phi$, while neglecting higher-derivative operators whose contributions are controlled in the large-charge expansion.
Given these assumptions, an effective \ac{ir} Wilsonian action for $\Phi$ living in $\setR\times\mani$  is given by
\begin{equation}\label{eq:LSU(N)}
  \Lag_{\ac{lsm}} =  \frac{1}{2} \Tr (\del_\mu \Phi \del^\mu \Phi) { - V(\Phi)},
\end{equation}
with the potential
\begin{equation}\label{LSM:ScalarPotential}
  V(\Phi) = {\frac{R}{16} \Tr(\Phi^2) +} g_1 \Tr( \Phi^6) + g_2 (\Tr\Phi^3)^2  + g_3 \Tr(\Phi^4)\Tr(\Phi^2) + g_4 (\Tr\Phi^2)^3,
\end{equation}
which has to be bounded from below, meaning  it cannot have a runaway behavior when $\Tr(\Phi^2)\rightarrow\infty$.
This amounts to a set of conditions on the couplings $g_i$.
For instance, when all $g_i\geq0$, $V(\Phi)$ is well bounded from below.
By trace cyclicity we readily see that the action is invariant under global SU(N) transformations acting on $\Phi$ via the adjoint map: 
\begin{align}
	V &\in SU(N),  & \Phi \rightarrow \Ad[V] \Phi = V \Phi V^{-1}.
\end{align}
The associated Noether current is given by
\begin{equation}
	J_\mu = i \left[\Phi,\partial_\mu \Phi\right].
\end{equation}
Assigning to the field operator the classical mass dimension $[\Phi]=1/2$, the action under consideration becomes also scale invariant.

The Euler--Lagrange \ac{eom} are found by varying the action with respect to $\Phi$:
\begin{equation}
\label{LSM:EulerLagrangeEOMs}
        \ddot \Phi = - V'(\Phi).
\end{equation}
Since $\Phi$ is Hermitian, we can diagonalize it, 
\begin{align}\label{PhiEigendecomposition}
	\Phi &= U A U^\dagger, & U &\in SU(N)/U(1)^{N-1},	
\end{align}
and  obtain the eigenvalue matrix
\begin{align}\label{LSM:EigenvalueMatrix}
	A&=\diag\left(a_1,\dots,a_N\right), & a_1 +\dots+a_N&=0.
\end{align}
Tracing both sides of Eq.~\eqref{LSM:EulerLagrangeEOMs} we deduce a necessary condition on the classical solution:
\begin{equation}
	\Tr(V'(\Phi_{\text{cl}}))  = 0.
\end{equation}
The homogeneous solution to the \ac{eom} Eq.~\eqref{LSM:EulerLagrangeEOMs} has two distinct branches, depending on the values of the Wilsonian parameters $g_i$, which can both be parametrized as 
\begin{equation}\label{LSM:ClassicalSolutionSCHEMA}
	\Phi_{\text{cl}} = \Ad[\exp\left(i \sum\nolimits_{j=1}^{n_h} \mu_j H^j\, t\right)]  \Phi_0,
\end{equation}
where $\Phi_0 \in su(N)$ denotes the time-independent part. 
The direction of the time dependence can always be chosen to be inside the Cartan sub-algebra of $su(N)$, generated by the $H^j$. %
Different choices of the fixed charges will correspond to appropriate values of the chemical potentials \(\mu_j\).
Modulo accidental enhancements at  special charge configurations,
the number $n_h$ of non-vanishing \(\mu_j\) gives the number of type-I Goldstones $\chi_j$ associated to the charges $Q_j$ in the low-energy theory. 

The two branches of the classical solution mentioned are associated to different 
fixed points of the \ac{rg} flow.
Quantizing the fluctuations on top of the corresponding vacua leads to distinct predictions for the low-energy spectrum and the scaling dimension of lowest charge-Q operators.
We will summarize here the classification of the fixed points appearing in~\cite{Loukas:2017hic}.
\begin{description}
\item[The special cases of SU(2) and SU(3).]
In the classification of fixed points there are two special cases.
The SU(2) matrix model is locally equivalent to the $O(3)$ vector model and hence its analysis follows immediately from Section~\ref{sec:O2n}.
Despite the SU(3) matrix model not being equivalent to any vector model, it turns out that this matrix theory gives qualitatively the same predictions as vector models~\cite{Loukas:2017lof}.
In $su(3)$, it is always possible to choose a basis of Casimirs to be spanned by $\Tr(\Phi^2)$ and $\Tr( \Phi^3)$.
This means that we can set $g_1=g_3=0$ in the potential~\eqref{LSM:ScalarPotential}.
Consequently, SU(3) falls automatically into the class of  Wilson--Fisher-like fixed points discussed next.
We can never fix more than one independent U(1) scale in the low-energy description of a model with global SU(3) symmetry and be still homogeneous in space.
From the SU(4) matrix model on, we will see new non-trivial behavior not present in the vector model.
\item[Wilson--Fisher-like fixed point.]
To understand the fixed point structure, we take a closer look at the role of the scalar potential \eqref{LSM:ScalarPotential} in the classical \acp{eom} for any SU(N) matrix model. %
Concentrating on the locus where $g_1= g_3 = g_4=0$ and $g_2$ is arbitrary, the scalar potential evaluated on the classical solution $\Phi_{\text{cl}}$ becomes
\begin{equation}
	\label{LSM:ScalarPotential_WilsonFisher}
V(\Phi_{\text{cl}}) = V(A_{\text{cl}}) = \frac{R}{16} \Tr(A^2_{\text{cl}}) + g_2 \left(\Tr(A^2_{\text{cl}})\right)^3.
\end{equation}
Then, the full action $S[\Phi_{\text{cl}}]$ has a $O(N^2-1)$ symmetry.
Consequently, this branch of the solution follows the pattern of the classical ground state constructed for $O(N^2-1)$ vector models.
The lowest-lying state of fixed charge admits only one charge scale given by $Q$, as there appears only one independent $\mu$ in Eq.~\eqref{LSM:ClassicalSolutionSCHEMA}, meaning $n_h=1$. Only one relativistic Goldstone arises.
\item[Multi-charge fixed point.]
In a SU(2k) or SU(2k+1) matrix model with $k\geq2$ there exists a fixed point for \textit{generic} values of the couplings $g_i$ in $V(\Phi)$ (well inside the allowed parameter range). 
This class of fixed points is generally characterized by $k$ different chemical potentials in the embedding of Eq.~\eqref{LSM:ClassicalSolutionSCHEMA}, \emph{i.e.} $n_h=k$.
This leads to $k$ relativistic Goldstones, enabling us to fix up to $k=\floor{N/2}$ different U(1) charges in the low-energy description, while still maintaining spatial homogeneity. In the large-charge expansion up to order one, we obtain qualitatively distinct predictions, compared to the vector models.
In~\cite{Loukas:2017hic}, this type of fixed point is referred to as ``multi-charge fixed point'', where ``multi-charge'' refers to the possibility to fix multiple U(1) scales in the low-energy theory around a homogeneous vacuum.
\end{description}

For general \(N\), the structure of the solutions becomes much richer and remains to be explored.
The situation is sensibly simpler in \(d = 3\), where the scale-invariant potentials are more constrained and it is possible to analyze the system in detail for generic \(N\).
This is discussed in the following section.

\subsection{An asymptotically safe CFT}
\label{sec:asymptotically-safe}

The next matrix model that we will consider is part of a four-dimensional gauge-fermion theory including a sector with a complex matrix-valued scalar field which in the appropriate limit is known to develop an asymptotically safe fixed point in four dimensions. Also this example is not a general treatment of charge-fixing in the matrix model, as we single  out one particularly simple charge-fixing pattern. It serves however to illustrate a number of points that have not yet arisen in the simpler models studied at large charge so far. One point is due to the large non-Abelian global symmetry of the model, which gives rise to a more interesting symmetry-breaking pattern and spectrum of light \ac{dof}. We will see in particular how the resulting type I and II Goldstone bosons appear in representations of the unbroken symmetry. 

The other point arises from the fact that our model contains not just scalars as in all the cases studied so far, but also fermions and gluons. We will see that at large charge, these sectors decouple from the low-energy physics encoded by the Goldstones as the fermions acquire large rest masses of order $\rho^{1/3}$.

Another difference is that in this case the linear sigma model is under perturbative control in the appropriate Veneziano limit and it can be used directly to study the large-charge behavior. 
\bigskip

We start from a \ac{cft} in four dimensions, containing $SU(N_C)$ gauge fields $A_\mu^a$, $N_F$ flavors of fermions $Q_i$ in the fundamental and an $N_F \times N_F$ complex matrix scalar field $H$ which is not charged under $SU(N_C)$. In the Veneziano limit of $N_F \to \infty, \ N_C \to \infty$ with the ratio $N_F/N_C$ fixed, this theory is asymptotically safe, as shown in~\cite{Litim:2014uca}.
Its Lagrangian is given by
\begin{equation}\label{eq:fullLag}
  \begin{aligned}
    \Lag ={}& -\frac{1}{2}\Tr (F^{\mu\nu}F_{\mu\nu}) + \Tr(\bar Q i\slashed{D} Q) + y \Tr(\bar Q_L H Q_R + \bar Q_R H^\dagger Q_L)\\
    &+ \Tr(\del_\mu H^\dagger \del^\mu H ) - u\Tr(H^\dagger H)^2 - v(\Tr H^\dagger H )^2- \frac{R}{6} \Tr( H^\dagger H) .
  \end{aligned}
\end{equation}
The trace runs over both color and flavor indices and $Q_{L/R} = \tfrac{1}{2} (1\pm \gamma_5)Q$.
The rescaled couplings of the model appropriate for the Veneziano limit are
\begin{align}\label{eq:couplings}
  \alpha_g & = \frac{g^2 N_C}{(4\pi)^2}, & \alpha_y & = \frac{y^2 N_C}{(4\pi)^2}, & \alpha_h & = \frac{u N_F}{(4\pi)^2}, & \alpha_v & = \frac{v N_F^2}{(4\pi)^2},
\end{align}
where $\alpha_g$ is the gauge coupling (as opposed to the original gauge coupling \(g\) in Eq.~\eqref{eq:fullLag}), $\alpha_y$ the Yukawa coupling, $\alpha_h$ the quartic scalar coupling and $\alpha_v$ the double-trace coupling.
We also introduce the control parameter 
\begin{equation}
	\epsilon = \frac{N_F}{N_C}-\frac{11}{2},
\end{equation}
which in the Veneziano limit is continuous and arbitrarily small.
As shown in~\cite{Litim:2014uca}, if \(0 \leq \epsilon \ll 1\), the system develops a \ac{uv} fixed point that has only one relevant direction, the other three being irrelevant.
At leading order in perturbation theory and properly respecting the Weyl consistency conditions~\cite{Antipin:2013sga} the couplings at the fixed point have the values
\begin{equation}
  \label{NNLOseries}
  \begin{aligned}
\alpha_g^*&= 0.4561\,\epsilon
+\order{\epsilon^2} &
\alpha_y^*&= 0.2105\,\epsilon
+\order{\epsilon^2} ,\\
\alpha_h^*&= 0.1998\,\epsilon
+\order{\epsilon^2}\,, &
\alpha_{v} ^*&=-0.1373\,\epsilon +\order{\epsilon^2}\,.
\end{aligned}
\end{equation}

\paragraph{Equations of motion and ground state}

We will first focus on the sector involving just the scalar field $H$, using the Lagrangian
\begin{equation}\label{eq:scalarL}
  \Lag_H = \Tr(\del_\mu H^\dagger \del^\mu H ) - u\Tr(H^\dagger H)^2 - v(\Tr H^\dagger H )^2 - \frac{R}{6} \Tr( H^\dagger H). 
\end{equation}
In a later step, we will show that indeed, all the fermions $Q_i$ will receive large masses from fixing the charge and, together with the gluons, decouple from the dynamics. 

In this sector, the model naively has a global \(U(N_F)_L \times U(N_F)_R\) symmetry at the classical level. We know however that the full model (given in Eq.~\eqref{eq:fullLag}) has an axial anomaly due to the Yukawa term. For this reason, we work directly in the quantum global symmetry group \(SU(N_F)_L \times SU(N_F)_R \times U(1)_B\) which is generated by the currents
\begin{align}
  J_L &=  \frac{i}{2} \pqty{\dd{H} H^\dagger - H \dd{H^\dagger}}, & J_R &= - \frac{i}{2}  \pqty{H^\dagger \dd{H} - \dd{H^\dagger} H}.
\end{align}
For the sake of simplicity, we will consider only the behavior of the system fixing the conserved charges to be proportional to the identity: \(Q_L = -Q_R = J \Id_{N_f/2} \otimes \sigma_3\). 
We make the homogeneous ansatz
\begin{equation}\label{eq:ansatz}
  H_0(t) = e^{2iMt}B, 
\end{equation}
with \(M^2 = \mu^2/4 \Id_{N_F}\) and \(B = b \Id_{N_F}\). Since $M$ is proportional to the charge matrix that lives in the algebra \(su(N)\), it must be traceless and it contains $N_F/2$ diagonal elements equal to $\mu/2$ and $N_F/2$ diagonal elements equal to $-\mu/2$. The \ac{eom} is
\begin{multline}
  \frac{\del}{\del t}\left[\frac{\del}{\del \dot H^*}\left( \Tr(\dot H^\dagger \dot H) \right)\right] + \frac{\del}{\del H^*}V(H, H^*) \\
  = \del^2_0 H  +2u (H^\dagger H)H + 2v\Tr(H^\dagger H) H+ \frac{R}{6}H = 0,
\end{multline}
and on the above ansatz reduces to
\begin{equation}
  \frac{\mu^2}{2} = \pqty{u + v N_F } b^2  + \frac{R}{12} , 
\end{equation}
with the condition 
\begin{equation}
  J =  V b^2 \mu .
\end{equation}
If for given fixed charges a solution exists which is homogeneous in space, this will be the solution of minimal energy in this sector.
In general it is not easy to identify in the strong-coupling regime the operator whose insertion corresponds to this charge configuration.
Progress beyond the simplest cases of vector models has been obtained recently in~\cite{Antipin:2020rdw}.
It is convenient to assume \(J\) to be large and expand in series. The natural expansion parameter is \(\mathcal{J}\):
\begin{equation}
  \Jexp = J \frac{(u+v N_F)}{8 \pi^2} = 2 J \frac{ \alpha_h + \alpha_v}{N_F} = 2 J_{\text{tot}} \frac{\alpha_h + \alpha_v}{N_F^2},
\end{equation}
where \(J_{\text{tot}} = J N_F/2 \) is the total charge.
The expansion requires \(\mathcal{J} \gg 1\) and, observing that at the fixed point both \(\alpha_h\) and \(\alpha_v\) are of order \(\epsilon\), we see that the expansion is consistent in the regime
\begin{equation}
  J_{\text{tot}} \gg  \frac{N_F^2}{ \epsilon}.
\end{equation}
This is a typical feature of the large-charge expansion, where the total charge has to be the dominant large parameter in the problem. In the case at hand, also the number of \ac{dof} $N_F^2$ and the inverse coupling $1/\epsilon$ are large.
Since there are no other dimensionful parameters in our problem, the energy scale \(\rho^{1/3}\) will control the tree-level and the quantum corrections to the energy of the ground state.
For our choice of charges, we obtain an expansion in \(\Jexp\) for the energy of the ground state, starting from \(\Jexp^{4/3}\):
\begin{equation}\label{eq:safe_semiclass_GSenergy}
  E = \frac{3}{2} \frac{N_F^2}{\alpha_h + \alpha_v} \pqty{\frac{2\pi^2}{V}}^{1/3} \bqty{ \Jexp^{4/3} + \frac{R}{36} \pqty{\frac{V}{2\pi^2}}^{2/3} \Jexp^{2/3} - \frac{1}{144} \pqty{\frac{R}{6}}^2 \pqty{\frac{V}{2 \pi^2}}^{4/3} \Jexp^0 + \order{\Jexp^{-2/3}}}.
\end{equation}
Specialized to the sphere ($V=2\pi^2 R_0^3$ and $R=6/R_0^2$), this yields again the semi-classical contribution to the scaling dimension of the lowest fixed-charge operator.
Since here we started from a calculable \ac{cft}, which is described by a trustworthy linear sigma model, the coefficients in front of $\Jexp^{4/3}$ and $\Jexp^{2/3}$ are trustworthy. This is in contrast to the earlier cases where \acp{cft} which are not perturbatively accessible were studied at large charge and their Wilsonian couplings could not be determined within the framework of effective field theory. The coefficient in front of $\Jexp^{0}$ will however receive a scheme-dependent contribution from the Casimir energy of the Goldstones.

\paragraph{Decoupling of the fermions and gluons}

Now that we have understood the effect of fixing the charge in the bosonic sector, we can discuss the fermionic and gauge sectors.
We will see that the fermions become massive, with the mass scale given by the charge density, and this in turn decouples also the gluons. The same mechanism is at play in the case of the supersymmetric $W=\Phi^3$ model discussed in Section~\ref{sec:Phi3}.
As expected, the large-charge, low-energy physics is therefore described completely by the Goldstone fields that result from the symmetry breaking.

We start with the fermionic part of the Lagrangian~\eqref{eq:fullLag}, 
\begin{equation}
  \Lag_{\text{f}} = \Tr(\bar Q i\slashed{D} Q) + y \Tr(\bar Q_L H Q_R + \bar Q_R H^\dagger Q_L).
\end{equation}
Expanded around the ground state \(H = H_0(t)\) given in Eq.~\eqref{eq:GS}, the action takes the form
\begin{equation}
  \Lag_{\text{f, GS}} = \Tr(\bar Q i\slashed{D} Q) + y \Tr(\bar Q_L e^{2i M t} B Q_R + \bar Q_R B e^{-2i M t}  Q_L).
\end{equation}
It is convenient to redefine the fermionic fields \(Q\) to eliminate the time-dependent coupling and trade it for a mass term.
A possible choice is
\begin{align}
  \psi_L &= e^{-iM t} Q_L, & \psi_R = e^{iM t} Q_R,
\end{align}
so that the Lagrangian reads
\begin{equation}
  \label{eq:fermionic-action-psi}
  \begin{aligned}
    \Lag_{\text{f, GS}} &= \Tr(\bar\psi i\slashed{D} \psi) - \Tr( \bar \psi_L \gamma^0 M \psi_L) + \Tr( \bar \psi_R \gamma^0 M \psi_R) + y\Tr(\bar\psi B \psi ) \\
    &= \Tr(\bar \psi i\slashed{D} \psi) - \Tr( \bar \psi \gamma^0  \gamma^5 M \psi)  + y\Tr(\bar\psi B \psi ).
  \end{aligned}
\end{equation}
The simplest way to see if the fermions actually become massive and decouple is to write down the inverse propagator.
The zero-momentum limit of its determinant gives the product of the masses of the fields:
\begin{equation}
  \eval{\det(D^{-1}(\omega, p))}_{\omega = 0, p = 0} = \prod_{\text{fields}} m_f^2.
\end{equation}
The inverse propagator corresponding to the action in Eq.~(\ref{eq:fermionic-action-psi}) is
\begin{equation}
  D^{-1}(\omega, p) = -\omega \gamma^0 + p_i \gamma^i - \gamma^0 \gamma^5 M + y B  ,
\end{equation}
and since $B$ and $M$ are diagonal matrices, we find
\begin{align}
	\det( {\begin{matrix}[c|c]
        y B & -M \\ \hline
        M & y B
      \end{matrix}})^2 &= \prod_{i=1}^{N_F/2} \det(
    {\begin{matrix}
        y b_i + \frac{\mu_i^2}{4y b_i} & 0\\
        0 & y b_i + \frac{\mu_i^2}{4y b_i}
      \end{matrix}})^2 \det( {\begin{matrix}
        y b_i & 0 \\
        0 & y b_i 
      \end{matrix}})^2 \\
    &= \prod_{i = 1}^{N_F/2} \pqty{ \frac{\mu_i^2}{4}  + y^2 b_i^2 }^4 .
\end{align}
We see that both the Yukawa term (via the term \(y^2 b_i^2\)) and the kinetic term (via the term \(\mu_i^2\)) contribute to the final expression.
If all the charges are equal, also all the fermions have the same mass which is given by
\begin{equation}
    m_\psi = \pqty{ \frac{\mu^2}{4} + y^2 b^2}^{1/2} = \pqty{\frac{2 \pi^2}{V}}^{1/3} \pqty{ 1 + 2 \frac{N_F}{N_c} \frac{\alpha_y}{\alpha_h + \alpha_v}}^{1/2} \Jexp^{1/3} + \order{\Jexp^{-1/3}}.
\end{equation}
We see that all the fermions are massive, with a mass fixed by the charge density, and they decouple from the rest of the theory.

Once all the fermions of the theory have decoupled, at scales below $m_\psi$ the pure gauge sector starts running towards lower energies as pure Yang--Mills theory. The resulting theory gaps with an estimated confining scale $\Lambda_{YM}$ 
\begin{equation}
\Lambda_{YM} = m_{\psi} \exp\left[ - \frac{3}{22 \alpha_g (m_\psi)} \right] \ . 
\end{equation}
Here $\alpha_g (m_\psi)$ is very close to the \ac{uv} fixed-point value that is of order $\epsilon$.
The gluons will generically modify all the terms in the large-charge expansion of the energy.
These contributions are however exponentially suppressed as \(\order{e^{-1/\epsilon}}\) and can be neglected in our approximation.
Below the scale \(\Lambda_{YM}\) we have the Goldstone excitations that we will discuss in the following. 

\paragraph{Symmetry-breaking pattern and Goldstone spectrum}

We will again discuss the symmetry-breaking in terms of the two steps separating the breaking leading to massive and massless Goldstones.
The \(M\) matrix acts like a chemical potential and gives rise to a term which breaks the symmetries explicitly (step 1), while the \(B\) matrix is akin to a ground state that breaks the remaining symmetry spontaneously (step 2). 

The explicit breaking only happens for the \(SU(N_F)_L\) symmetry, which is reduced to the subgroup \(C(M)_L \subset SU(N_F)_L \times U(1)_B\) that commutes with \(M\) which is diagonal, with half the elements equal to \(\mu/2\) and half equal to \(-\mu/2\).
It follows that
\begin{equation}
	C(M) = SU(N_F/2) \times SU(N_F/2) \times U(1)^2.
\end{equation}
\(B\) is proportional to the identity, so the spontaneous breaking preserves a group \(C(M)\) embedded ``diagonally'' in \(C(M)_L \times SU(N_F)_R \times U(1)\), in the sense that \(B\) remains invariant under the adjoint action of \(C(M)\).
The full symmetry-breaking pattern is thus
\begin{equation}\label{eq:symmBrJeq}
  SU(N_F)_L \times SU(N_F)_R \times U(1) \to C(M)_L \times SU(N_F)_R \leadsto C(M) .
\end{equation}
By Goldstone's theorem, the low-energy dynamics is described by \(\dim(SU(N_F)) = N_F^2 - 1 \) \ac{dof}.
By construction, the Goldstone fields will arrange themselves into representations of the unbroken group \(C(M)\).

To study their spectrum, we need to expand the fields at second order in the fluctuations \(\Phi\) around the ground state \(H_0(t)\):
\begin{equation}
  H(t,x) = \exp[ i \mu t
    \begin{pmatrix}[c|c]
      \Id & 0 \\ \hline
      0 & - \Id
    \end{pmatrix}] \pqty{b \begin{pmatrix}[c|c]
    \Id & 0 \\ \hline
    0 &  \Id
  \end{pmatrix} + \Phi(t,x)}.
\end{equation}
Since we are only interested in the leading behavior at small momenta, we can neglect the term proportional to the curvature and write the Lagrangian as:
\begin{equation}\label{eq:quadrLag}
  \begin{aligned}
    \Lag ={}& \Tr[ \del_\mu H^\dagger \del_\mu H ] - u \Tr[ H^\dagger H H^\dagger H] - v \Tr[H^\dagger H]^2 \\
    ={}&
    \Tr[\del_\mu \Phi^\dagger \del_\mu \Phi] - i \Tr[  \mu  (\Phi^\dagger \del_0 \Phi - \del_0 \Phi^\dagger \Phi)] +  \mu^2 \Tr[ \Phi^\dagger \Phi ]\\
    &- u \Tr[ (b + \Phi ) (b + \Phi )^\dagger (b + \Phi ) (b + \Phi )^\dagger ] - v \Tr[(b + \Phi ) (b + \Phi )^\dagger]^2 .
  \end{aligned}
\end{equation}
The fluctuation \(\Phi(t,x )\) can always be expanded as
\begin{equation}
  \Phi(t,x) = \sum_{A = 0}^{N_F^2 - 1 } (h_A(t,x) + i p_A(t,x)) T^A,
\end{equation}
where \(T^0 = 1/\sqrt{2N_F} \Id\) and the \(T^a\) are the generators in the generalized Gell--Mann basis of \(SU(N_F)\), which satisfy the identity
\begin{equation}
	T_aT_b = \frac{1}{2}\left(\frac{\delta_{ab}}{N_F}+(d_{abk} + i f_{abk})T^k\right),
\end{equation}
where $f_{abc}$ are the structure constants and $d_{abc}$ is the totally symmetric tensor of $SU(N_F)$.
The second-order expansion of the potential in terms of the \(h_A\) and \(p_A\) then takes the form
\begin{equation}
  V^{(2)} = \frac{1}{2} M^2_{h_A h_B} h_A h_B + \frac{1}{2} M^2_{p_A p_B} p_A p_B  + \frac{1}{2} M^2_{p_A h_B} p_A h_B, 
\end{equation}
where the \(M^2\) are those in Eq.(A.8) in~\cite{Abel:2017ujy} for the case of the background field state being \(H = b \Id =  \sqrt{2N_F} b T^0\):
\begin{align}
  M^2_{h_0 h_0} &= -  \mu^2 + 6 b^2 (u + N_F v ) = 2 \mu^2, &
  M^2_{h_a h_b} &= 2 \mu^2 \frac{\alpha_h}{\alpha_h + \alpha_v} \delta^{ab}, \\
  M^2_{p_A p_B} &= M^2_{p_A h_B} = 0.
\end{align}

Let us pause a moment and see what we have found.
The effective mass matrix is diagonal.
Imposing the \ac{eom}, we find that there are \(N_F^2\) massless and \(N_F^2\) massive \ac{dof}.
This is not consistent with the symmetry breaking pattern \(C(M) \times SU(N_F) \to C(M)\).
The reason for this discrepancy is that we have decided not to consider the anomalous \(U(1)\) axial symmetry that is broken by quantum effects.
This means that one of the massless \ac{dof} found here is actually spurious, and we are left with the expected \(N_F^2 -1\).

We now need to identify the type I and type II Goldstones in the spectrum. This information is encoded in the term linear in \(\mu\) with one time derivative appearing in Eq.~\eqref{eq:quadrLag}.
Remember that in our parametrization all the \(h_A\) are massive, while all the \(p_A\) do have a vanishing quadratic term \(M_{p_A p_B}^2\).
A cross term between a \(h_a\)  mode and a \(p_a \) mode gives a type-I Goldstone and a massive field as first observed in the O(2) model, see Eq.~\eqref{eq:confGoldO2}.
\begin{itemize}
\item For \(h_0\), the inverse propagator reads
  \begin{equation}
    \Delta^{-1}_{h_0 p} =
    \begin{pmatrix}
      \frac{1}{2}(\omega^2 - p^2)  -  \mu^2 &  i \mu \omega \\
      - i \mu \omega & \frac{1}{2}(\omega^2 - p^2) 
    \end{pmatrix},
  \end{equation}
  corresponding to the expected conformal Goldstone with velocity \(1/\sqrt{3}\).
\item For \(h_a\), the inverse propagator reads
  \begin{equation}
    \Delta^{-1}_{h_a p} =
    \begin{pmatrix}
      \frac{1}{2}(\omega^2 - p^2)  -  \frac{\alpha_h}{\alpha_h+ \alpha_v } \mu^2 &  i \mu \omega \\
      - i \mu \omega & \frac{1}{2}( \omega^2 - p^2 )
    \end{pmatrix},
  \end{equation}
  corresponding to a massless and a massive mode with dispersion relations
  \begin{align}
    \omega &= \sqrt{\frac{\alpha_h}{3 \alpha_h + 2 \alpha_v}} p + \dots & \omega &= \sqrt{\frac{2 \pqty{3 \alpha_h + 2 \alpha_v}}{\alpha_h + \alpha_v}} \mu + \order{p^2}.
  \end{align}
  In this case the velocity is not fixed by scale invariance, but we have the constraint \(0< \alpha_h/\pqty{3 \alpha_h + 2 \alpha_v} \le 1\) from causality, which implies \(\alpha_h + \alpha_v > 0\). This constraint is satisfied at the fixed point since, using Eq.~\eqref{NNLOseries}, \(\alpha_h + \alpha_v = 0.6991 \alpha_h > 0\).  
\item A type-II Goldstone arises when the fields \(p_A\) and \(p_B\) are related by a linear term proportional to \(\dot p_A p_B - p_A \dot p_B\).
In this case, the corresponding inverse propagator is
\begin{equation}
  \Delta^{-1}_{p_A p_B} =
  \begin{pmatrix}
    \frac{1}{2}(\omega^2 - p^2)  & i \mu \omega \\
    -i \mu \omega & \frac{1}{2}(\omega^2 - p^2)
  \end{pmatrix} .
\end{equation}
\end{itemize}
How are these Goldstone fields organized into representations of the unbroken group \(C(M) = SU(N_F/2) \times SU(N_F/2) \times U(1)^2\)?
The original \(SU(N_F)\) global symmetry is explicitly broken to \(SU(N_F/2) \times SU(N_F/2)\).
The \ac{dof} \(p_a\) in the adjoint of \(SU(N_F)\) decompose into the sum of two adjoints of $SU(N_F/2)$, a pair of bifundamentals (that together form a single bifundamental type-II Goldstone) and a singlet (the conformal Goldstone), see Table~\ref{tab:Goldstone-spectrum}.
We can verify that this distribution accounts for all the expected low-energy \ac{dof}:
\begin{multline}
	\dim(\mathbf{1},\mathbf{1}) + \dim(\mathbf{N_F^2/4 - 1}, \mathbf{1}) + \dim(\mathbf{1}, \mathbf{N_F^2/4 - 1} ) + 2 \dim(\mathbf{N_F/2}, \mathbf{N_F/2}) \\= N_F^2 - 1
	 = \dim(SU(N_F)).
\end{multline}
The contribution of the Casimir energy of the Goldstones to the ground-state energy is scheme-dependent in $3+1$ dimensions, as discussed in Section~\ref{sec:conformal-goldstone}, so the semi-classical term at order $\Jexp^0$ in Eq.~\eqref{eq:safe_semiclass_GSenergy} receives a contribution which is not calculable.

 \begin{table}
   \centering
   \ytableausetup{centertableaux,boxsize=0.5em}
   \begin{tabular}{L{4cm}cccc}
    \toprule
        type & I & I & I & II \\
     \ac{dof} & \(1\) & \(N_F^2/4 -1 \) & \(N_F^2/4 -1 \) &  \(2 \times N_F^2/4 \) \\
     velocity & \(1/\sqrt{3}\) & \(\sqrt{\frac{\alpha_h}{3 \alpha_h + 2 \alpha_v}}\) & \(\sqrt{\frac{\alpha_h}{3 \alpha_h + 2 \alpha_v}}\) & n/a\\[1em]
     \(SU(N_F/2) \times SU(N_F/2) \) representation & \((\mathbf{1}, \mathbf{1})\) & {  (                                                                                                         \begin{ytableau}                                                                                                          { } &{ }  \\                                                                                                         { } \\  \none\\                                                                                                        \none[\scriptscriptstyle  \vdots]\\ \none\\                                                                                                          { }                                                                                                         \end{ytableau}, \(\mathbf{1}\) )} &                                                                                                                                             {    (\(\mathbf{1}\),                                                                                                                                             \begin{ytableau}                                                                                                                                               { } &{ }  \\                                                                                                                                               { } \\ \none  \\                                                                                                                                               \none[\scriptscriptstyle\vdots]\\ \none \\                                                                                                                                               { }                                                                                                                                             \end{ytableau})} &
                                                                       { (\ydiagram{1}, \ydiagram{1})} \\[1em]
     \bottomrule
   \end{tabular} 
   \caption{The Goldstone spectrum resulting from fixing the charges in the sector  \(\set{J, \dots, J, -J, \dots, -J}\). The \(N_F^2 - 1 \)
     DOF stemming from the breaking of the global symmetry are arranged into a singlet (the conformal Goldstone), two adjoints of \(SU(N_F/2)\) and a pair of bifundamentals (that together form a single bifundamental type-II Goldstone).
     The type-I Goldstones contribute to the zero-point energy according to their velocities.
     The type-II Goldstone has a quadratic dispersion relation and has zero velocity.
     Not represented here is the spurious singlet corresponding to the anomalous axial symmetry.
   }
   \label{tab:Goldstone-spectrum}
 \end{table}

\section{Supersymmetric models}\label{sec:susyModels}

A large and well-studied class of \acp{cft} to which we can apply the large-charge expansion is \acf{scft}. \acp{scft} naturally come with a continuous global symmetry, namely R-symmetry and have the advantage that we can make use of the constraining features of \ac{susy}, which in combination with the large-charge expansion allows us to push our analysis further that in the non-supersymmetric cases. It furthermore also allows us to compare large-charge predictions to existing results obtained via localization techniques. In the cases, where this was possible, we found excellent agreement (see~\cite{Hellerman:2018xpi,Hellerman:2020sqj}). 

The subject of \ac{scft} at large R-charge has not yet been exhaustively developed. We present here two cases with qualitatively different behaviors.

We begin by studying a simple model in three dimensions without a moduli space, where we find that it displays the same characteristic behavior as the O(2) model~\cite{Hellerman:2015nra}.

The generic case in \ac{scft} can however involve a non-trivial moduli space of vacua, which drastically changes the behavior at large charge.
Of this kind, we present a simple example in three-dimensions, the $XYZ$ model~\cite{Hellerman:2017veg}, and $\mathcal{N}=2$ theories in four dimensions with a one-dimensional moduli space~\cite{Hellerman:2017sur,Hellerman:2018xpi,Hellerman:2020sqj}.\footnote{While we were finishing this work, a paper appeared discussing an interpolation between these two behaviors~\cite{Sharon:2020mjs}.}

\subsection{The supersymmetric $W=\Phi^3$ model}\label{sec:Phi3}

The simplest supersymmetric example to consider is an $\mathcal{N}=2$ model in $2+1$ dimensions with a single chiral superfield $\Phi = (\phi, \psi, F)$ and
\begin{align}
	K&=\Phi^\dagger\Phi, & W&= \frac{1}{3}\Phi^3.
\end{align}
This model is known to flow to an interacting superconformal fixed point and has no small parameters. It is special in that it contains no marginal deformations. At the fixed point, $\Phi$ has R-charge $=2/3$. We choose however to normalize the charge $Q$ to be $3/2$ times the R-charge, so that $\phi$ carries the charge $+1$.
We use a convention in which $W$ is unit normalized and has dimension $[\text{mass}]^2$. Based on the fact that the superpotential does not renormalize, we find that at the fixed point, $\Phi$ has engineering dimension $\Phi \propto [\text{mass}]^{2/3}$. Since the \ac{ir} Lagrangian must be classically scale invariant, so the Kähler potential has dimension 1 in $d=2$, and we pick
\begin{equation}
	K = \frac{16 b_K}{9} \abs{\Phi}^{3/2}.
\end{equation}
For $\phi$, this results in the kinetic term and potential
\begin{align}
	\Lag_{\text{kin}} &= b_K \abs{\phi}^{-1/2} \del_\mu \phi \del^\mu \bar\phi, & V&= \frac{1}{b_K}\abs{\phi}^{9/2}.
\end{align}
Independently of the Kähler potential, there is also a Yukawa coupling
\begin{equation}
	\Lag_{\text{Yuk}} = i\phi \psi_\uparrow\psi_\downarrow + (\text{h.c.}).
\end{equation}
As before, we now want to write the leading terms of the large-Q effective action in terms of $a=\abs{\phi}$ and $\chi=\arg{\phi}$:
\begin{align}
	\Lag_{\ac{lsm}} &= b_K a^{3/2}\del_\mu \chi\del^\mu\chi + b_K\frac{(\del a)^2}{a^{1/2}} -V(a)+ (\text{higher-derivative terms}) + (\text{fermions}).
\end{align}
As in the O(2) model, the action is minimized for $a = v =\text{const.}$ in which case the \ac{eom} for $a$ becomes
\begin{equation}
	\frac{3}{2}b_k a^{1/2}\del_\mu \chi\del^\mu\chi-\frac{9}{2b_K}a^{7/2}+\dots=0,
\end{equation}
which is solved by 
\begin{equation}
	\del_\mu \chi\del^\mu\chi = \frac{3}{b_k^2}a^3+ (\text{higher-derivative terms}) + (\text{fermions}).
\end{equation}
Eliminating $a$ classically gives the same structure for the leading-order \ac{nlsm} Lagrangian as in the O(2) model, plus fermionic terms:
\begin{equation}
	\Lag_{\ac{nlsm}} = k_0 (\del_\mu \chi\del^\mu\chi)^{3/2} + (\text{lower-order terms in }(\del_\mu \chi\del^\mu\chi)^{1/2} )+ (\text{fermions}).
\end{equation}
It turns out that the only real difference to the O(2) model at large charge, namely the fermionic terms, do not play a role in the effective description: just like in the case of the asymptotically safe \ac{cft} discussed in Section~\ref{sec:asymptotically-safe}, the fermions decouple. They acquire a large rest mass via the Yukawa coupling and the kinetic term.
This means in particular, that the ground-state solution breaks \ac{susy} completely and \ac{susy} is realized non-linearly by a massive Goldstino, \emph{i.e.} the fermionic mode that must have a rest energy proportional to $\abs{\del\chi}=\mu\propto\sqrt{Q}$.

The resulting scaling dimension associated to the ground state has again the form
\begin{equation}
	\Delta(Q) = c_{3/2}Q^{3/2} + c_{1/2}Q^{1/2} + \dots
\end{equation}
On general grounds, we expect for \ac{bps} states the behavior 
\begin{equation}
	\Delta(Q) = Q + S, 
\end{equation}
where $S$ is the spin of the state.
The fact that the ground state at large charge is thus not a \ac{bps} state but breaks \ac{susy} completely might be surprising at first sight.  But, in this model, one finds via a partition function calculation\footnote{We thank Richard Eager for sharing his unpublished results with us.}
that at fixed charge, no scalar \ac{bps} states exist and $S \propto Q^2$:
the first such state is parametrically heavier than the ground state at large charge.

\subsection{The $XYZ$ model}\label{sec:XYZ}

The models that we have encountered up to this point have one common feature.
They describe isolated fixed points in the phase diagram.
The physics is qualitatively different in the presence of a moduli space of Lorentz-invariant vacua on flat space.
This is for example the case for supersymmetric theories with an infinite chiral ring.
Having a degenerate spectrum when the curvature of the space vanishes means that the curvature is always relevant in the relationship between energy and charge and in general the low-energy states will not anymore satisfy a simple relation of the type \(E \propto Q^{(d+1)/d}\).
For a supersymmetric model with a moduli space the lowest state always saturates the \ac{bps} bound, fixing the conformal dimension to be equal to the charge \(\Delta = Q\).
Equivalently, the energy of the ground state on the sphere must be \(E = (R/(d(d-1)))^{1/2} Q\). We find a correspondence between the chiral ring operators and the helical solutions of the \ac{eft} at large charge. This creates a dictionary between the physics of the moduli space and the \ac{ope} coefficients of the \ac{cft}.

The simplest interacting theory with a moduli space is the three-dimensional \(XYZ\) model.
It describes three chiral superfields \(X\), \(Y\) and \(Z\) with superpotential \(W = g X Y Z\).
It is known that for \(g \gg 1\) (or, more precisely at scales \(E \ll g^2\)), the theory flows to an interacting fixed point where all dimensions are of order \(\order{1}\) and there are no perturbative parameters.
There is exactly one marginal operator and the moduli space consists of three branches, freely generated by \(X\), \(Y\) and \(Z\) respectively.
There are three independent \(U(1)\) global symmetries, rotating the three fields independently.
The \(R\)-charge is the linear combination of the three corresponding charges
\begin{equation}
  \label{eq:R-symmetry-and-U1}
  Q_R = \frac{2}{3} \pqty{Q_X + Q_Y + Q_Z} .
\end{equation}
A scalar superconformal primary is in the chiral ring if it satisfies the \ac{bps} bound \(\Delta = Q_R\).
We want to see how this algebraic constraint is realized in the large-R-charge \ac{eft}~\cite{Hellerman:2017veg}.

First observe that we can consistently choose one of the three branches, where the \acp{vev} of two of the fields (say \(Y\) and \(Z\)) vanish and the fields have masses above the cutoff of the \ac{eft}.
The \ac{rg} analysis of the possible terms is parallel to the one that we have discussed in Sec.~\ref{sec:top-down} for the \(O(2)\) model, with classical terms that now have the superspace form
\begin{equation}
 \Op = \int \dd[2]{\theta} \dd[2]{\bar \theta} \mathcal{I} ,
\end{equation} 
where \(\mathcal{I}\) is an operator that must satisfy the following conditions
\begin{itemize}
\item \(\mathcal{I}\) must have scaling dimension \(1\);
\item \(\mathcal{I}\) must be invariant under the Weyl transformation that transforms \(X\) into \(X \to e^{2\sigma/3}X\) (and its fermionic superpartner as \(\psi\to e^{7\sigma/6} \psi\));
\item \(\mathcal{I}\) must be compatible with supersymmetry.
\end{itemize}
It is convenient to do a field redefinition and introduce the field \(\phi = X^{3/4}\), that has dimension \([\phi]= 1/2\).
This transformation is well-behaved at large values of \(X\) (where the \ac{eft} is defined) and it is singular only at the origin of field space.
This becomes however relevant, together with charge quantization, in fixing the allowed number of quantum \(\phi\) excitations.

The first term that we can write is the usual kinetic term \(\del_\mu \phi^* \del^\mu \phi\), together with the usual conformal coupling to the curvature and its supersymmetrization.
There is no potential term, since \(\abs{\phi}^6\) is not compatible with supersymmetry.
The interactions are realized by higher-derivative terms.
The leading one is the \emph{unique Weyl-invariant operator with four derivatives}, the \ac{ftpr} term~\cite{Fradkin:1981jc,Fradkin:1982xc,Riegert:1984kt}:
\begin{equation}
  \Op_{\ac{ftpr}} = \frac{1}{\bar \phi} \pqty{ \del_\mu \del^\mu \del_{\nu}\del^\nu+ \del_\mu \pqty{\frac{5}{4} g^{\mu \nu} R - 4 R^{\mu\nu}} \del_\nu - \frac{1}{8} \Laplacian R + R^{\mu\nu}R_{\mu\nu} - \frac{23}{64} R^2  } \frac{1}{\phi} 
\end{equation}
and its super-Weyl-invariant completion, that in flat space takes the simple form~\cite{Kuzenko:2015jda}
\begin{equation}
  \Op_{\ac{ftpr}} = \int \dd[2]{\theta}  \frac{\del_\mu \Phi \del^\mu \Phi}{(\Phi \bar \Phi)^2} , 
\end{equation}
where  \(\Phi = \phi + \sqrt{2} \theta \psi + \dots \) is the chiral superfield.
It has been pointed out in~\cite{Adams:2006sv} that such a term can only appear with a positive sign in the effective action for a massless field. A negative sign would give rise to superluminal signal propagation, as well as unitarity violation in moduli scattering, within the regime of validity of the \ac{eft}.
The component decomposition on \(\setR \times S^2 \) can be found in~\cite{Hellerman:2017veg}.

Fixing the R-charge does not break supersymmetry completely.
It follows that the operator of lowest dimension saturates the bound \(\Delta = Q_R\).
In turn this implies that the ground state on the cylinder must have energy \(E_0 = Q_R/ R_0\).
Generically this helical solution will preserve a linear combination of time translations and R-symmetry (see the discussion after Eq.~\eqref{eq:symmBreakingO2}) and its angular frequency is \(\omega = \dv*{E}{Q_R} = 1/R_0\).
Given the relationship between R-symmetry and the \(U(1)\) symmetry that rotates the field \(X \) in Eq.~\eqref{eq:R-symmetry-and-U1} it follows that the ground state in the X-branch must have the form
\begin{equation}
  \bar X = X_0 e^{2i t/(3 R_0)} \Rightarrow \bar \phi = \phi_0 e^{i t/(2R_0)}.
\end{equation}
The amplitude \(\phi_0\) is not fixed by the symmetries but depends on the Wilsonian coefficients in the \ac{eft}.
However, since the \ac{eft} cannot have dimensionful parameters,  it must be that
\begin{equation}
  \phi_0 \propto \pqty{\frac{Q_R}{R_0} }^{1/2} .
\end{equation}
Just like in the non-supersymmetric \ac{eft} discussed in the previous sections, this implies that the derivative and curvature expansion of the effective Lagrangian is actually an expansion in inverse powers of \(Q_R\).
The leading term is the two-derivative kinetic one, which controls the leading contribution to the amplitude \(\phi_0\).
Explicitly, if the bosonic component of the Lagrangian is
\begin{equation}
  \Lag = k \pqty{\del_\mu \phi^* \del^\mu \phi - \frac{R}{8} \phi^* \phi } + \dots
\end{equation}
and the R-charge acts on \(\phi \) as
\begin{equation}
  \phi \to e^{i \epsilon /2}\phi ,
\end{equation}
the ground state solution at fixed charge \(Q_R\) is
\begin{equation}
  \bar \phi = \pqty{\frac{Q}{2 \pi k R_0} }^{1/2} e^{i t/(2R_0) }.
\end{equation}
The contribution of the \ac{ftpr} term vanishes when evaluated on the ground state, which means that there are no corrections at least up to order \(1/Q_R^2\).
It should be possible to prove that all the corrections cancel order-by-order in \(1/Q_R\), since this state is protected by supersymmetry.

Apart from realizing explicitly the properties of the ground state, which is one of the basic predictions of superconformal invariance, the \ac{eft} can also be used to study near-\ac{bps} states which satisfy \(\abs{\Delta - Q_R} \ll Q_R\), in the spirit of~\cite{Berenstein:2002jq,Gross:2002su}.
This is done in~\cite{Hellerman:2017veg} and we refer the reader to the original literature for this discussion.

\subsection{$4d$ $\mathcal{N}=2$ models with 1d moduli space}%
\label{sec:4d-1dmoduli}

In the previous example we have seen how the large-charge \ac{eft} reproduces the \ac{bps} bound from supersymmetry.
While this tells us that the \ac{eft} is consistent, it would be more interesting to extract results that could not be predicted by using the constraints from the algebra alone.
This is for example possible in four-dimensional \(\mathcal{N}=2\) theories with a dimension-one Coulomb branch~\cite{Hellerman:2017sur,Hellerman:2018xpi,Hellerman:2020sqj}.
The \ac{eft} for these theories is strongly constrained by supersymmetry and, together with an integrability condition, this will let us resum completely the large-charge expansion and obtain results only corrected by non-perturbative terms, exponentially suppressed by the R-charge.

Just like in the previous example, the states of lowest fixed R-charge are \ac{bps} and their conformal dimension is fixed to be \(\Delta = Q\).
We will use the \ac{eft} instead to compute the three-point function coefficients.
Let \(\Op(x)\) be the operator that generates the chiral ring.
The only non-vanishing three-point functions are those of the form
\begin{align}
  \ev{\Op^{n_1}(x_1) \Op^{n_2}(x_1') \bar \Op^{n_3}(x_2)}, && n_1 + n_2 = n_3 .
\end{align}
The \ac{ope} of two chiral-ring operators is regular, so we can take the limit of \(x_1' \to x_1\) where the three-point function reduces to a two-point function:
\begin{equation}
  \ev{\Op^{n_1}(x_1) \Op^{n_2}(x_1') \bar \Op^{n_3}(x_2)} \xrightarrow[x_1' \to x_1  ]{} \ev{\Op^n(x_1) \bar \Op^n(x_2)} = \frac{C_{\Op^n}}{\abs{x_1 - x_2}^{2n \Delta}} .
\end{equation}
In this way, computing the normalization of the  two-point function we recover the fusion coefficients \(\lambda_{\Op}\).
Note that this calculation will depend on the normalization of the field \(\Op(x)\).
It follows that it will be convenient to compute directly the two-point function in terms of insertions in the path integral, as opposed to via the state-operator correspondence, which is not sensitive to this normalization.
In turn, this introduces a new issue.
The large-charge \ac{eft} is only well-defined on a compact space.
No matter how large the charge is, far away from the insertions, the charge density will vanish and the \ac{eft} breaks down.
For this reason, even if we are interested in the theory on \(\setR^4\), in order to use the \ac{eft} we need to apply a Weyl transformation and work on \(S^4\).

In path-integral terms, the two-point function of interest is
\begin{equation}
  \label{eq:2pt-fn-N=2-d=4}
  \ev{\Op^n(x_1) \bar \Op^n(x_2)} = \frac{1}{Z} \int \DD{\Op} \Op^n(x_1) \bar \Op^n(x_2) \exp[- \int \Lag_{\acs{eft}}[\Op]].
\end{equation}
In the limit of large charge, \(n \to \infty\), we expect the integral to localize around its saddle point.
To see this explicitly, it is convenient to exponentiate the insertions and minimize the action,
\begin{equation}
  S_{\text{full}} = S_{\acs{eft}} + S_{\text{sources}} = \int \dd[4]x \bqty{ \Lag_{\ac{eft}} + n \log(\Op(x)) \delta(x-x_1) - n \log(\Op(x)) \delta(x-x_2)}.
\end{equation}
The path integral will localize around the saddle in the limit \(n \to \infty\) or equivalently, in terms of the R-charge, \(Q = n \Delta \to \infty\).

How do we write the \ac{eft}?
By assumption the \ac{ir} theory contains a single vector multiplet, whose lowest component is a complex scalar \(\varphi\).
The kinetic term for \(\varphi\) must describe a flat moduli space and take the free-field form
\begin{equation}
  S_{\text{kin}}[\varphi] = \int\dd[4]{x} \frac{\Im(\tau)}{4 \pi} \del_\mu \bar \varphi \del^\mu \varphi .
\end{equation}
It is actually convenient to reabsorb the parameter \(\tau\) and define a new field \(\phi\) with a canonically-normalized kinetic term,
\begin{equation}
  S_{\text{kin}}[\phi] = \int\dd[4]{x} \del_\mu \bar \phi \del^\mu \phi .
\end{equation}
In turn, the generator of the chiral ring \(\Op\) must be related to \(\phi\) by
\begin{equation}
  \Op^n(x) = N_{\Op}^Q \phi^Q(x) ,
\end{equation}
where \(N_{\Op}\) is a normalization that has to be fixed independently.
Using the free action for \(\phi\) and the source term, we find that the full action has a saddle for
\begin{align}
  \phi(x) &= \frac{e^{i \beta_0} \abs{x_1 - x_2}}{2 \pi (x - x_2)^2} Q^{1/2}, & \bar \phi(x) &= \frac{e^{-i \beta_0} \abs{x_1 - x_2}}{2 \pi (x - x_1)^2} Q^{1/2} ,
\end{align}
and the \ac{vev} of the action is
\begin{equation}
  \ev{S_{\text{full}}} = Q \pqty{- \log(Q) + 2 \log(\abs{x_1 - x_2}) + 2 \log(2 \pi) + 1} .
\end{equation}
The \(\log(\abs{x_1 - x_2})\) term reproduces the expected position dependence of the two-point function, while the \(Q \log(Q)\) term is the leading contribution to the fusion coefficient.

Of course a free action cannot be the end of the story.
For one, the \ac{eft} must reproduce the \(a\) anomaly of the theory that we want to describe.
This is possible if we add an interaction in the form of a four-derivative \acl{wz} term, with a coupling \(\alpha\) chosen to compensate for the difference in the anomaly between our single-vector multiple \ac{eft} and the full \ac{cft}.
The \ac{wz} term is a function of the dilaton \(\tau\), normalized so that \(e^{- \tau}\) transforms as a scalar of dimension one.
This must be the modulus \(\abs{\phi(x)}\) of the field that spontaneously breaks the scale invariance as we have seen for example in Section~\ref{sec:bottom-up-approach}.
It follows the identification
\begin{equation}
  \tau = - \log(\frac{\abs{\phi}}{\mu} ),
\end{equation}
where \(\mu \) is some mass scale.

We have seen that the \ac{vev} of \(\abs{\phi}\) is at leading order proportional to \(Q^{1/2}\).
The leading contribution of the \ac{wz} term is proportional to \(\tau\), and so it will scale as \(\log(Q)\).
More precisely,
\begin{equation}
  \eval{\Lag_{\acs{wz}}}_{\log(Q)} = - \tau \Delta a E_4,
\end{equation}
where \(\Delta a\) is the difference in the \(a\)-anomaly coefficients between the \ac{eft} and the \ac{cft} and \(E_4\) is the Euler density.
In the case of interest, in the \(S^4\) frame, we find that the contribution to the action at the saddle is
\begin{equation}
  \eval{\ev{S_{\acs{wz}}}}_{\log(Q)} = \alpha \tau ,
\end{equation}
where \(\alpha = 2 (a_{\acs{cft}} - a_{\ac{eft}})\) and \(a_{\ac{eft}}\) is the \(a\)-coefficient of the \ac{eft} of massless fields in moduli space\footnote{We use the conventions in~\cite{Anselmi:1997ys} for the normalization of the \(a\) anomaly.}.
Putting it all together, we find that the leading expression in $Q$ for the two-point function in Eq.~(\ref{eq:2pt-fn-N=2-d=4}) is
\begin{equation}
  \int \DD{\phi} \phi^Q(x_1) \bar \phi^Q(x_2) e^{-S_{\acs{eft}}} = \pqty{\frac{N_\Op}{2 \pi \abs{x_1 - x_2}} }^{2Q} Q! Q^\alpha \pqty{1 + \order{Q^{-1}}}.
\end{equation}

A key difference between the theories that we have described in the previous section and the one at hand is that the \ac{eft} made of the kinetic and \ac{wz} term is all there is at tree level.
In a theory of rank one there are no other possible superconformally-invariant higher-derivative F-terms.
All higher-order contributions in \(1/Q\) to the two-point function must come from quantum effects due to the interaction described by the \ac{wz} term and will be functions of \(\alpha\) alone even if the theory has a marginal coupling.
It is convenient to define the quantities \(q_n\) as
\begin{equation}
  \ev{\Op^{n}(x_1) \bar \Op^n(x_2)} = \frac{e^{q_n - q_0}}{\abs{x_1 - x_2}^{2n\Delta}} 
\end{equation}
and recast our leading-order large-charge result in the form
\begin{equation}
  \label{eq:qn-EFT-expansion}
  q_n = A Q + B + Q \log(Q) + \pqty{\alpha +  \frac{1}{2}} \log(Q) + \sum_{m=1}^{\infty} \frac{k_m(\alpha)}{Q^m} ,
\end{equation}
where \(A\) and \(B\) depend on the normalizations, and the coefficients \(k_m(\alpha)\) can be computed perturbatively using the effective action.
$k_1$ is for example given by
\begin{equation}
  \label{eq:k1-in-N=2}
  k_1(\alpha) = \frac{1}{2} \pqty{\alpha^2 + \alpha + \frac{1}{6} } .
\end{equation}
\acused{sqcd}
If the theory has a marginal coupling \(\tau\), as it is for example the case for \(\mathcal{N}=2\) \(SU(2)\) \ac{sqcd},
the coefficients \(A\) and \(B\) will depend on \(\tau\), but the \(k_m(\alpha)\) will not.
This is important because in this case one can show that the \(q_n\) obey a Toda-lattice equation~\cite{Papadodimas:2009eu,Baggio:2014ioa,Baggio:2014sna,Baggio:2015vxa,Gerchkovitz:2016gxx}:
\begin{equation}
  \del_\tau \del_{\bar \tau} q_n(\tau, \bar \tau) = e^{q_{n+1}(\tau, \bar \tau) - q_n(\tau, \bar \tau)} - e^{q_n(\tau, \bar \tau) - q_{n-1}(\tau, \bar \tau)} .
\end{equation}
Plugging the \ac{eft} expansion into Eq.~(\ref{eq:qn-EFT-expansion}) and using the one-loop result in Eq.~(\ref{eq:k1-in-N=2}) as a boundary condition, the recursion relation can be solved up to two functions of \(\tau\) and \(\bar \tau\):
\begin{equation}
  q_n(\tau, \bar \tau)= Q A(\tau, \bar \tau) + B(\tau, \bar \tau) + \log(\Gamma(Q + \alpha + 1)),
\end{equation}
where \(A\) and \(B\) are solutions to a Liouville equation~\cite{Hellerman:2020sqj}.
Finally, we can write the two-point function in a compact form:
\begin{equation}
  \ev{\Op^{n}(x_1) \bar \Op^n(x_2)} = c_n(\tau, \bar \tau) \frac{\Gamma(n \Delta + \alpha + 1)}{\abs{x_1 - x_2}^{2n \Delta}}  ,
\end{equation}
where
\begin{equation}
   c_n(\tau, \bar \tau) = e^{n A + B} + \order{e^{-k \sqrt{n}}} .
\end{equation}
This \emph{completely resums the large-charge expansion}, up to exponentially-suppressed terms.

Note that the \(\tau\)-dependence enters only via the coefficient \(c_n\), which also has a precise dependence in \(n\).
In particular, the second variation of \(q_n\) with respect to \(n\) is \(\tau\)-independent.
This means that, even though we have computed the explicit form of \(q_n\) using the Toda-lattice equation, the value that we have obtained for this second variation is valid for \emph{any rank-one theory}, including the non-Lagrangian Argyres--Douglas theories~\cite{Argyres:1995jj} in the classification of \cite{Argyres:2015ffa}.
Explicitly, we can define
\begin{equation}
  \lambda_{\Op^{2n}} = \exp(q_{n+1} + q_{n-1} - 2 q_{n}) = \frac{\Gamma((n-1) \Delta + \alpha + 1) \Gamma((n+1) \Delta + \alpha + 1)}{\Gamma(n \Delta + \alpha + 1)^2} .
\end{equation}
This same quantity for \(n =1 \) has been recently the object of study using bootstrap techniques (see~\cite{Gimenez-Grau:2020jrx}) and the numerical values obtained with the \ac{eft} approach are consistent with the bootstrap bounds with an error of order \(1\%\), which can be attributed to the exponentially-suppressed  non-perturbative corrections.

\begin{figure}[b]
  \begin{tabular}{lr}
    \includegraphics[width=.45\textwidth]{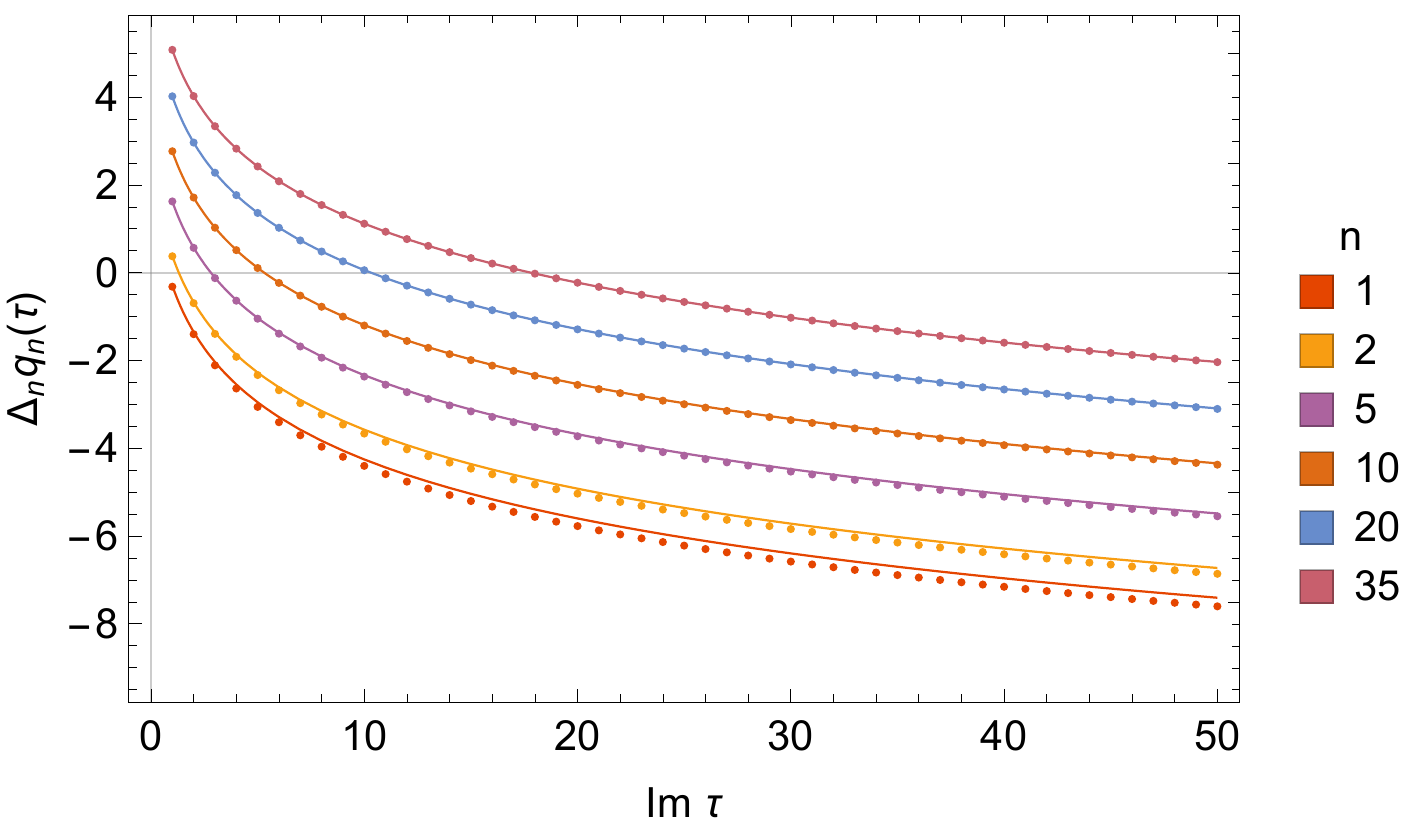} &
                                                                 \includegraphics[width=.45\textwidth]{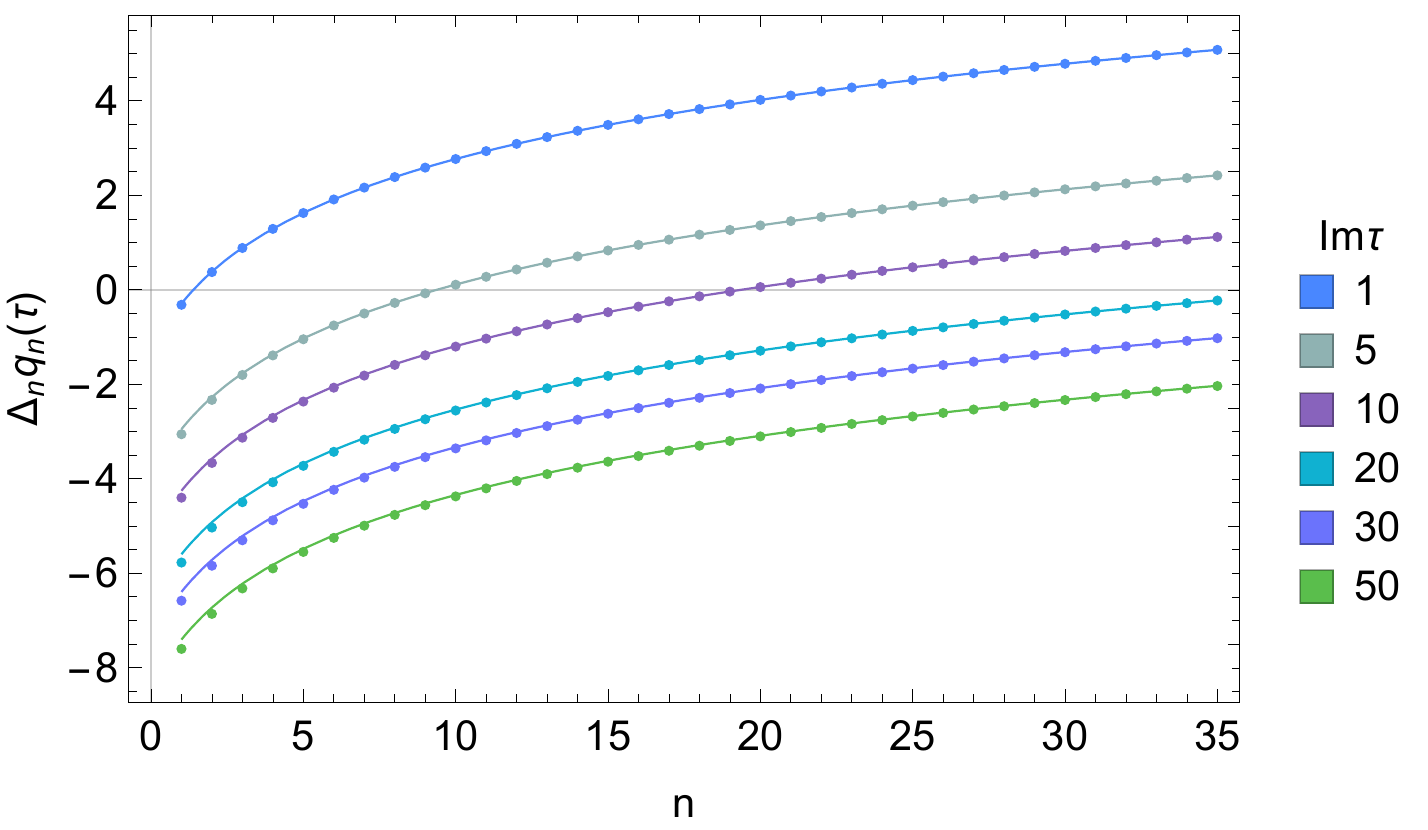}
  \end{tabular}
  \caption{First variation \(\Delta_n q_n(\tau) = q_{n+1}(\tau) - q_n(\tau)\) as function of \(\Im(\tau)\) for fixed values of \(n\) (left) and as function of \(n\) for fixed values of \(\Im(\tau)\) (right). The dots are numerical values estimated from localization, the continuous lines are the \ac{eft} prediction in~\cite{Hellerman:2020sqj}.}
  \label{fig:DeltaQ-prediction}
\end{figure}

In the case of \ac{sqcd}, where $\tau$ is the marginal coupling, we can choose a scheme that preserves explicit S--duality invariance, which becomes a boundary condition for the Liouville equation satisfied by $A$ and $B$~\cite{Hellerman:2020sqj}. This allows a direct comparison to the numerical estimates for 
the coefficients \(\lambda_{\Op^{2n}}\) from localization~\cite{Gerchkovitz:2016gxx}.
Figure~\ref{fig:DeltaQ-prediction} shows the remarkable agreement between these estimates and the \ac{eft} results.

\section{Alternative approaches}
\label{sec:alternative}

Apart from the examples described in great detail in this review, there are also a number of other approaches related to the large-charge expansion that have appeared in the literature and deserve mention.
In this section, we single out lattice simulations that give an independent confirmation for some large-charge predictions put forward here, %
and then 
we list a number of other topics related to the large-charge expansion, a detailed discussion of which would go beyond the scope of this already sizable review.

\subsection{Lattice results}
\label{sec:numerics}

We have seen that the large-charge expansion is self-consistent.

Since we have mostly dealt with strongly-coupled systems which are notoriously hard to access, an independent verification of the large-charge predictions has however been hard to come by (an exception is given by the \ac{sqcd} model discussed in Section~\ref{sec:4d-1dmoduli}, for which a comparison to the results of supersymmetric localization computations was possible).
Moreover, by construction, the large-charge expansion depends on a set of parameters (for example the coefficients \(c_i\) in the expression for the conformal dimension in Eq.~\eqref{eq:EVLapSphere}) that enters as input and cannot be computed within the \ac{eft}.\footnote{If we introduce an extra controlling parameter as in Section~\ref{sec:O2n-largeN}, their direct determination is in principle possible.}
It is therefore important to pursue alternative, non-perturbative, techniques to verify the \ac{eft} predictions at large charge.

Obvious candidates to verify our predictions are numerical simulations that make it possible to work directly at strong coupling. The conformal dimensions for the lowest operators of charge $Q$ have been computed in the cases of the \(O(2)\) and \(O(4)\) vector models in~\cite{Banerjee:2017fcx,Banerjee:2019jpw} via lattice simulations.
The main outcome of these works is that the predictions for the conformal dimension as a function of the charge are confirmed to high precision and that --- surprisingly --- in these models, the large-charge approximation is still viable and reproduces the non-perturbative results even for charges of order one. Moreover, the two-parameter fit of our prediction to the numerical data has yielded numerical values for the parameters $c_{1/2}$ and $c_{3/2}$.

The detailed description of the lattice techniques goes beyond the scope of this review.
We will however highlight some of the most salient points and invite the interested reader to consult the original literature for details.

\begin{itemize}
\item The two-point function in the \(O(2)\) model has been computed in~\cite{Banerjee:2017fcx} generalizing the Worm algorithm of~\cite{PhysRevD.81.125007} to include local sources.
  The method is based on the decomposition of the partition function in terms of Bessel functions, in order to sidestep sign problem issues~\cite{Chandrasekharan:2008gp}.
\item The energy of the torus for the \(O(2)\) model was again computed using a worm algorithm, but this time introducing a chemical potential for the global symmetry and computing the energy difference \(E(Q) - E(Q-1)\) by tuning the value of \(\mu\)~\cite{Banerjee:2017fcx}.
\item In the case of the \(O(4)\) model it was not possible to use directly the model given by the Lagrangian in Eq.~(\ref{eq:lsm_ON}) because it leads to sign problems. Instead, in~\cite{Banerjee:2019jpw} the authors used a vicious walkers model (non-intersecting paths) that had been originally introduced in~\cite{PhysRevD.77.014506} as a model for pion physics in two-color \ac{qcd} and was shown to have the same critical exponents as the \(O(4)\) model.
\end{itemize}

\begin{figure}
  \begin{tabular}{lr}
    \includegraphics[height=12em]{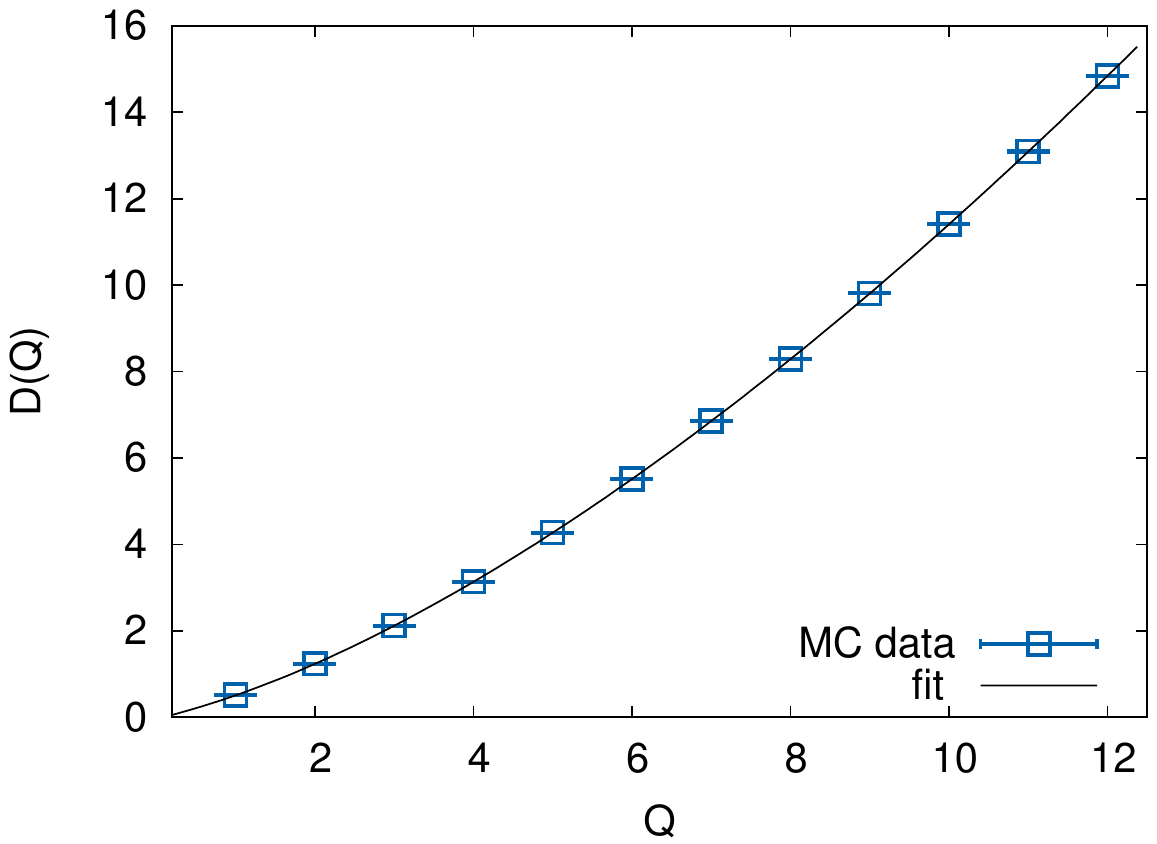} &
                                                                 \includegraphics[height=12.25em]{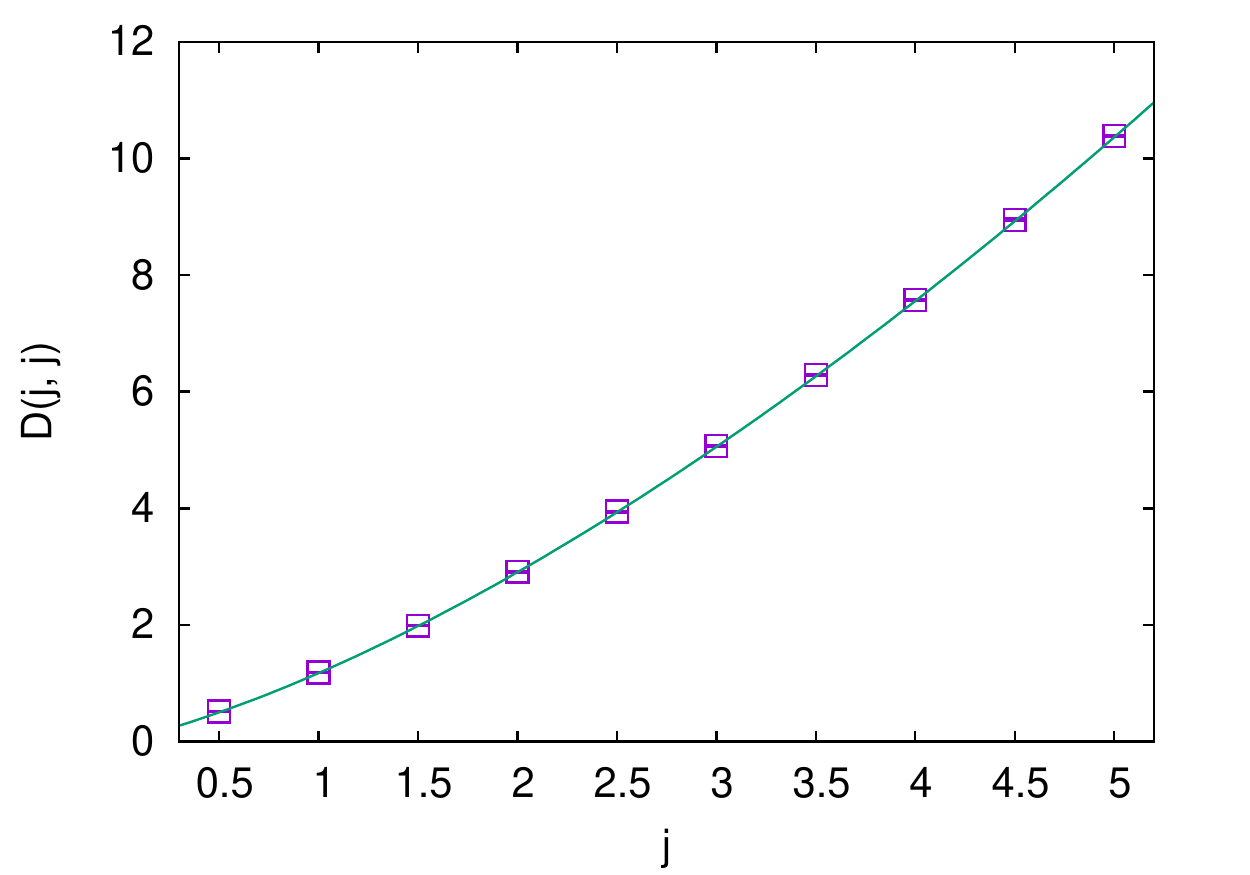}
  \end{tabular}
  \caption{Plot of the values of the conformal dimension of the lowest operator as function of the fixed charge extracted from the Monte Carlo calculations at the $O(2)$ Wilson--Fisher fixed point~\cite{Banerjee:2017fcx} (left) and at the $O(4)$ Wilson--Fisher fixed point~\cite{Banerjee:2019jpw} (right), together with the plot of the large-charge predictions in Eq.~(\ref{eq:EVLapSphere}) (solid line)}
  \label{fig:lattice-dimensions}
\end{figure}

The results of the simulations are reported in Figure~\ref{fig:lattice-dimensions}.
In both examples we see that the large-charge prediction is extremely successful at reproducing the numerical estimates, even for small values of the charge.

\subsection{Other topics not included in this review}
\label{sec:not-included}

In this review, we have mainly focused on the most basic principles of the large-charge expansion exemplified in the O(2) model, and the interesting effects arising from a non-Abelian global symmetry group as evidenced in the O(2n) model. As the literature on the large-charge expansion is growing continuously, there is a number of very interesting topics that we have not discussed here, partially due to limitations in space and partially due to the limits of our own expertise.
While these topics are being actively explored and deserve mentioning, we must confine ourselves to referring the reader to the original literature.
\begin{itemize}
\item One of the main features of the large-charge approach is that it can be used for strongly-coupled systems that are not otherwise perturbatively accessible.
  However, if the model at hand has a small control parameter \(\epsilon\), one can consider a double-scaling limit in which, schematically, \(\epsilon \to 0\) and \(Q \to \infty\), but their product \(\kappa = \epsilon Q\) is kept fixed.
  We have seen for example this strategy in action in the case of the large-\(N\) limit in Section~\ref{sec:O2n-largeN} where \(\epsilon = 1/N\), and in the case of the asymptotically free model of Section~\ref{sec:asymptotically-safe} where \(\epsilon = 11/2 - N_f/N_c\). The advantage is that one can then combine the semiclassical large-\(Q\) analysis with other, more standard, perturbative techniques for \(\epsilon \to 0\).
  This has been done recently in a series of works, among which~\cite{Watanabe:2019pdh,Arias-Tamargo:2019xld,Badel:2019oxl,Arias-Tamargo:2019kfr,Antipin:2020abu} for vector models in \(d = 4 - \epsilon\) dimensions, \cite{Badel:2019khk} in \(d = 3 - \epsilon\) dimensions, \cite{Arias-Tamargo:2020fow} in \(d = 6-\epsilon\), \cite{Antipin:2020rdw} for \(U(N) \times U(N)\) non-Abelian Higgs theories in \(d = 4 - \epsilon\) dimensions, and in~\cite{Bourget:2018obm,Beccaria:2018xxl,Grassi:2019txd,Beccaria:2020azj} for \(\mathcal{N} = 2\) \ac{sqcd} in four dimensions.
  In all these cases, the double-scaling parameter \(\kappa\) plays a role akin to the 't Hooft coupling and allows bridging between the weakly-coupled \ac{uv} regime and the \ac{ir} fixed point.
  Since we cannot do justice to this rapidly-evolving and promising line of research, we refer the interested reader to the original literature.
\item Since in a sector of large charge we have analytic control over a strongly-coupled \ac{cft}, it would be very interesting to explore the holographic duals of these models. First attempts in this direction have been made in~\cite{Loukas:2018zjh,delaFuente:2020yua,Nakayama:2020dle,Liu:2020uaz}, but this topic clearly deserves a lot more attention.
\item More in general, dual descriptions of the same physics can lead to interesting insights.
  The authors of~\cite{Cuomo:2017vzg,Cuomo:2019ejv} for example have used the particle-vortex duality to access the higher-spin fixed-charge operators in the \(O(2)\) model.
  In the same vein, the study of Chern--Simons-matter dualities has been initiated in~\cite{Watanabe:2019adh}.
\item Much of the progress made in recent years in \acp{cft} in $d>1$ is due to bootstrap methods~\cite{Simmons-Duffin:2016gjk}. A comparison of large-charge and bootstrap results would thus be very useful. A technical obstacle is that the regime typically accessed via bootstrap is not same where the large-charge \ac{eft} is well-defined. The only case which allows direct comparison are the $\mathcal{N}=2$ theories discussed in Section~\ref{sec:4d-1dmoduli}, where we could completely resum the large-charge expansion. \\
  Some early large-charge predictions have been recovered via bootstrap methods in~\cite{Jafferis:2017zna}.
\item Non-relativistic systems with Schrödinger symmetry (also called non-relativistic \acp{cft}) share many similarities to \acp{cft}, such as strong constraints on the correlation functions and a state-operator correspondence (albeit involving a harmonic potential instead of working on a sphere)~\cite{Nishida:2007pj}. Such theories are of special interest as they can be experimentally realized by a cold Fermi gas in a trap tuned to unitarity~\cite{randeria2012bcs,Son:2005rv}. It turns out that also this class of theories lends itself to working at large charge. This avenue has been explored in~\cite{Favrod:2018xov,Kravec:2018qnu,Kravec:2019djc}.
\end{itemize}

\subsection*{Acknowledgments}

D.O. and S.R. would like to thank Debasish Banerjee, Shailesh Chandrasekharan, Gabriel Cuomo, Richard Eager, Andrew Gasbarro, Simeon Hellerman, Zohar Komargodksi, Riccardo Rattazzi, Francesco Sannino, Hersh Singh and Masataka Watanabe for enlightening discussions and Ioannis Kalogerakis, Rafael Moser and Vito Pellizzani for comments on the manuscript.
L.A.--G. would like to thank Miguel A. Vazquez--Mozo for discussions on \ac{ssb} and constrained systems. 

The work of S.R. is supported by the Swiss National Science Foundation under grant number \textsc{pp00p2\_183718/1}.
D.O. acknowledges partial support by the \textsc{nccr 51nf40--141869} ``The Mathematics of Physics'' (Swiss\textsc{map}).
D.O. and S.R. would like to thank the Simons Center for Geometry and Physics for hospitality during part of this work.

\newpage

\appendix

\section{Mathematical background}
\label{sec:MathematicalBackground}

We collect here a number of mathematical results that are useful in the construction of effective actions.
Most of the results can be found in standard texts on differential geometry (see for instance~\cite{doi:10.1142/11058}).
The Lie derivative with respect to vector fields on a manifold is defined on the full tensor algebra and satisfies the Leibnitz rule
\begin{equation}
  \Lie(\comm{X}{Y}) = \comm{\Lie(X)}{\Lie(Y)},
\end{equation}
where \(\comm{X}{Y}\) is the standard Lie bracket between vector fields.
We will be considering also the action of \(\Lie(X)\) on differential forms on manifolds.
Given a Riemannian manifold \(M\) of dimension \(n\), we can consider the space of \(p\)-forms \(\Lambda^p(M)\) that generate the De Rham complex.
The exterior derivative 
\begin{equation}
\dd : \Lambda^p(M) \to \Lambda^{p+1 }(M)
\end{equation}
satisfies \(\dd^2 = 0\) and it is a graded differential (or antiderivation) of degree one.
For any vector field \(X\) we define the interior product 
\begin{equation}
	i(X): \Lambda^p(M) \to \Lambda^{p-1}(M).
\end{equation}
It is the unique antiderivation of degree \(-1\) on the exterior algebra and such that on one-forms \(i(X)\omega = X^a \omega_a\), and it obeys a graded Leibnitz rule in the exterior algebra.
After introducing \(\dd\) and \(i(X)\) it is not difficult to prove that
\begin{equation}
  \label{eq:CWZ-12}
  \Lie(X) = i(X)\dd + \dd i(X).
\end{equation}
Since we did not define \(\Lie(X)\) in general, we can take Eq.~(\ref{eq:CWZ-12}) as its definition on differential forms.
Then, on \(p\)-forms the following identities hold:
\begin{align}
  \comm{\dd}{\Lie(X)} &= \comm{i(X)}{\Lie(X)} = 0, \\
  i(\comm{X}{Y}) &= \comm{i(X)}{\Lie(Y)} = \comm{\Lie(X)}{i(Y)} .
\end{align}

Next recall the adjoint action of the group and the Lie algebra.
For the Lie algebra defined with the bracket
\begin{equation}
  \label{eq:CWZ-14}
  \comm{X_A}{X_B} = f\indices{_A_B^C}X_C,
\end{equation}
the adjoint action is
\begin{equation}
  \ad(X) Y = \comm{X}{Y}
\end{equation}
and the Jacobi identity becomes
\begin{equation}
  \ad(\comm{X}{Y}) = \comm{\ad(X)}{\ad(Y)} .
\end{equation}
In terms of structure constants
\begin{equation}
  (T_A)\indices{_B^C} = -f\indices{_A_B^C}
\end{equation}
satisfies Eq.~(\ref{eq:CWZ-14}).
We will use a scalar product \(\ev{T_A T_B} = \delta_{AB}\) for simple compact groups.
The adjoint representation of the group is given by
\begin{equation}
  \Ad(g)T_A =   g^{-1} T_A g = D_{AB}(g) T_B .
\end{equation}

\section{Finite volume}
\label{sec:finite-volume}

Two crucial ingredients of the large-charge expansion are the compactification on a sphere (or a more general manifold \mani) and the fact that the low-energy dynamics is dictated by Goldstone's theorem.
But there is a tension.
\ac{ssb} can only happen for infinite systems. 
In this section we will see how finite-volume effects are exponentially controlled in the large-charge limit for \(d > 1\).
The \(d = 1\) case (two spacetime dimensions) must be discussed separately.\footnote{In this appendix we only address one particular aspect of finite-size effects on symmetry breaking. We invite the interested reader to consult~\cite{koma1994symmetry} and references within for a more comprehensive study.}

The standard argument against \ac{ssb} in finite systems is that in this case, all the states in the theory live in a unique separable Hilbert space,
the symmetry is realized by a unitary operator acting on this space and it relates all possible minima to the unique ground state.
This is to be contrasted with the situation in an infinite-volume system.
Now the states of the theory live in a non-separable Hilbert space and the algebra of observables has a family of inequivalent irreducible representations.
Each representation has its own (separable) Hilbert space and its own ground state.
The symmetry maps those representations into each other but it is not a well-defined unitary operator, so we have a family of ground states and (one or more) Goldstone bosons that relate them.

To make this more explicit and see the implications on the large-charge expansion, consider the fluctuations over the fixed-charge ground state of a system with \(U(1)\) symmetry. 
The Lagrangian takes the form
\begin{equation}
  \Lag = \frac{1}{2} \del_\mu \chi \del^\mu \chi ,
\end{equation}
where \(\chi\) is a real field.
The \(U(1)\) symmetry is non-linearly realized as \(\chi \mapsto \chi + f \epsilon \), where \(f \) is a dimensionful constant that depends on the \ac{vev} of the initial field.
In the case of a \ac{cft} with broken symmetries due to working at fixed charge, \(f\) can only depend on the scale fixed by the charge and must be given by
\begin{equation}
  f^2 = c_f \Lambda_Q^{d-1},
\end{equation}
where \(c_f\) is a dimensionless parameter, characteristic of the \ac{cft} at hand.
We decompose \(\chi(t,\mathbf{x})\) into modes,
\begin{equation}
  \chi(t,\mathbf{x}) = \chi_0 + \pi_0 t + \sum_{k\neq 0} \frac{1}{\sqrt{2V k}} \pqty{ a(\mathbf{k}) e^{-i k t + i \mathbf{k} \mathbf{x} } + a^\dagger(\mathbf{k}) e^{i k t - i \mathbf{k} \mathbf{x} }},
\end{equation}
and we concentrate on the zero modes, which only appear in finite volume.
Promoting the field to an operator and imposing the standard equal-time commutation relations,
\begin{equation}
  \comm{\chi(t, \mathbf{x})}{\del_0 \chi(t, \mathbf{y})} = i \delta(\mathbf{x} - \mathbf{ y}),
\end{equation}
we see that the zero modes satisfy
\begin{equation}
  \comm{\chi_0}{\pi_0} = \frac{i}{L^d} ,
\end{equation}
where \(L = V^{1/d}\) is the scale fixed by the geometry.
As usual, it is convenient to express the modes in terms of ladder operators
\begin{equation}
  \begin{cases}
    a = \frac{1}{\sqrt{2}} L^{(d-1)/2} \pqty{\chi_0 + i L \pi_0} ,\\
    a^\dagger = \frac{1}{\sqrt{2}} L^{(d-1)/2} \pqty{\chi_0 - i L \pi_0} .
  \end{cases}
\end{equation}
They satisfy the usual commutation relation
\begin{equation}
  \comm{a}{a^\dagger } = 1 ,
\end{equation}
which we can use to generate a highest-weight representation of the algebra, starting from the vacuum \(\ket{0}\).

Using the fact that the non-zero modes integrate to zero, the charge operator is
\begin{equation}
  Q = \int \dd{x} f \fdv{\Lag}{\del_0 \chi}  = f \int \dd{x} \del_0 \chi = f V \pi_0 = \frac{f L^{(d-1)/2}}{\sqrt{2} i} (a - a^\dagger) .
\end{equation}
For any real constant \(\xi \), we introduce  \(\ket{\xi}\) as the state obtained by the action of \(Q\) on \(\ket{0}\):
\begin{equation}
  \ket{\xi} = e^{i \xi Q} \ket{0}.
\end{equation}
Given the expression of \(Q\), \(\ket{\xi}\) is a coherent state with parameter \(\alpha = \xi f L^{(d-1)/2}/\sqrt{2}\),
\begin{equation}
  \ket{\xi} = e^{\alpha a^{\dagger} - \alpha^* a} \ket{0}.
\end{equation}
By the usual properties of coherent states, its overlap with the vacuum is
\begin{equation}
  \braket{0}{\xi} = \ev{e^{\alpha a^{\dagger} - \alpha^* a}}{0} = e^{- \xi^2 f^2 L^{d-1}/4 },
\end{equation}
which we can rewrite in terms of the charge \(Q = (\Lambda_Q L)^d\):
\begin{equation}
  \braket{0}{\xi} = \exp[ - \xi^2 \frac{c_f}{4} Q^{(d-1)/d}] .
\end{equation}
This is precisely what we wanted.
Starting from the true vacuum \(\ket{0}\) and acting with the charge (that commutes with the Hamiltonian) we can construct a family of vacua \(\ket{\xi}\).
For a finite system all these vacua have a non-zero overlap, so eventually the system will relax to the true vacuum \(\ket{0}\). In an infinite system, \(Q\) is not a well-defined operator, so each vacuum belongs to a different Hilbert space and we have \ac{ssb}.
The large-charge limit is somewhat in between.
Different vacua have a non-zero overlap, but it is exponentially suppressed if \(d > 1\).
So, for all practical purposes, up to exponential corrections we can treat the system as if there was \ac{ssb} and use Goldstone's theorem.

\section{Fixed $\mu$ vs. fixed $Q$}
\label{sec:finite-mu-vs-Q}

We briefly compare the analogies and differences between finite chemical potential and finite charge.
For some relevant references see~\cite{Kapusta:1981aa,Schafer:2001bq,Miransky:2001tw,Andersen:2006ys}.
While the equations for fixed charge and fixed chemical potential look similar, their interpretation is different. The two cases correspond to the canonical and grand-canonical ensembles in thermodynamics. When the charge is fixed, the chemical potential is a function of the charge and only some values of $\mu$ are allowed.
Choosing other values of $\mu$ leads to behaviors that cannot occur at fixed charge.

As an example will study here the same \(O(2n)\) theory as in Section~\ref{sec:O2n}, but with a fixed chemical potential.
The Hamiltonian density is
\begin{align}
  \Ham &= \frac{1}{2} \pi_a^2 + \frac{1}{2} (\nabla \phi_a)^2  + V(\phi^2/2),  & a = 1, \dots, 2n.
\end{align}
The charges
\begin{equation}
  Q_I = \int \dd[d]{x} \pqty{ \phi_{2I-1} \pi_{2I} - \phi_{2I} \pi_{2I -1 }}
\end{equation}
generate the Cartan subalgebra and act on the  complex fields \(\varphi^I = (\phi_{2I-1} + i \phi_{2I})/\sqrt{2}\) as
\begin{equation}
  e^{i \alpha Q_I} \varphi_I e^{-i \alpha Q_I} = e^{i \alpha} \varphi_I .
\end{equation}
A chemical potential can be introduced for each \(Q_I\).
For simplicity we concentrate only on the \(I=1\) sector, and later introduce more fields.
The Hamiltonian density is then
\begin{equation}
  \Ham_\mu = \Ham + \mu J^0,
\end{equation}
and the Hamiltonian \ac{eom} are given by
\begin{align}
  \dot \phi_1 &= \pi_1 - \mu \phi_2, & \dot \phi_2 &= \pi_2 + \mu \phi_1.
\end{align}
The Lagrangian density yields
\begin{equation}
  \Lag_\mu = \frac{1}{2} \pqty{\dot \phi_1 + \mu \phi_2}^2 + \frac{1}{2} \pqty{\dot \phi_2 - \mu \phi_1}^2 + \dots = \abs{D_0 \varphi_1}^2 - \abs{\nabla \varphi_1}^2 - V(\abs{\varphi_1}), 
\end{equation}
where
\begin{align}
  D_0 \varphi &= ( \del_0 - i \mu ) \varphi, & V(\abs{\varphi}) = m^2 \abs{\varphi}^2 + \frac{\lambda}{2} \abs{\varphi}^4.
\end{align}
If we introduce more fields, for each pair \((\phi_{2I-1}, \phi_{2I})\) we can choose a different \(\mu_j\), but to compare to the homogeneous solution with finite \(Q\) we take them to be all equal, \(\mu_I = \mu\), and think of \(\varphi\) as a complex \(n\)-vector field:
\begin{equation}
  \Lag_\mu = \del_\mu \bar \varphi_I \del^\mu \varphi_I - (m^2 - \mu^2) \bar \varphi_I \varphi_I - \frac{\lambda}{2} (\bar \varphi_I \varphi_I)^2 + i \mu \pqty{\bar \varphi_I \del_0 \varphi_I - \del_0 \bar \varphi_I \varphi_I}.
\end{equation}
There are three cases:
\begin{enumerate}
\item If \(m^2 > \mu^2\), there is no \ac{ssb}. For every complex field the determinant of the kinetic \(2 \times 2\) matrix in momentum space is
  \begin{equation}
    \pqty{ \pqty{E + \mu}^2 - \omega_p^2} \pqty{ \pqty{E - \mu}^2 - \omega_p^2} = 0,
  \end{equation}
  with \(\omega_p^2 = p^2 + m^2\). The four zeros are at \(\pm \mu \pm \omega_p\) and the positive-energy excitations are
  \begin{align}
    E_p^{(1)} &= \omega_p + \mu = m +  \mu +  \frac{p^2}{2m} + \dots, \\
    E_p^{(2)} &= \omega_p - \mu = m - \mu + \frac{p^2}{2m} + \dots,
  \end{align}
  which are both massive.
\item In the limiting case \(m = \mu \), we have
  \begin{align}
    E_p^{(1)} &= 2\mu + \frac{p^2}{2\mu}  + \dots, \\
    E_p^{(2)} &=  \frac{p^2}{2\mu}  + \dots.     
  \end{align}
  This is the case closest to the fixed-charge analysis, where \(\mu\) is the same for all fields and we have \(n\) type-II Goldstone bosons.
\item Finally, take \(\mu^2 > m^2\). This leads to \ac{ssb}. The minimum of the potential satisfies
  \begin{equation}
    \ev{\bar \varphi \varphi} = \frac{\mu^2 - m^2}{\lambda} = A_0^2 .
  \end{equation}
  With a \(U(n)\) rotation we can write
  \begin{equation}
    \varphi =
    \begin{pmatrix}
      \eta_1 \\ \vdots \\ \eta_n
    \end{pmatrix} +
    \begin{pmatrix}
      0 \\ \vdots \\ A_0
    \end{pmatrix},
  \end{equation}
  and, similar to the finite-\(Q\) case, the low-energy excitations include \((n-1)\) type-II Goldstone bosons with energy \(E = p^2/(2\mu) + \dots\), and a type-I one with dispersion relation
  \begin{equation}
    E = \frac{2 \lambda A_0^2}{4 \mu^2 + 2 \lambda A_0^2}p^2 + \dots .
  \end{equation}
\end{enumerate}
So we see that cases (ii) and (iii) are close to the case of fixed charge, but there are more free parameters.
On general grounds it makes sense that fixed \(\mu\) and fixed \(Q\) have such similar properties, because classically, they are related by a Legendre transformation.

\printbibliography

\end{document}